\documentclass[twocolumn]{aastex62}
\usepackage{graphicx}   
\usepackage{multirow}

\graphicspath{{./}{figures/}}

\received{XXX}
\revised{XXX}
\accepted{XXX}
\submitjournal{AAS journal}

\shorttitle{Barred galaxies in the Illustris-1 and TNG100 simulations}
\shortauthors{Zhou et al.}

\begin{document}

\title{Barred galaxies in the Illustris-1 and TNG100 simulations: A comparison study}

\author{Ze-Bang, Zhou}
\affil{School of Physics and Astronomy, Sun Yat-Sen University, Zhuhai campus, No. 2, Daxue Road \\
Zhuhai, Guangdong, 519082, China}

\author{Weishan, Zhu}
\affil{School of Physics and Astronomy, Sun Yat-Sen University, Zhuhai campus, No. 2, Daxue Road \\
Zhuhai, Guangdong, 519082, China}

\author{Yang, Wang}
\affil{School of Physics and Astronomy, Sun Yat-Sen University, Zhuhai campus, No. 2, Daxue Road \\
Zhuhai, Guangdong, 519082, China}
\nocollaboration

\author{Long-Long, Feng}
\affil{School of Physics and Astronomy, Sun Yat-Sen University, Zhuhai campus, No. 2, Daxue Road \\
Zhuhai, Guangdong, 519082, China}
\affiliation{Purple Mountain Observatory, CAS, No.10 Yuanhua Road, Qixia District, Nanjing 210033, China}



\begin{abstract}
We carry out a comparison study on the bar structure in the Illustris-1 and TNG100 simulations. At $z=0$, 8.9\% of 1232 disc galaxies with stellar mass $>10^{10.5}M_{\odot}$ in Illustris-1 are barred, while the numbers are 55\% of 1269 in TNG100. The bar fraction as a function of stellar mass in TNG100  agrees well with the survey $S^4G$. The median redshift of bar formation are $\sim 0.4-0.5$ and $\sim 0.25$ in TNG100 and Illustris-1 respectively. Bar fraction generally increases with stellar mass and decreases with gas fraction in both simulations. Barred galaxy had higher gas fraction at high redshift tend to form bar later. When the bars were formed, the disc gas fractions were mostly lower than 0.4. The much higher bar fraction in TNG100 probably have benefit from much lower gas fraction in massive disc galaxies since $z\sim3$, which may result from the combination of more effective stellar and AGN feedback. The latter may be the primary factor at $z<2$. Meanwhile, in both simulations, barred galaxies have higher star formation rate before bar formation, and stronger AGN feedback all the time than unbarred galaxies. The properties of dark matter halos hosting massive disc galaxies are similar between two simulations, and should have minor effect on the different bar frequency. For individual galaxies under similar halo environment cross two simulations, different baryonic physics can lead to striking discrepancy on morphology. The morphology of individual galaxies is subject to combined effects of environment and internal baryonic physics, and is often not predictable.


\end{abstract}



\section{Introduction} \label{sec:intro}
Stellar bars are present in the inner regions of many disc galaxies in the local and high redshifts universe. The reported frequency of bars declines from $30\% \sim 70\%$ at $z = 0$ to $10\% \sim 20\%$ at $z=0.8$ in different observational studies(e.g., \citealt{2008ApJ...675.1194B}, \citealt{2008ApJ...675.1141S}, \citealt{2009A&A...495..491A}, \citealt{2010ApJS..186..427N}, \citealt{2011MNRAS.411.2026M}, \citealt{2014MNRAS.438.2882M}, \citealt{2015ApJS..217...32B}, \citealt{2016A&A...587A.160D}, \citealt{2017MNRAS.464.4420S}, \citealt{2017MNRAS.464.4176W}). Many of these observations suggest that massive and gas-poor disc galaxies are more likely to host bars than low mass and gas-rich galaxies(e.g. \citealt{2011MNRAS.411.2026M}, \citealt{2013ApJ...779..162C}), while some studies indicate the opposite trend(e.g. \citealt{2008ApJ...675.1194B},\citealt{2010ApJS..186..427N}), or argue that there is no difference(\citealt{2018MNRAS.474.5372E}). The bar plays important role in driving the secular evolution of disc galaxies by redistributing the gas, stars and even dark matter, as well as altering the angular momentum associated to these components(see \citealt{2013seg..book....1K}, for a review). For instance, bars could induce gas flowing into the galaxy central region and contribute to the formation of pseudo-bulges and bulges. On the other hand, the origin, growth and destroy of bars is a key piece of the galaxy evolution puzzle, and many details are still unclear. The answer to this issue is crucial to explain the presence or absence of bars in disc galaxies with different properties. 

Bar formation can be triggered either by internal secular evolution or by external processes, including merge and tidal effects of nearby galaxies. Early theoretical and N-body simulation studies suggested that massive cold stellar discs are highly vulnerable to instability, and bars can grow quickly in these stellar discs(e.g., \citealt{1973ApJ...186..467O}, \citealt{1977ARA&A..15..437T}, \citealt{1981seng.proc..111T}). However, this scenario is unable to account for the origin of bars in realistic galaxies, because these studies did not take account of several factors, such as dark matter halos, gas component, baryonic physics including cooling, star formation and feedback, and the impact of external processes (\citealt{2013MNRAS.429.1949A}; \citealt{2013seg..book....1K}).

Later idealized simulations including halos, gas component and gas physics showed that bar formation may be a gradual process, and both the dark matter halos and gas play important roles(e.g. \citealt{1998MNRAS.300...49B}, \citealt{2000ApJ...543..704D}, \citealt{2002MNRAS.330...35A}, \citealt{2002ApJ...569L..83A}, \citealt{2003MNRAS.341.1179A}, \citealt{2004MNRAS.347..220B}, \citealt{2007ApJ...666..189B}, \citealt{2010ApJ...719.1470V}, \citealt{2013MNRAS.429.1949A}, \citealt{2019ApJ...872....5S}).  \cite{2002ApJ...569L..83A} demonstrated that halo would firstly delay bar formation, but then can strengthen the bar during secular evolution by absorbing the angular momentum of stars. The strength of bars in simulated isolate galaxies was found to correlate with the amount of angular momentum absorbed by halos, and depend on the halo central concentration(\citealt{2002MNRAS.330...35A}, \citealt{2003MNRAS.341.1179A}). Many simulations showed that  gas would obstruct the growth of bar by giving angular momentum to bar. Consequently, bars will form much later and are much weaker in gas-rich disc galaxies(e.g., \citealt{2004MNRAS.347..220B}; \citealt{2005MNRAS.364L..18B}; \citealt{2007ApJ...666..189B}; \citealt{2013MNRAS.429.1949A}).  

These idealized simulations, however, usually study the formation and evolution of bars in isolated disc galaxies. Those disc galaxies were set up at the beginning of simulations by assuming varies models, but not resulting from self-consistent evolution. In addition, the effect of tidal force,  if included, was modelled in simplified ways. To overcome these two limitations, several works have investigated the origin and development of bars in more realistic environment using cosmological zoom-in simulations(e.g. \citealt{2012ApJ...757...60K}, \citealt{2012MNRAS.425L..10S}, \citealt{2015MNRAS.447.1774G} \citealt{2016MNRAS.459.2603B}, \citealt{2018MNRAS.473.2608Z}). These simulations show that bars can emerge naturally in the concord $\Lambda$CDM cosmology, and most of the bars became easily observable only after $z\sim 0.4-0.5$.

In addition, the roles of baryon physics, such as gas content, star formation, feedback from supernovae and super massive black hole(SMBH), in the evolution of bars are explored in the literature. Based on the cosmological zoom-in simulations Eris and ErisBH, \cite{2016MNRAS.459.2603B} shows that the AGN feedback in ErisBH can lower the gas content and star formation in the central region of disc galaxy, which results in a smaller bulge and larger disk, and the formation of a bar.  \cite{2019MNRAS.488.1864Z} further shows that different implementation of sub-grid physics can lead to bars with very different properties in the simulations ErisBH and Eris2k. They find that stronger effective stellar feedback can remove low angular momentum gas more efficiently, and helps to develop a stronger and longer bar in Eris2k. However, the impact of the difference on AGN feedback is not presented in their work. Meanwhile, whether the bar would induce quenching or enhance star formation in the inner region of discs(e.g, \citealt{2017MNRAS.465.3729S}, \citealt{2017ApJ...838..105L},\citealt{2017ApJ...845...93K},\citealt{2020MNRAS.492.4697N}), and whether it would enhance black hole activity or not are still controversial in different works(e.g., \citealt{2013A&A...549A.141A}, \citealt{2013ApJ...776...50C}). 

Recently, galaxy formation and evolution in cosmic volume, up to a cubic of 100 Mpc, and high resolution, down to kpc-100pc, have been studied in state-of-art cosmological hydrodynamical simulations such as Illustris-1, EAGLE, and IllustrisTNG \citep{2014MNRAS.444.1518V, 2015MNRAS.446..521S, Nelson:2018uso}. The properties of bars in the disc galaxies in these simulations, including their frequency, origin and correlation with gas fraction and stellar mass have been examined(\citealt{2017MNRAS.469.1054A}, \citealt{2019MNRAS.483.2721P}).  \cite{2017MNRAS.469.1054A} found that 20\% of the disc galaxies in EAGLE have strong bars, and another 20\% have weak bars, result in a total bar frequency consisting with the observation. They also found that the bar strength is correlated with the stellar mass, and stronger bars tend to locate in less gas-rich systems. Similar trends are found in Illustris-1. In contrast, the bar fraction in Illustris-1 is much lower, $\sim 21\%$, and increases slightly with increasing redshifts, which is in contradiction to the observed trend(\citealt{2019MNRAS.483.2721P}). 

These discrepancies over the bar fraction between simulations may be partially attributed to the different baryon physics, such as the feedback from star formation and AGN. As the Illustris-1 and TNG simulations share the same initial conditions, it would be worthwhile to carry out a comparison study on barred galaxies in them. Such a comparison study would find out the impact of baryon physics on the formation and evolution of bars. Note that, during the preparation of this work, \cite{2020MNRAS.491.2547R} publish their analysis on the bar fraction in the TNG100 simulations, which is 40\% in the stellar mass range $M_*=10^{10.4-11.0}M_{\odot}$ and is much higher than the fraction in the Illustris-1 simulation reported by \cite{2019MNRAS.483.2721P}. Moreover, \cite{2020MNRAS.491.2547R} shows that the star formation and black hole activity in the barred galaxies are stronger than that in unbarred galaxies in TNG100.

This paper is organised as follows. We introduce the simulations and galaxies samples in Section \ref{sec:samples}. The overall features of barred galaxies such as the bar fractions, and their origins in two simulations are shown in Section 3. We explore properties such as gas fraction, star formation, black hole and feedback, and the mass and shape of host dark matter halos during the evolution of bars in Section 4. A comparison of bar properties between matched galaxies pairs in the Illustris and TNG100 simulations is presented in Section 5. We summarize our findings, and compare them with previous works and discuss the results in Section 6.

\section{SIMULATIONS AND Galaxies Samples} \label{sec:samples}
\subsection{The Illustris-1 and TNG100 simulations}
In this paper, we make use of public released data from the Illustris-1 and TNG100 simulations. These two simulations use the same initial conditions, except for some adjustments in TNG100 for updated cosmology. There are some differences in the cosmological parameters and baryon physics between these two simulations.

The Illustris project \citep{2014MNRAS.444.1518V} is a set of large hybrid N-body/hydrodynamic simulations of galaxy formation, using the moving-mesh code AREPO \citep{2010MNRAS.401..791S}. The cosmological parameters of the Illustris simulations are set to the latest Wilkinson Microwave Anisotropy Probe (WMAP9) measurements : $\Omega_m =\Omega_{dm} + \Omega_b = 0.2726$, $\Omega_{\Lambda}$ = 0.7274, $\Omega_b$ = 0.0456, $\sigma_8$ = 0.809, $n_s$ = 0.963, and $H_0$ = 100$h \ \rm{km}\ {s}^{-1} \rm{Mpc}^{-1}$ with $h$ = 0.704 \citep{2013ApJS..208...19H}. The IllustrisTNG project \citep{Nelson:2018uso} is the successor of the original Illustris project. It updates galaxy formation models, which include new physics and numerical improvements, as well as refinements to the original Illustris simulations. The IllustrisTNG simulations are normalized by the recent Planck constraints: $\Omega_m = 0.3089$, $\Omega_b$ = 0.0486, $\Omega_\lambda$ = 0.6911, $\sigma_8$ = 0.8159, $n_s$ = 0.9667, $h$ = 0.6774\citep{Ade:2015xua}.

The Illustris-1 and TNG100 simulations have the same box size ($75h^{-1} Mpc^3$), and both use $2 \times 1820^3$ dark matter and gas particles. The mass of each dark matter and gas particle are $6.3 \times 10^6 M_{\odot}$ and $1.6 \times 10^6 M_{\odot}$ in Illustris-1, and $6.3 \times 10^6 M_{\odot}$ and $1.4 \times 10^6 M_{\odot}$ in TNG100, respectively. Both simulations were evolved from redshift $z = 127$ to the present time $z = 0$. The Illustris-1 simulation outputs 134 snapshots, while TNG100 has 100 snapshots.

The differences of baryonic physics between IllustrisTNG and the original Illustris have been described in detail in \cite{2017MNRAS.465.3291W} and \cite{2018MNRAS.473.4077P}. In addition to some key numerical improvements, the major updates in recipes of galaxy physics include three aspects: the evolution and feedback of super-massive black hole, galactic winds driving by star formation, and the stellar evolution. In IllustrisTNG, the seed mass of SMBH is increased by a factor of 8, and the thermal 'bubble' model in Illustris has been replaced by a kinetic model when the accretion rate is low. Model of isotropic galactic winds with velocity floor is implemented in TNG, instead of a 'bipolar' winds model without velocity floor in Illustris. In addition, the velocity of galactic winds in TNG is assumed to be redshift dependent. 

In the TNG models, the impact of AGN feedback and galactic winds on star formation, black hole growth, and other galaxy properties have been studied in \cite{2018MNRAS.473.4077P} and \cite{2018MNRAS.479.4056W}. \cite{2018MNRAS.473.4077P} demonstrated that the TNG models can suppress the star formation more effectively than Illustris-1 for halos with masses of $2\times10^{11}-2\times10^{13}M_{\odot}$. Galactic winds in TNG are faster than in Illustris-1, and can reduce the star formation efficiency more effectively, especially in systems with stellar mass of $M_*<3\times10^{10}M_{\odot}$, or halo mass less than $10^{12} M_{\odot}$ at $z=0$. For more massive galaxies/halos, AGN feedback dominates the suppression of the stellar mass. They demonstrated that the TNG models can address the main shortcomings of the Illustris-1 models in confrontation with observations, including the cosmic star formation rate at $z<1$, stellar content and sizes of low mass galaxies, and gas fraction in massive halos. \cite{2018MNRAS.479.4056W} revealed the relative importance of the AGN and stellar feedback in TNG in detail. They found that the stellar feedback dominates in overall mass range of galaxies at high redshifts, and even keeps its dominant role at low redshifts for galaxies with $M_*<\sim 10^{10}M_{\odot}$. On the other hand, the thermal AGN feedback becomes important for galaxies with $M_*>10^{10}M_{\odot}$. They also found that the kinetic AGN feedback dominates in massive galaxies with $M_*>10^{10.5}M_{\odot}$ since $z\sim 2$, which is coincident with the quenching of these massive galaxies(see their figure 1). For more detail comparison of the models of baryonic physics and their impacts between the two simulation projects, we refer the readers to \cite{2017MNRAS.465.3291W}, \cite{2018MNRAS.473.4077P}, and \cite{2018MNRAS.479.4056W}. These differences on feedback models should have contributed to the differences on the bar structure between the two simulations.

\subsection{Galaxies samples}
We make use of the Illustris-1 and TNG100 public released data, including the Subfind Subhalo catalog \citep{Springel:2000qu} and the SubLink merger trees, which enable us to track the evolution of galaxies and host halos along with time. In these projects, galaxy is identified as stellar component in subhalo, and dark matter halo is named as Halo. 

We basically follow the same schemes as in \cite{2019MNRAS.483.2721P} to identify barred galaxies in simulation samples. In order to study the bars, one should first locate the disc galaxies. \cite{2019MNRAS.483.2721P} uses two parameters provided by the simulations to specify disc galaxies, i.e., the stellar circularities $\epsilon$, and the flatness of galaxies. We first set the position of the most bounded stellar particle in a galaxy as its center, and then take the plane perpendicular to the angular momentum vector of stellar component as the galaxy plane.
Stellar particles belonging to galaxy disc are expected to have circularity parameter $\epsilon$ close to 1. Illustris and IllustrisTNG projects provide the fractional mass of stellar component with $\epsilon > 0.7$, i.e., $f(\epsilon>0.7)$ for each galaxy, which was first given by \cite{2015ApJ...804L..40G} as a measure of the fraction of stellar mass in the disc component. 
If more than 20\% of a galaxy's stellar mass have $\epsilon > 0.7$, i.e., $f(\epsilon>0.7)>0.2$, which means this galaxy having more than 20\% of their stellar mass to behave kinematically
as disc component, this galaxy would be taken as a disc galaxy candidate. Then, we use the three eigenvalues of the stellar mass tensor $M_1$, $M_2$, and $M_3$ to calculate the flatness of disc galaxy candidates, which is defined by $M_1/\sqrt{M_2M_3}$, $(M_1 < M_2 < M_3$). Finally, for a disc galaxy candidate, it will be confirmed as a disc galaxy if its flatness is smaller than 0.7.


Using this method, we find 2658 disc galaxies with stellar masses above $10^{10} M_\odot$, and 1269 disc galaxies with more than 40000 stellar particles ($M_* > 10^{10.5} M_\odot$) 
in the TNG100 samples at redshift $z = 0$. While in Illustris-1, we find 1232 disc galaxies with over 40000 stellar particles. Further, we use the $A_2$ parameter \citep{2013MNRAS.429.1949A} to identify whether these disc galaxies have bar or not. $A_2$ is defined by the two Fourier components\citep{2013MNRAS.429.1949A}: 
\begin{equation}
    a_m(R) = \sum_{i}^{N_R}M_{i}cos(m\phi_i)
\end{equation}
\begin{equation}
    b_m(R) = \sum_{i}^{N_R}M_{i}sin(m\phi_i)
\end{equation}
, where $N_R$ is the number of star particles within a given cylindrical radius R, $M_i$ is the $i$-th star particle's mass, and $\phi_i$ is its azimuthal angle. $A_2(R)$ is a function of cylindrical radius R, defined as
\begin{equation}
    A_2(R) = \frac{\sqrt{a_{2}^{2} + b_{2}^{2}}}{a_0}.
\end{equation}
We measure the bar strength by the maximum value of $A_2(R)$, i.e., $A_{2,max}$.

We calculate the $A_{2,max}$ parameter over all disc galaxies, and place a threshold value of 0.15 for $A_{2,max}$ parameter to preliminary determine whether a galaxy is barred or not. Next, we inspect the stellar surface density maps of galaxies with $A_{2,max}>0.15$ by naked eyes to confirm that this non-axisymmetric feature is indeed due to a bar. 
For these candidate galaxies, we visualize their morphological features, and exclude those actually exhibiting no bar structure, but have chaotic and distorted structures. This inspection procedure is somewhat crude, but the final bar fraction estimated thereby are consistent with previous works on bars of these two simulations(\citealt{2019MNRAS.483.2721P},\citealt{2020MNRAS.491.2547R}). We define disc galaxies with $A_{2,max} > 0.15$ but visually do not looks like barred galaxies as false positive samples. There are 21 and 70 false positive cases in Illustris-1 and TNG100 respectively, and they are not included in our barred galaxies sample. Finally, we identify 110 and 871 barred galaxies in Illustris-1 and TNG100 respectively at z=0. 

\begin{figure}[htbp]
\begin{center}
\includegraphics[width=0.5\textwidth,trim=50 70 10 30,clip]{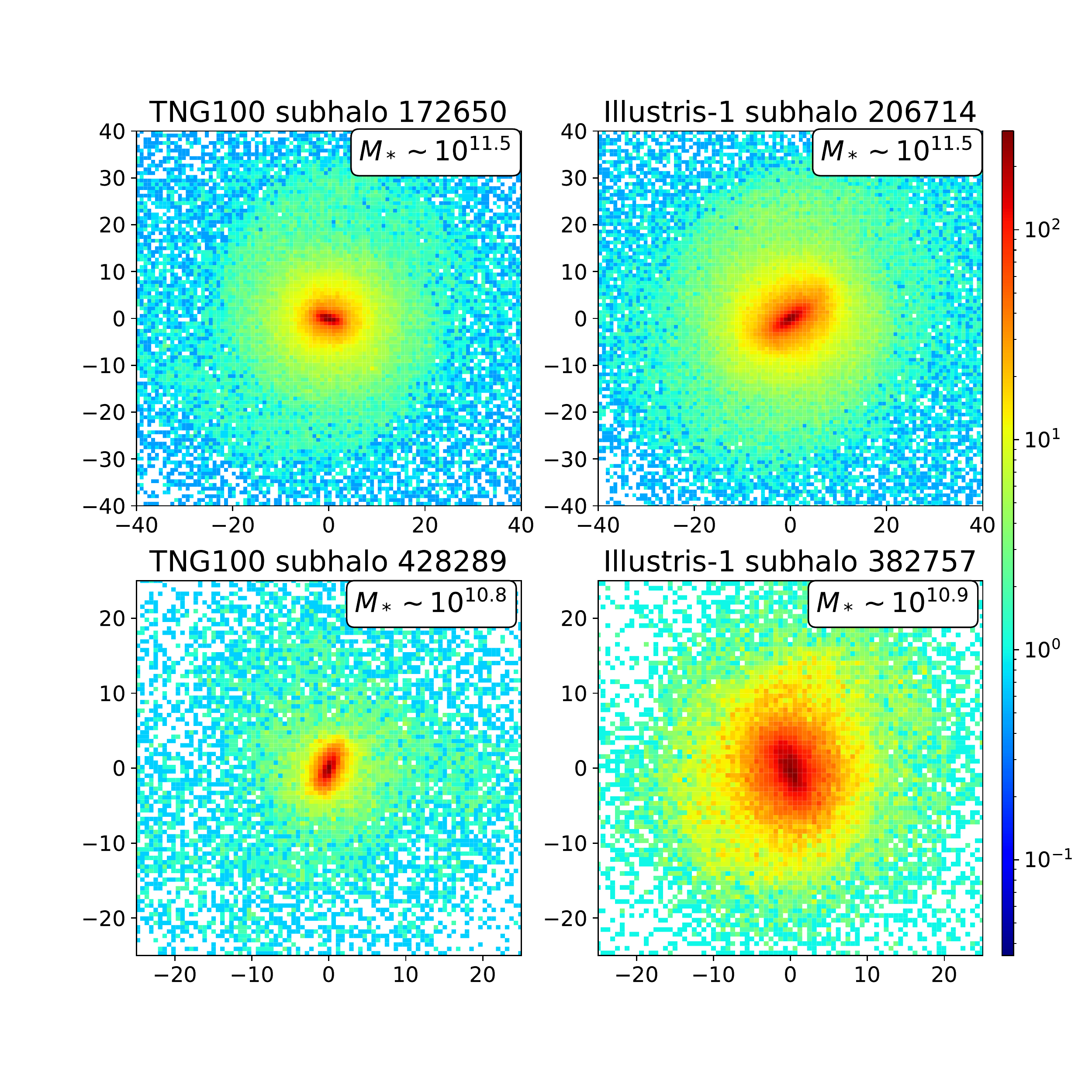}
\caption{Images of barred galaxies in TNG100(left) and Illustris-1(right). The top row shows two galaxies with stellar mass $M_*=10^{11.5}M_{\odot}$. The bottom row shows two galaxies with stellar mass $M_*=10^{10.8-10.9}M_{\odot}$. }
\end{center}
\label{fig:galaxy_image}
\end{figure}

\begin{figure}[htbp]
\begin{center}
\includegraphics[width=0.5\textwidth,trim=70 70 10 30,clip]{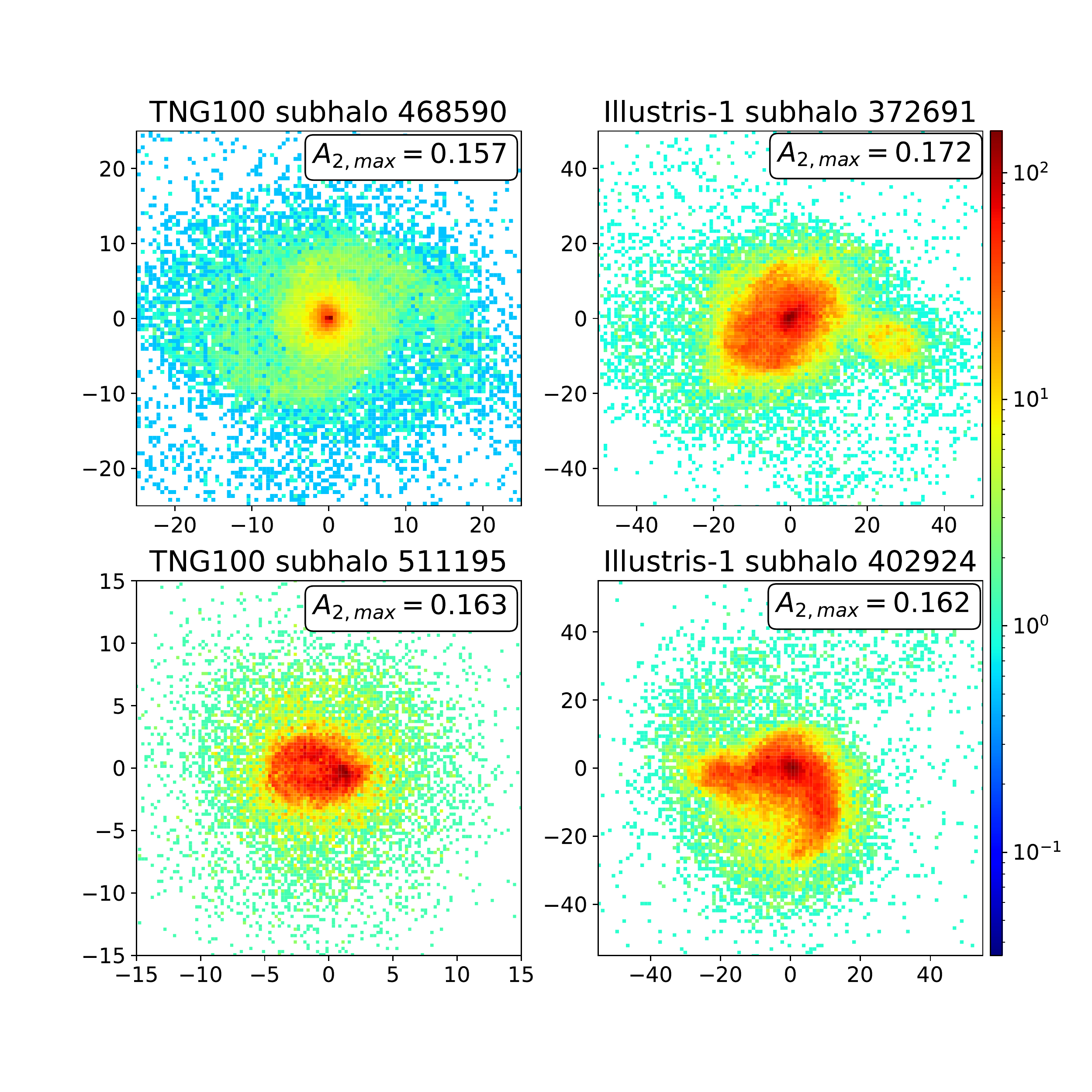}
\caption{Samples of disc galaxies with $A_{2,max} > 0.15$ but visually do not looks like barred galaxies.}
\end{center}
\label{fig:fp_image}
\end{figure}

We display images of two examples of barred galaxies from each simulation in Fig.~\ref{fig:galaxy_image}. The top and bottom row show galaxies with stellar mass of $M_*=10^{11.5}M_{\odot}$ and $M_*=10^{10.8-10.9}M_{\odot}$ respectively. In both simulations, the bar features are manifest for galaxies with stellar masses around  $10^{11} M_{\odot}$, and are still visible with stellar mass $\sim 10^{10.8-10.9} M_{\odot}$.
Fig.~\ref{fig:fp_image} shows four samples of disc galaxies with $A_{2,max} > 0.15$, but without bar structure, i.e., false positive samples. Their non-axisymmetry are mainly caused by complex substructures or arms. As mentioned above, these false positive samples are not identified as barred galaxies. 


\section{Bars in disc galaxies} 
\subsection{Bar fraction} 

\begin{figure}[htbp]
\begin{center}
\includegraphics[width=1.0\columnwidth]{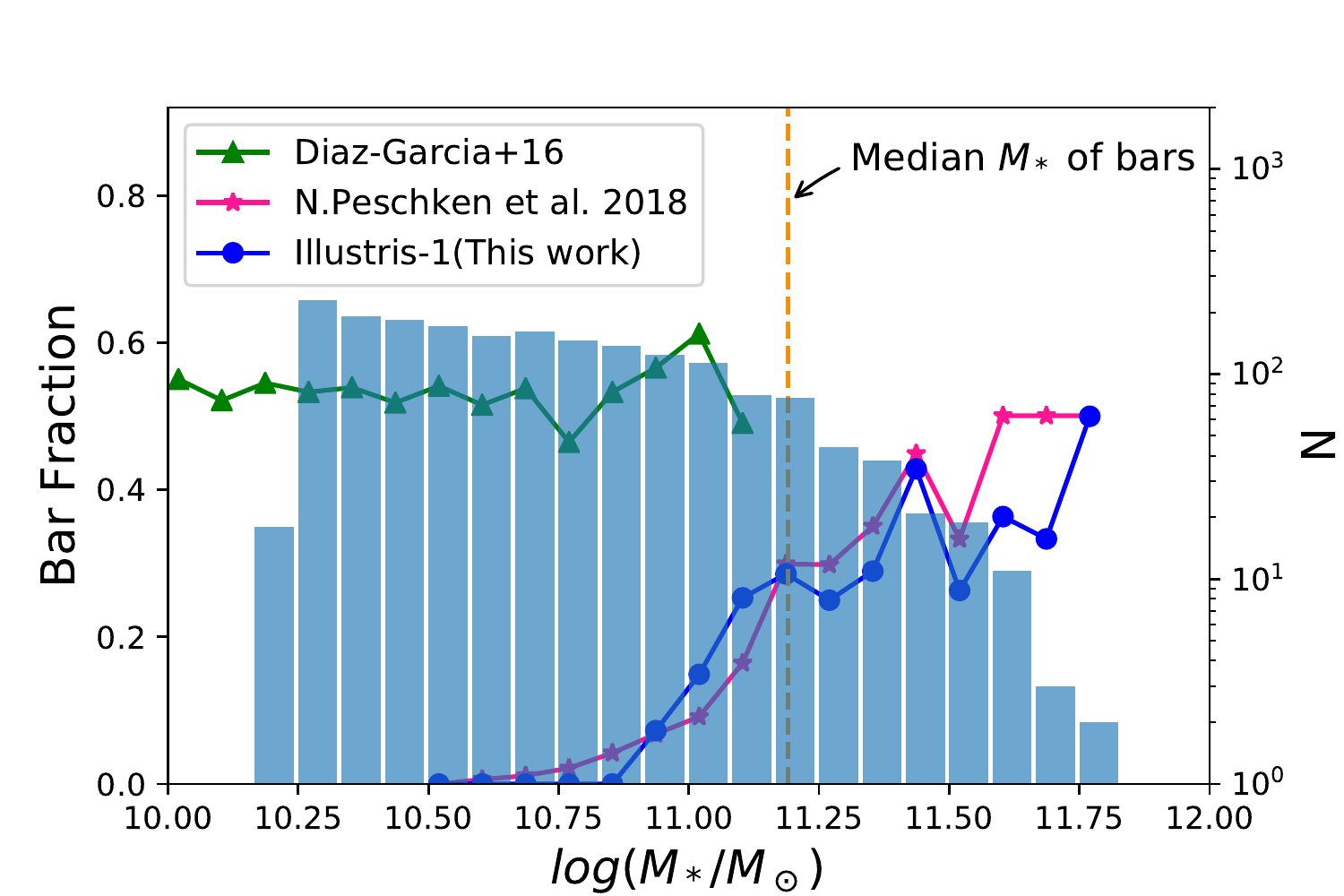}
\includegraphics[width=1.0\columnwidth]{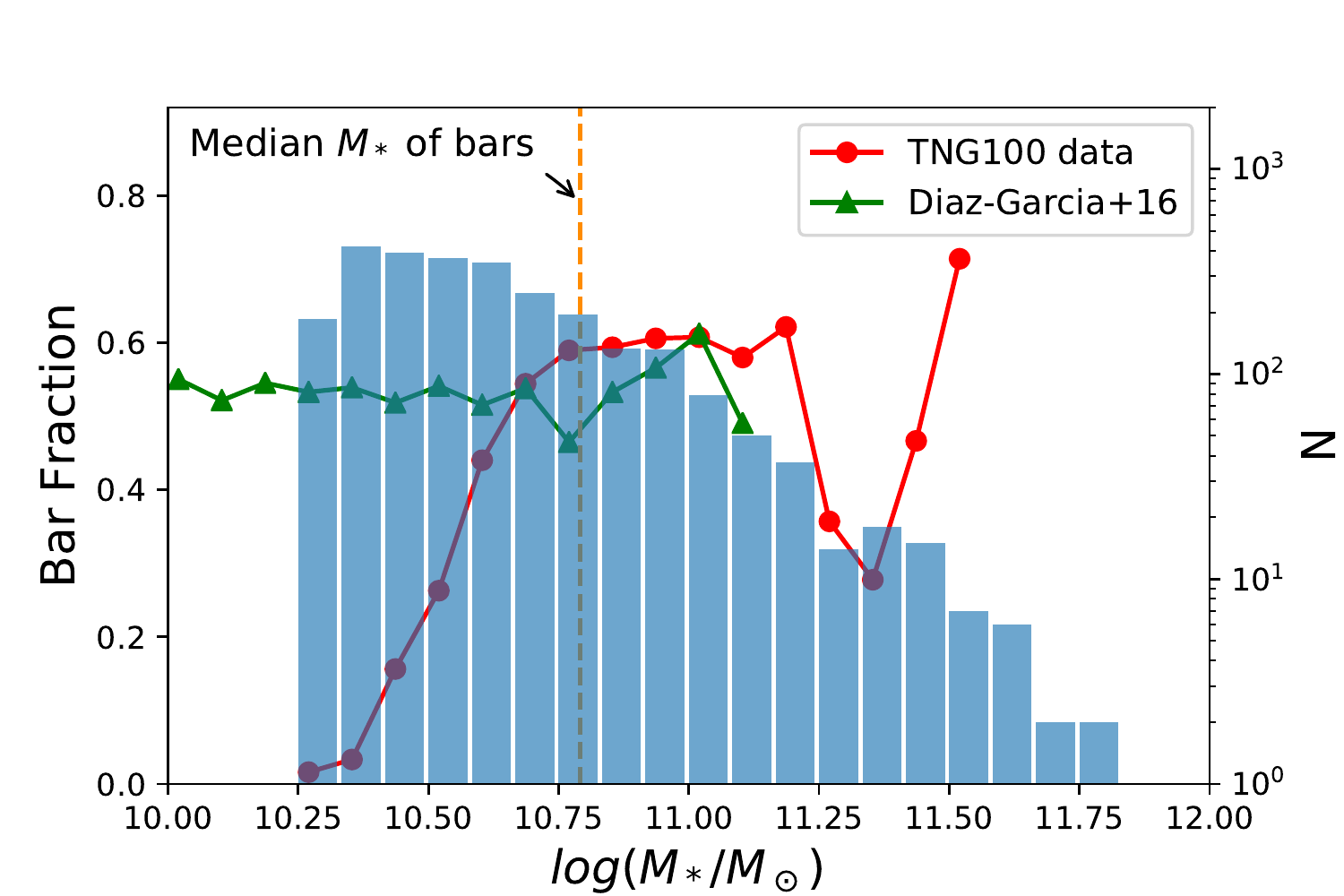}
\caption{Blue(red) dots in the top(bottom) plots indicate bar fraction(left axis) as a function of stellar mass in the Illustris-1(TNG100) simulation.  The red stars in the top plot is the bar fraction in Illustris-1 identified by \cite{2019MNRAS.483.2721P}. Green triangles in two plots are the bar fraction in observational work \cite{2016A&A...587A.160D}. The vertical yellow dashed lines indicate the median stellar mass of barred galaxies. Histogram in both figures are the stellar masses function of disc galaxies with bin size 0.083. Number of disc galaxies in each bin can be read from the right axis.}
\end{center}
\label{fig:overall_bar_fraction}
\end{figure}

We measure the overall bar fraction of disc galaxies as a function of stellar mass $M_*$ at redshift $z=0$ in the TNG100 and Illustris-1 simulations respectively, and plot the results in Fig.~\ref{fig:overall_bar_fraction}. In Illustris-1, the bar fraction increases gradually from $\sim 0\%$ in the stellar mass bin $M_*=10^{10.50-10.58}M_{\odot}$ to $\sim 10\%$ in the bin $M_*=10^{11.00-11.08}M_{\odot}$, and then grows rapidly to $30\% \sim 40\%$ for galaxies more massive than $M_*=10^{11.25}M_{\odot}$. The bar fraction in Illustris-1 identified by our procedures is basically in agreement with \cite{2019MNRAS.483.2721P}, although there are some slight differences in some mass bins. These differences may result from that \cite{2019MNRAS.483.2721P} recomputed the stellar masses of some interacting galaxies, but we use the original estimations of stellar masses provided by the Illustirs-1 project. 

The bar fraction in TNG100 is much higher than that in Illustris-1 in the stellar mass range from $\sim 10^{10.5}$ to $\sim 10^{11.25}M_{\odot}$. It increases rapidly from $\sim 0\%$ in the  bin $M_*=10^{10.25-10.33}M_{\odot}$ to $\sim 30\%$ in the bin $M_*=10^{10.50-10.58}M_{\odot}$, and then to $\sim 50\%$ for galaxies in the mass range $M_*=10^{10.66-11.25}M_{\odot}$. For galaxies more massive than $M_*=10^{11.25}M_{\odot}$, the bar fraction has a significant scatter because of the very limited number of very massive disc samples. Different feedback models are expected to be responsible for the discrepancy on the bar fraction between the two simulations. The properties of gas content, stellar component, SMBH and halo are influenced by feedback models, and should have significant effects on bar formation(\citealt{2016MNRAS.459.2603B}, \citealt{2019MNRAS.488.1864Z}). We will investigate these properties and their relations with bar formation in the next section. On the other hand, the bottom plot of Fig.~\ref{fig:overall_bar_fraction} indicates that the bar fraction of disc galaxies with stellar mass $M_*=10^{10.66-11.25}M_{\odot}$ in TNG100 is well consistent with the result of the local survey $S^4G$ \citep{2016A&A...587A.160D}.

Unless specified otherwise, in the following comparison study we will only include disc galaxies containing more than 40000 stellar particles, i.e.,$M_*>10^{10.50} M_{\odot}$ so as to obtain reliable results. Disc galaxies with and without bar will be named as barred and unbarred respectively. The same threshold has been applied in \cite{2019MNRAS.483.2721P}. Above this mass threshold, there are 1269 disc galaxies, of which 698 galaxies are identified as barred in TNG100. The median stellar mass of these 698 barred galaxies is $M_*=10^{10.80}M_{\odot}$. In illustris-1, there are 1232 disc galaxies more massive than $10^{10.50} M_{\odot}$, of which 110 galaxies are barred and the median stellar mass of barred galaxies is $M_*=10^{11.20}M_{\odot}$. In contrast, \cite{2020MNRAS.491.2547R} identified 270 disc galaxies at $z=0$ within the range $M_*=10^{10.4-11.0}M_{\odot}$ and 107 of them are barred in TNG100. They used the kinematic bulge-to-disc composition algorithm and additional limitation on the stellar disk/bulge-to-total mass ratio to identify bar. In \cite{2019MNRAS.483.2721P}, very few bars are found in low-mass galaxies($3.3 \times 10^{10} M_\odot<M_*<8.3 \times 10^{10} M_\odot$), and also, 109 out of 509 disc galaxies that more massive than $8.3 \times 10^{10} M_\odot $ are found to be barred at $z=0$ in Illustris-1. 

\begin{figure}[htbp]
\begin{center}
\includegraphics[width=0.85\columnwidth]{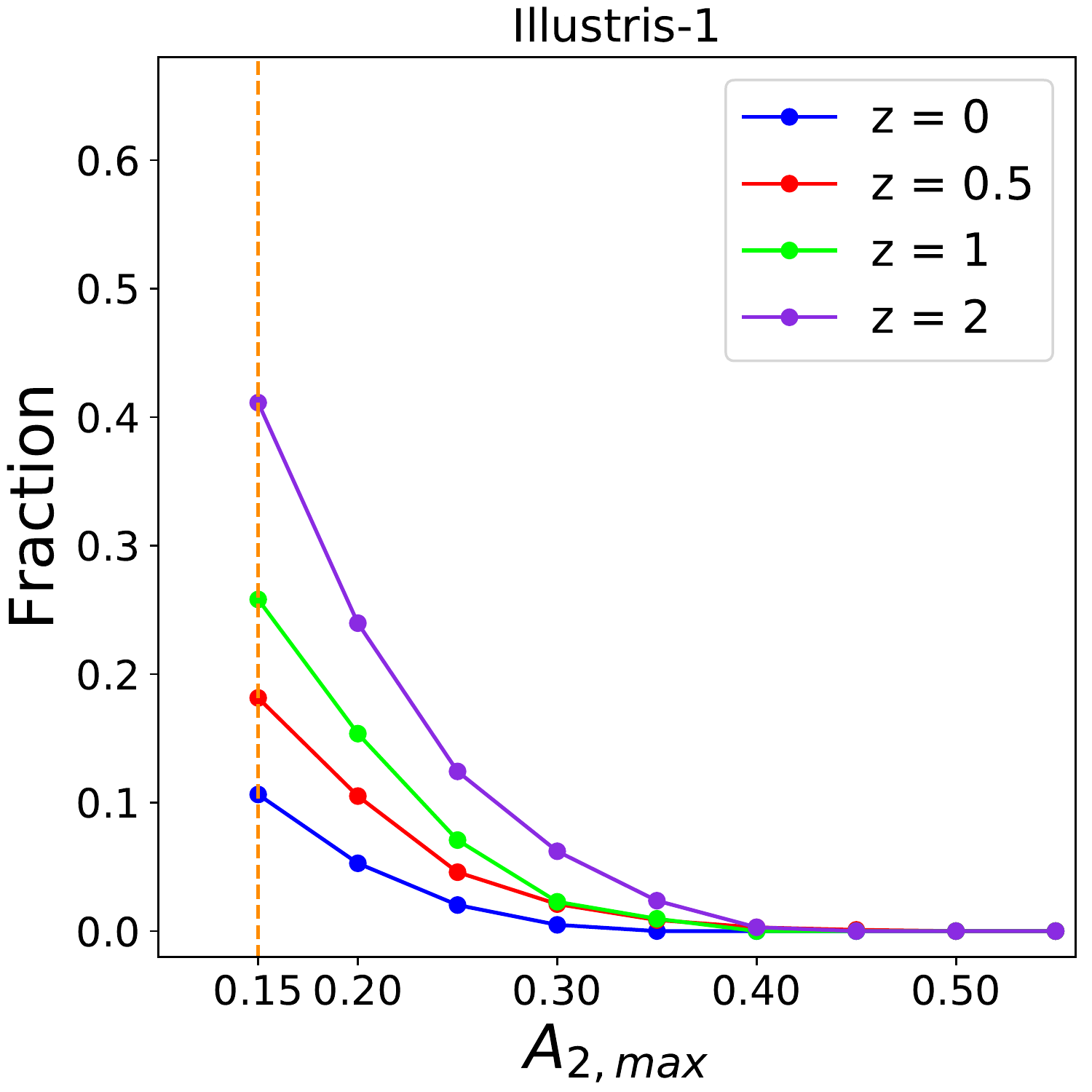}
\includegraphics[width=0.85\columnwidth]{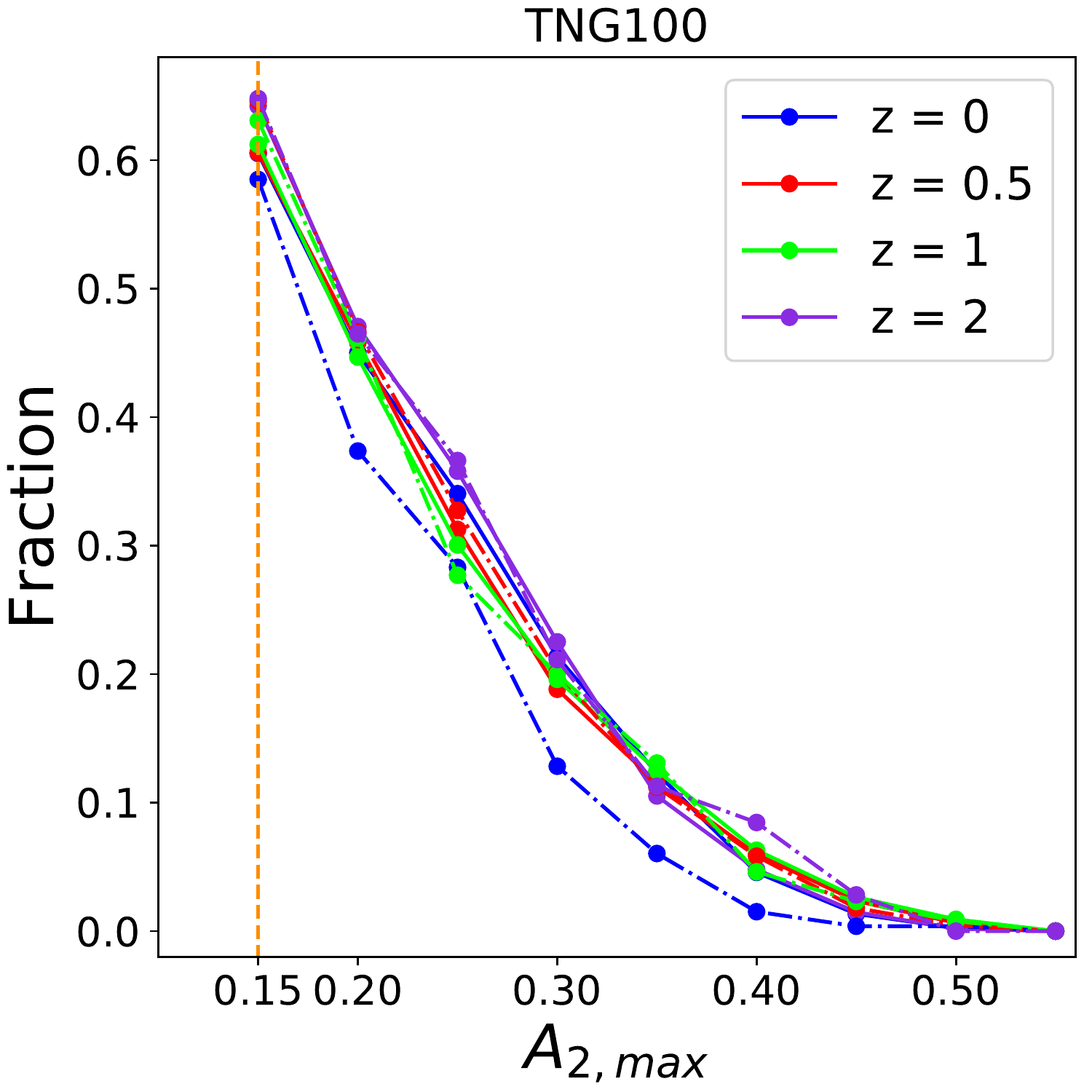}
\caption{Cumulative fraction of $A_{2}^{max}$ parameter of disc galaxies with more than $40 000$ stellar particles at different redshifts. Top: Illustris-1 data set; Bottom: TNG100 data set, dotted-dashed line indicate disc galaxies that have more than $100 000$ stellar particles.}
\end{center}
\label{fig:allsnap_bar_fraction}
\end{figure}

We also track disc galaxies at different redshifts and calculate the corresponding bar fractions. Some of the discs at $z=0$ might not be disc at high redshifts, and some disc galaxies at high redshifts may evolve to non-disc galaxies at $z=0$. Fig.~\ref{fig:allsnap_bar_fraction} shows that in Illustris-1, the fraction of disc galaxies having $A^{max}_2>0.15$ increases with redshift, which agrees with \cite{2019MNRAS.483.2721P}, but is inconsistent with many observations(e.g. \citealt{2008ApJ...675.1141S}, \citealt{2014MNRAS.438.2882M}). \cite{2019MNRAS.483.2721P} argued that, the observed trend that bar fraction decreases with increasing redshift holds for low-mass galaxies, but for massive galaxies the bar fraction is roughly constant or even increases with redshift. Nevertheless, the bottom panel of Fig.~\ref{fig:allsnap_bar_fraction} shows that TNG100 has a roughly constant fraction of $A^{max}_2>0.15$ at different epochs, partly relieve the conflict. Moreover, the bar fraction of those disc galaxies that more massive than $M_*=10^{10.83} M_{\odot}$, i.e, the mass threshold of default sample in \cite{2019MNRAS.483.2721P}, evolves slowly with redshift in TNG100. The bar fraction of massive disc galaxies at $z=0$ decreases slightly in comparison with high redshifts. 
\label{subsec:tables}

\subsection{Formation time and origin of bars}
\begin{figure}[htbp]
\begin{center}
\includegraphics[width=1.0\columnwidth,trim=3 15 30 10,clip]{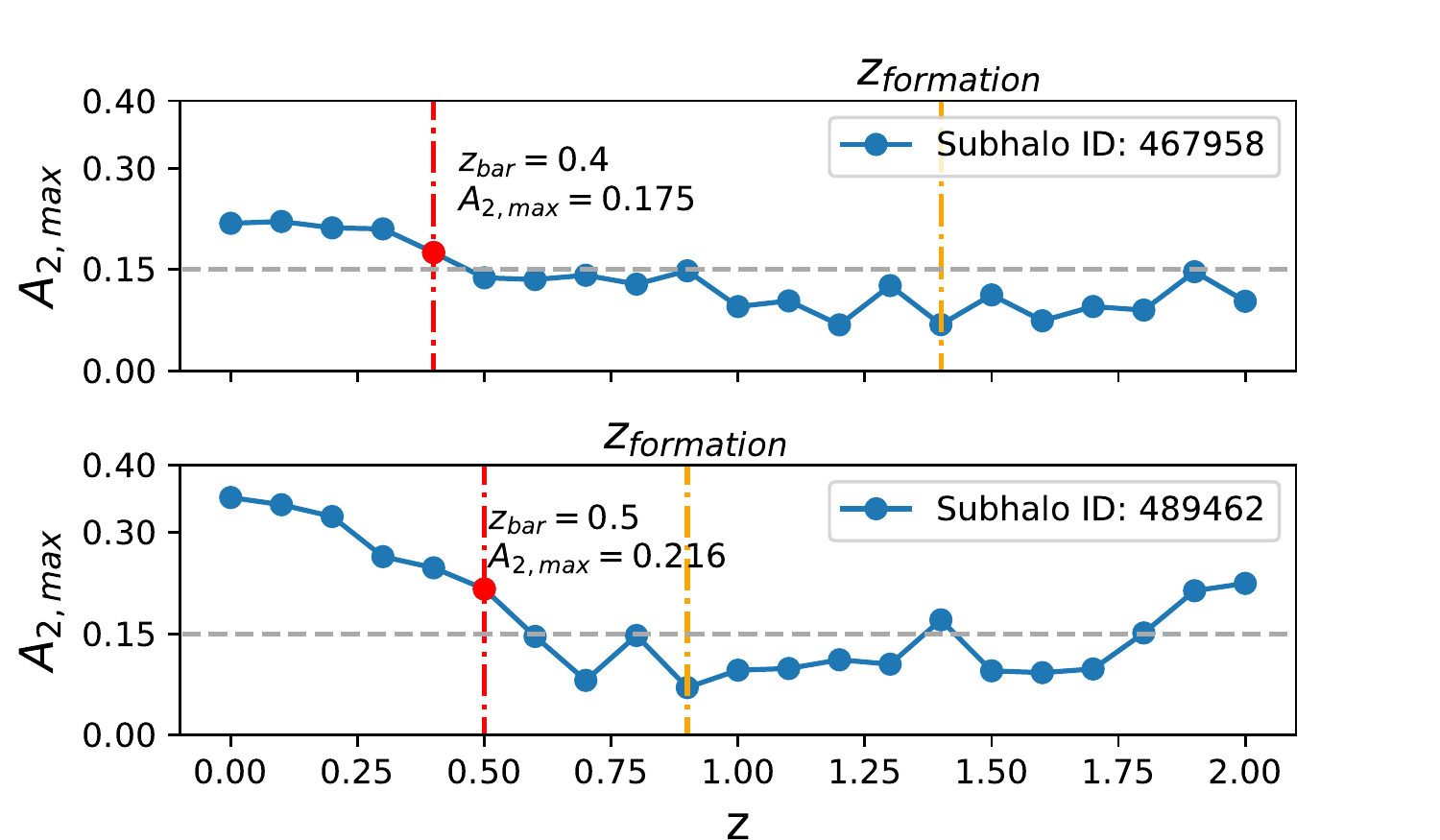}
\vspace{0.2cm}
\caption{Examples of the $A_{2,max}$ as a function of redshifts for two barred galaxies in the TNG100 simulation. Red vertical lines indicate the bar formation redshift(see the text for the identification procedure), and yellow lines indicate the redshift that galaxies had assembled $50\%$ of their stellar mass at $z=0$.}
\end{center}
\label{fig:a2_rs}
\end{figure}

To find out the formation time of bars in our samples, we track the evolution history of barred galaxies to identify the formation redshift of bars, denoted as $z_{bar}$. For each barred disc galaxy at $z=0$, we calculate the value of its $A_{2}^{max}$ at different redshifts. The solid lines in Fig.~\ref{fig:a2_rs} show the evolution of $A_{2}^{max}$ for two sample galaxies. Then, we determine $z_{bar}$ of each barred galaxy by the following procedure. Firstly, we determine the redshift at which one galaxy's $A_{2}^{max}$ is above 0.15 for the first time while evolving from high to low redshifts; if $A_{2}^{max}$ keeps above 0.15 thereafter, then this redshift will be marked as a candidate of $z_{bar}$ for this galaxy; otherwise, we will keep searching whenever $A_{2}^{max}$ is crossing upward above 0.15 toward lower redshift, till we find the candidate of $z_{bar}$.
Secondly, this candidate will be defined as this galaxies's $z_{bar}$ if it satisfies the following condition, 
\begin{equation}
\frac{|(A_{2}^{max}(z) - A_{2}^{max}(z + {\Delta}z)|}{A_{2}^{max}(z)} < 0.4
\label{eqn:z_bar}
\end{equation}
where ${\Delta}z$ is the redshift gap between the snapshot corresponding to the candidate redshift and its previous snapshot at higher redshift. Eqn.~\ref{eqn:z_bar} is applied to ensure the bar is stable and avoid violent fluctuations. Similar measure is adopted in \cite{2020MNRAS.491.2547R}. If the candidate redshift does not meet the condition of Eqn.~\ref{eqn:z_bar}, we will track toward lower redshift to find the next candidate, and check whether it fulfill Eqn.~\ref{eqn:z_bar} or not. The procedure is repeated till $z_{bar}$ of this galaxy is found. We perform this kind of searching throughout all the barred galaxies at $z=0$ to find their $z_{bar}$. The vertical red dotted-dashed lines marked in Fig.~\ref{fig:a2_rs} indicate the detected bar formation time of two example galaxies.

\begin{figure}[htbp]
\begin{center}
\includegraphics[width=1.0\columnwidth]{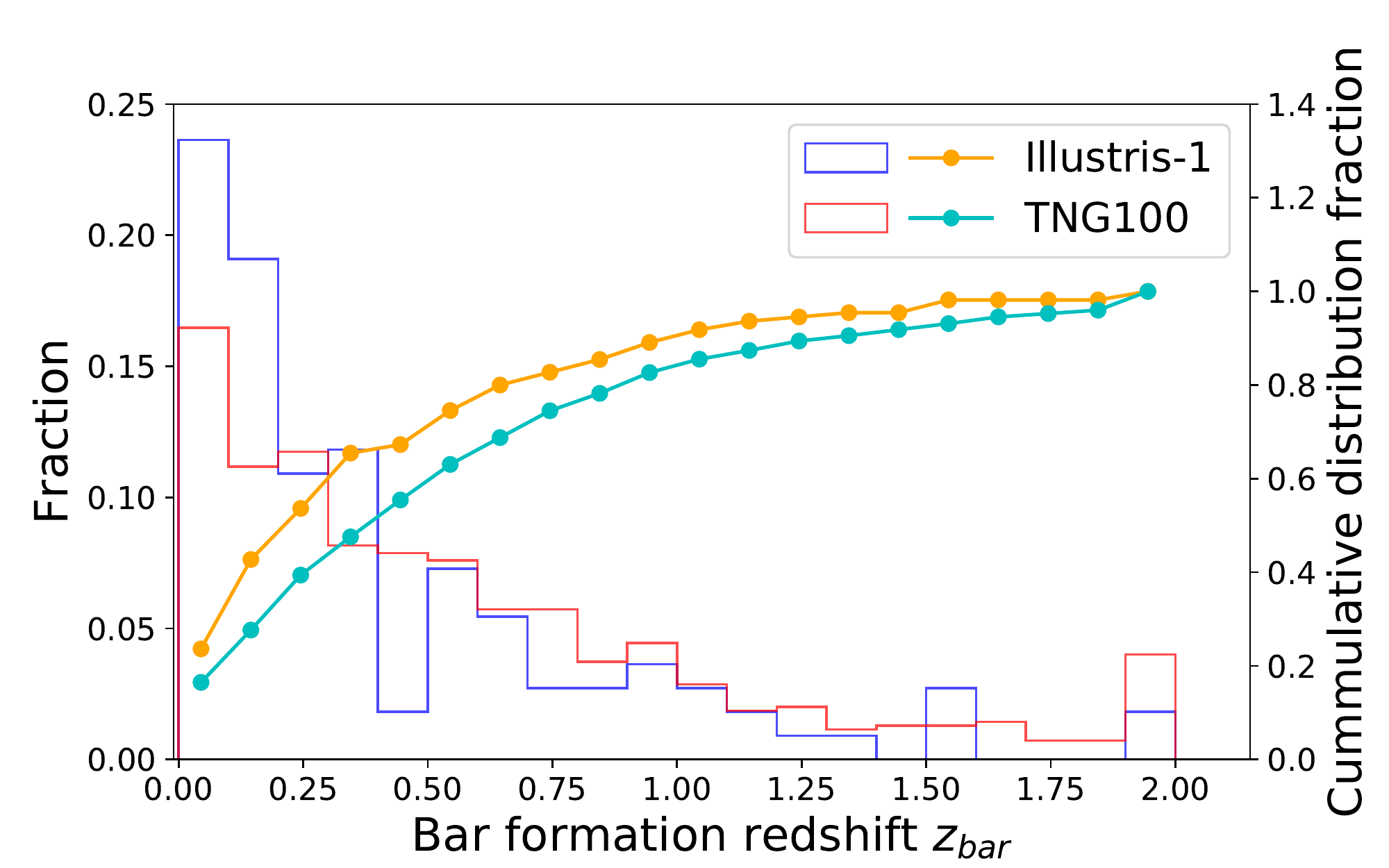}
\includegraphics[width=1.0\columnwidth,trim=0 0 20 20,clip]{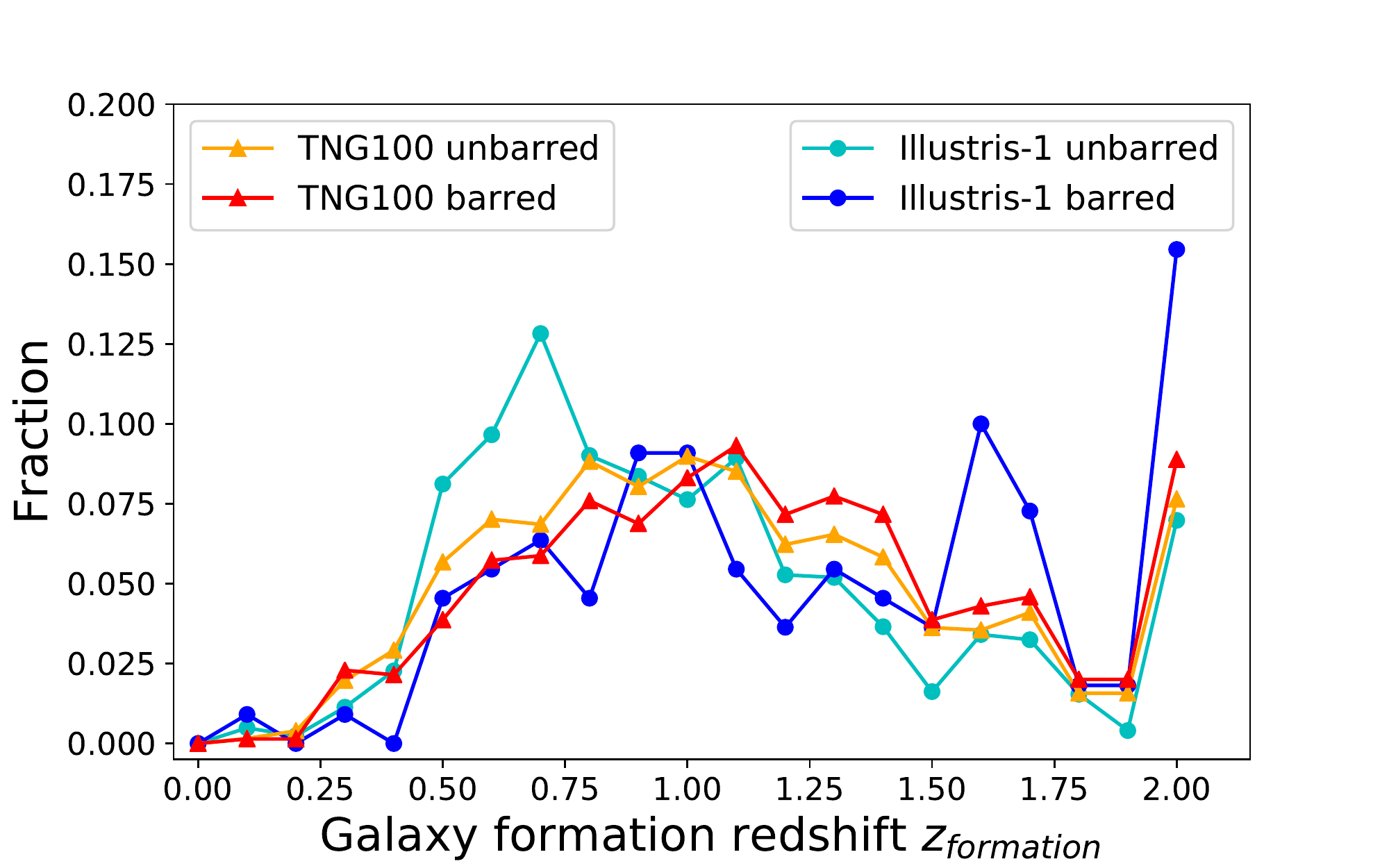}
\caption{Top: Histogram of bar formation redshift, $z_{bar}$, in TNG100(red) and Illustris-1(blue). The solid yellow and cyan dotted lines indicate the cumulative distribution function of $z_{bar}$ in Illustris-1 and TNG100 respectively. Bottom: Distribution of galaxy formation redshift, i.e., the epoch galaxy accumulate $50 \%$ of $M_*(z=0)$, in Illustris-1 and TNG100. }
\end{center}
\label{fig:z_form}
\end{figure}

Fig.~\ref{fig:z_form} shows the redshift distribution of bar formation. We find the median of 
$z_{bar}$ is $\sim 0.4-0.5$ in TNG100, and is $ \sim 0.25$ in Illustris-1. 
\cite{2019MNRAS.483.2721P} and \cite{2020MNRAS.491.2547R} used some different ways to figure out the formation time of galaxy bars, which is
about $z=0.5$ in TNG100 and $z=0.3$ in Illustris-1. Our results are in agreement with theirs. Note that, our sample size is larger than \cite{2020MNRAS.491.2547R} due to a wider mass range and different sample selecting method. Fig.~\ref{fig:z_form} also shows the distribution of time when disc galaxies had accumulated $50\%$ of their stellar mass at $z=0$. Generally, barred galaxies reach this milestone at higher redshifts than unbarred galaxies. For most barred galaxies, the bar structures emerge after most of stars have formed or assembled. In Fig.~\ref{fig:z_form}, we have assigned galaxies with a bar/galaxy formation redshift equal to or higher than $z=2$ to the bin of $z=2$, result in a peak at $z=2$.

    \begin{table}[htbp]
\begin{tabular}{|c|c|c|c|c|}
\hline
                     & \textbf{Merge} & \textbf{Flyby} & \textbf{Secular} & \textbf{Sum} \\ \hline
\textbf{TNG100}     & 400            & 119            & 179              & 698          \\ \hline
\textbf{TNG100($>10^{10.83}M_{\odot}$)} & 83             & 23             & 46               & 152     \\ \hline
\textbf{TNG100($<10^{10.83}M_{\odot}$)} & 317             & 96             & 133               & 546    \\ \hline
\textbf{Illustris-1} & 53             & 45             & 12               & 110          \\ \hline
\end{tabular}
\caption{Processes that associated to the bar formation in the TNG100 and Illustris-1 simulations}
\label{table:origins}
\end{table}

There are generally three types of processes that can drive the formation of galaxy bars, i.e., galaxy merger, flyby interaction and secular evolution. To figure out the roles of these processes, we check the barred disc galaxies visually, probing their images to see if there is a merger or flyby when bar was just formed. More specifically, we determine the origin of galaxy bar by first check the halo's merger history around $z_{bar}$, i.e., within the nearest two time snapshots before and after $z_{bar}$. If there was a merger event occurred around $z_{bar}$, this bar is defined to be associated with merger. Otherwise, we inspect the three orthographic views of the distribution of stellar particles within the radius of 50 times of one barred galaxy's half stellar radius, $r_{50}$, in the same time ranges. If there are many other galaxy’s particles locating within the sphere of radius $50r_{50}$ around a barred galaxy, and this bar galaxy also exhibits features of interaction, i.e., disturbed stellar distribution, we define this bar to be associated with “flyby” event. The rest galaxy bars are classified as "Secular evolution".

Table ~\ref{table:origins} lists the frequency of each process in the two simulations. In Illustris-1, $48.2\%$, $40.9\%$ of the bars are associated with merge, flyby events respectively, and the rest $10.9\%$ of bars result from secular evolution. These fractions basically agree with \cite{2019MNRAS.483.2721P}, despite our procedure is somewhat crude. In TNG100, the fractions of merge, flyby and secular are $57.3\%$, $17.0\%$, and $25.7\%$ respectively. We further explore the origin of bars in disc galaxies more/less massive than $M_*=10^{10.83}M_{\odot}$, i.e., 100000 stellar particles, in TNG100. The corresponding frequency are $54.6\%$, $15.1\%$ and $30.3\%$ for relatively massive disc galaxies. Thus, these frequencies depend weakly on galaxy stellar mass in TNG100. The fraction of bars due to secular evolution in TNG100 is roughly triple of that in Illustris-1, but the fraction of flyby related bars decreases by a factor of $\sim 2.4$. 


\section{Impact of gas fraction, star formation, black hole and dark matter halo} 

Previous studies using idealized simulations and cosmological hydrodynamical simulations found that the gas component, star formation, feedback from supernovae and AGN, and properties of dark matter halo play important roles in bar formation and evolution. Meanwhile, the star formation history and growth of super massive black hole should be different between Illustris-1 and TNG100 due to the implemented different models, which would influence the bar formation in disc galaxies. We examine these factors in this section. 
\begin{figure*}[htbp]
\begin{center}
\includegraphics[height=0.25\textwidth, width=0.35\textwidth,trim=3 0 20 20,clip]{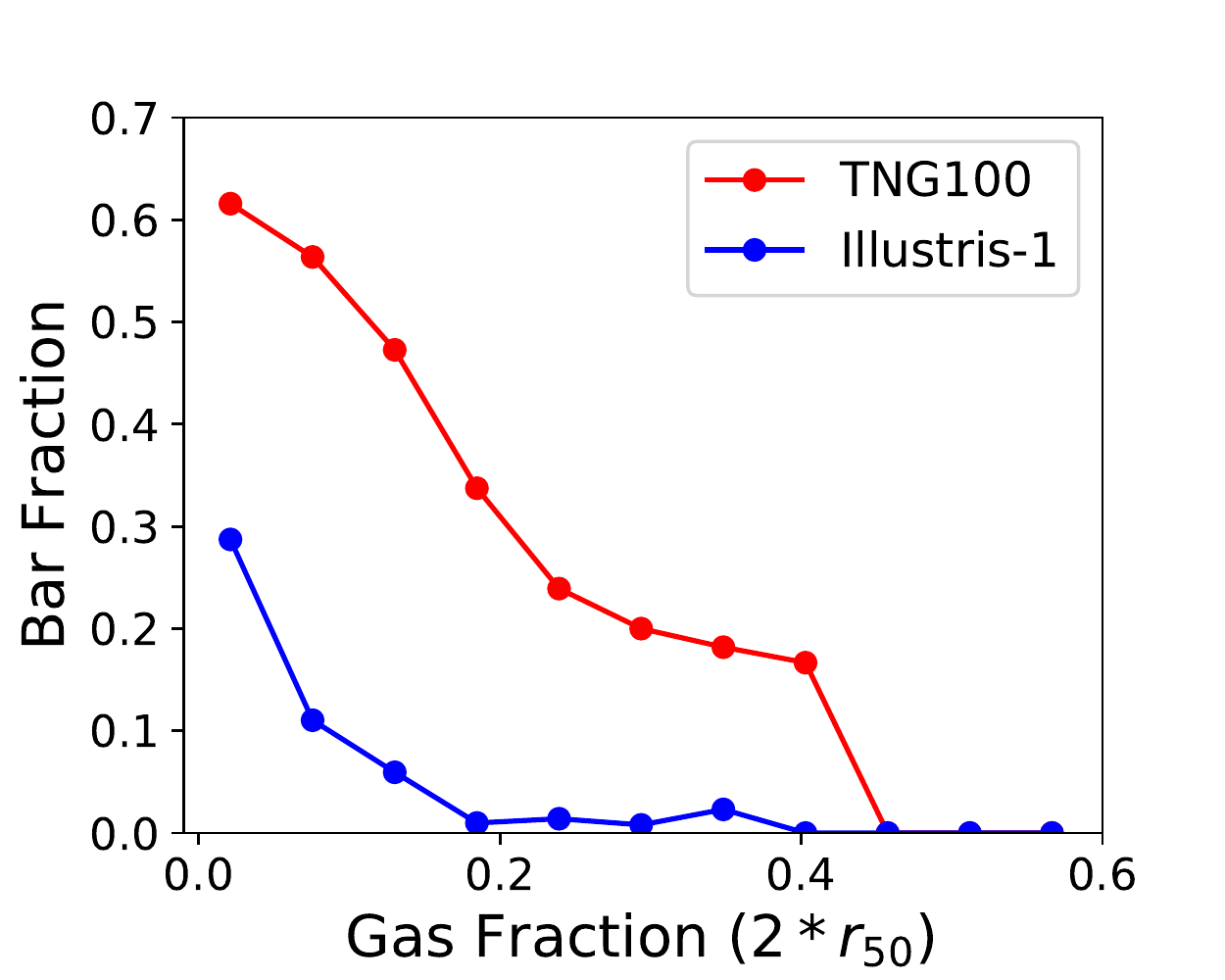}
\includegraphics[height=0.25\textwidth,width=0.35\textwidth,trim=3 0 20 20,clip]{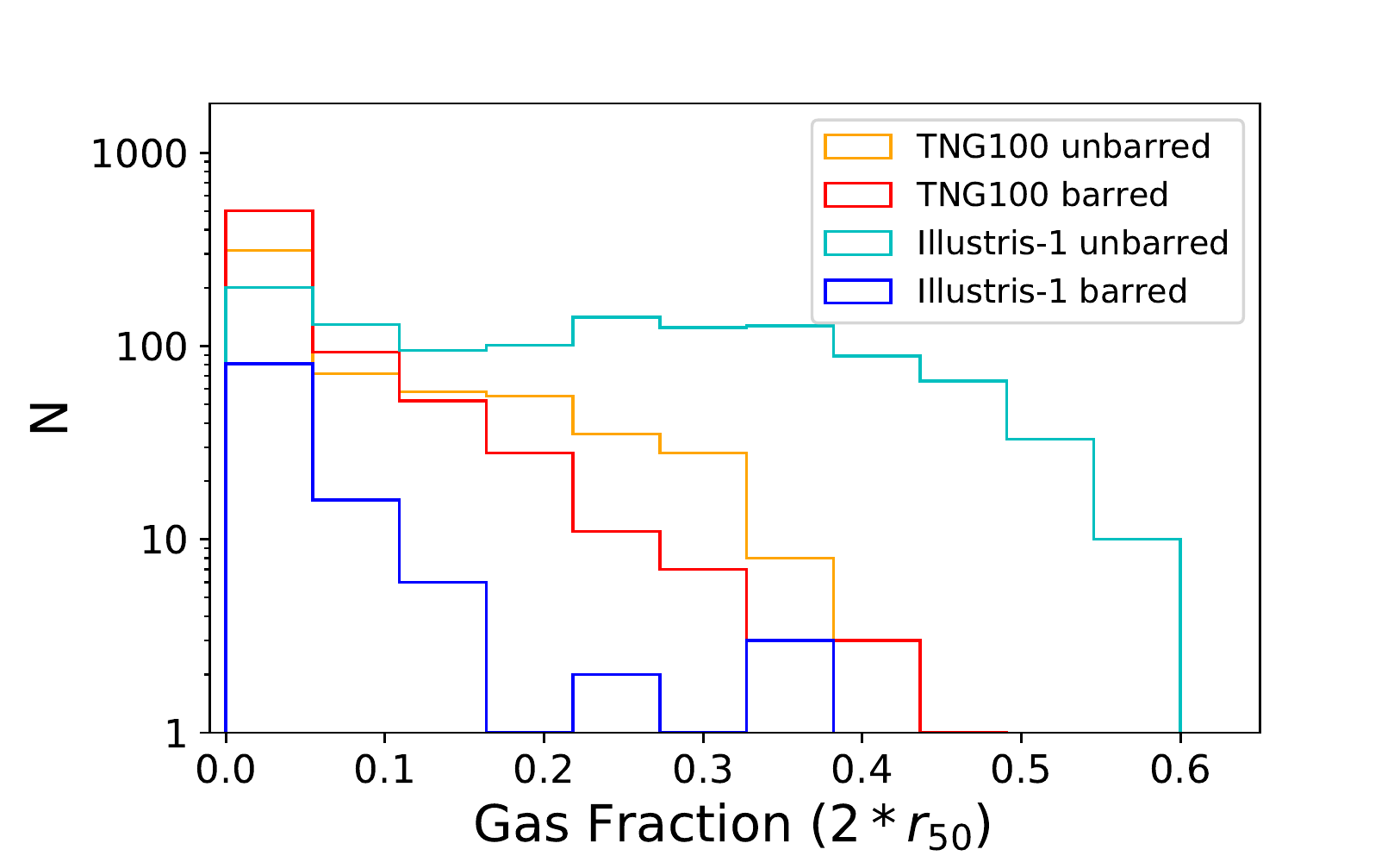}
\includegraphics[width=0.28\textwidth,trim=3 0 20 20,clip]{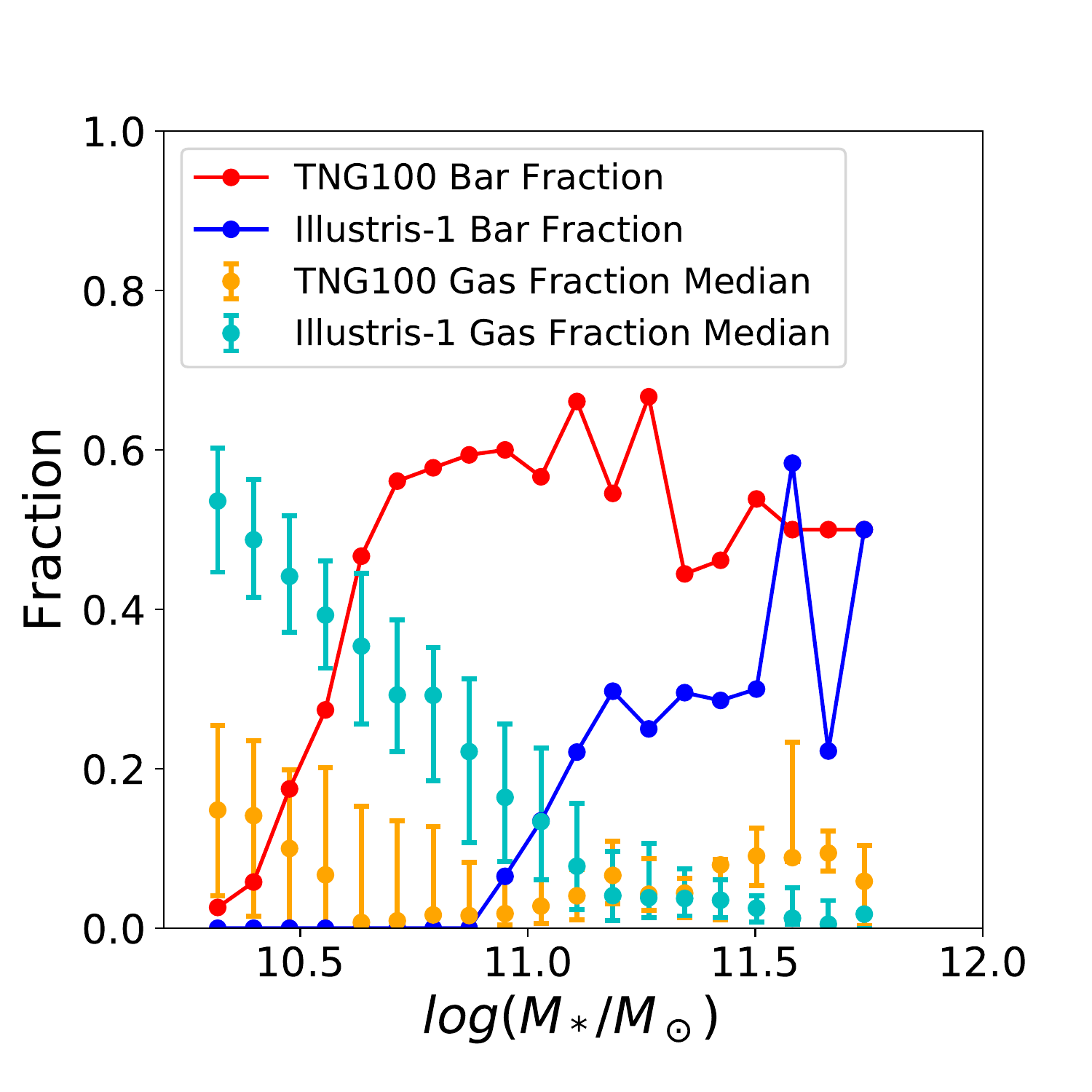}  
\includegraphics[height=0.25\textwidth, width=0.35\textwidth,trim=3 0 20 20,clip]{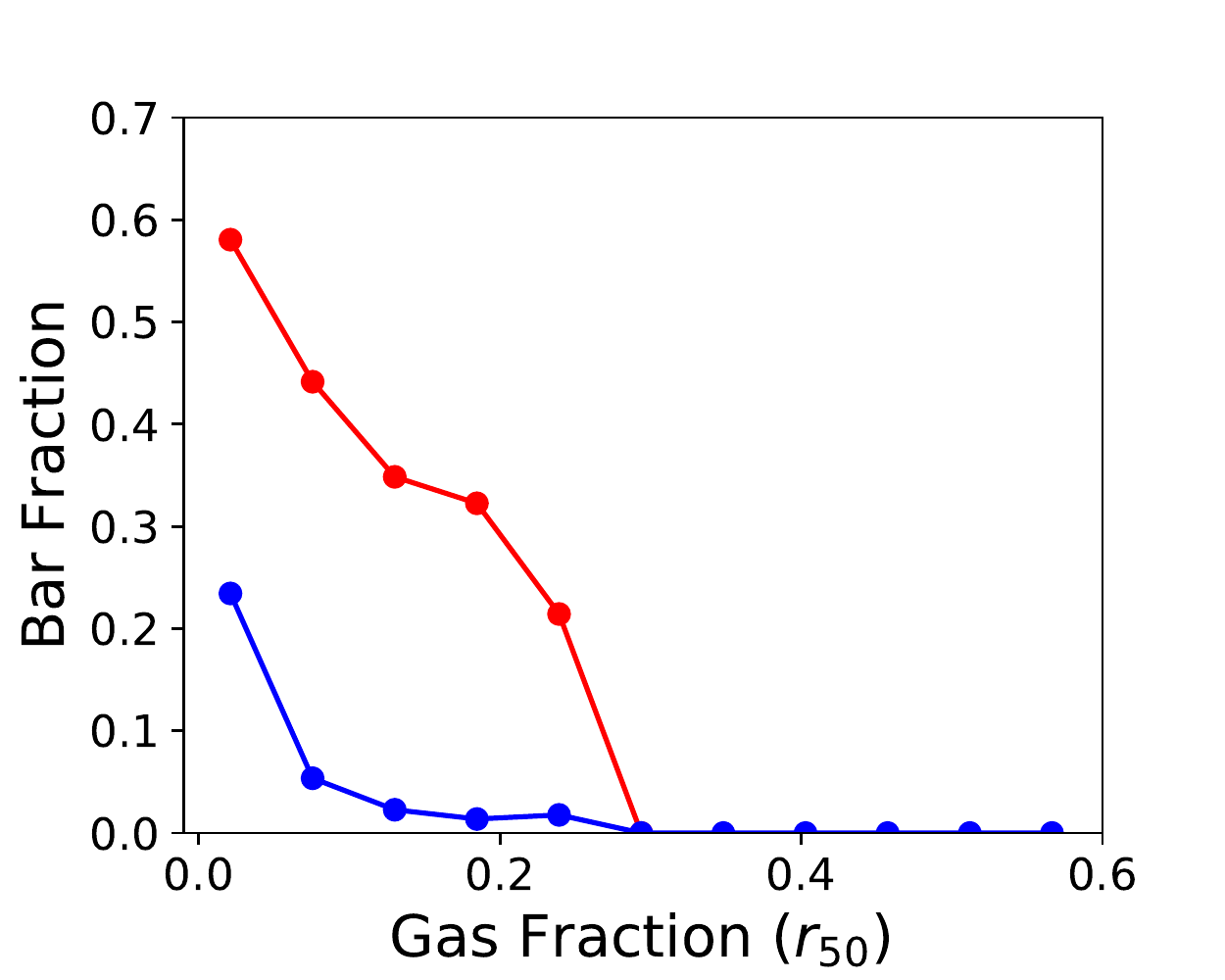}
\includegraphics[height=0.25\textwidth,width=0.35\textwidth,trim=3 0 20 20,clip]{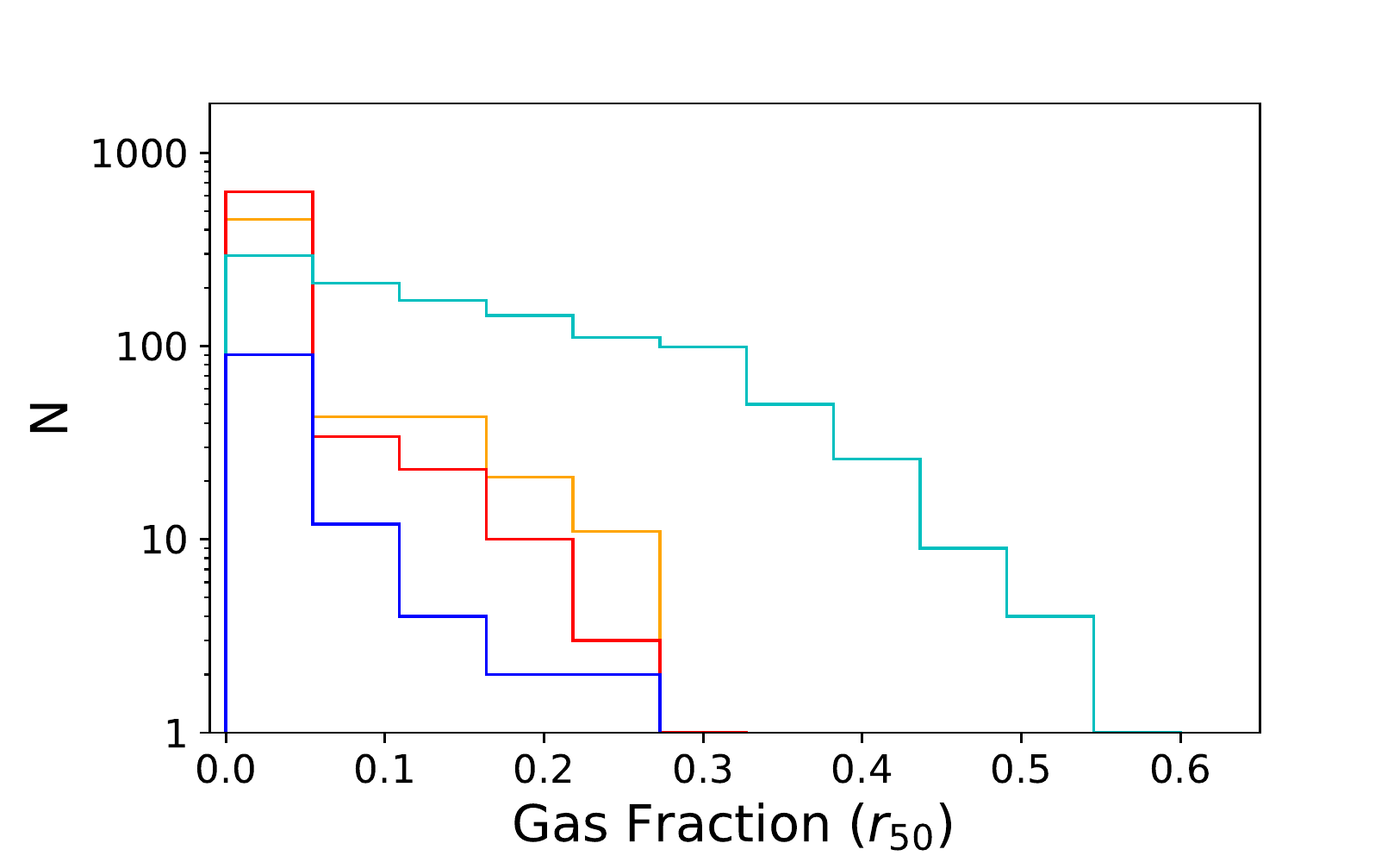}
\includegraphics[width=0.28\textwidth,trim=3 0 20 20,clip]{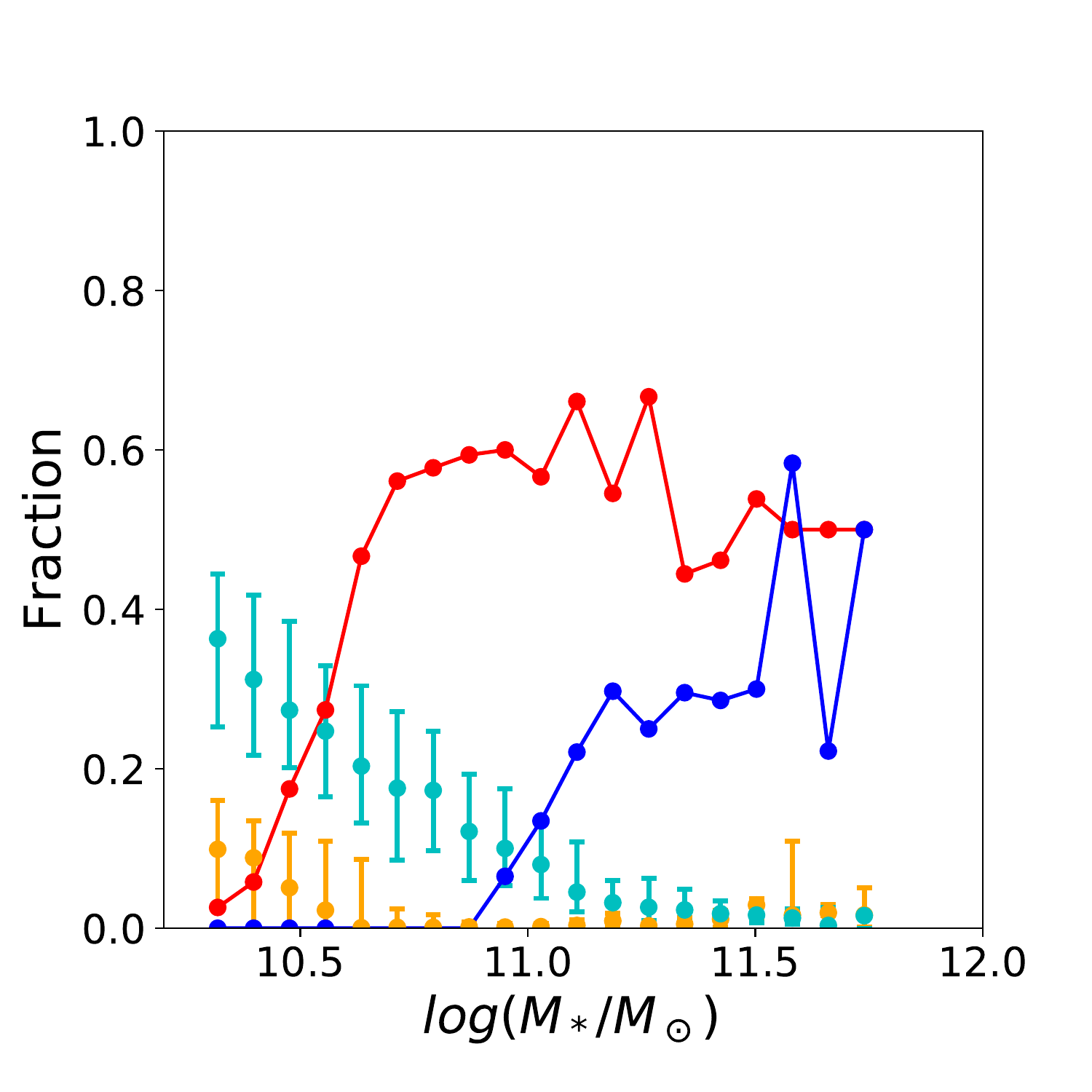}  
\caption{Left column: Bar fraction as a function of gas fraction in disc galaxies. Middle column: Histogram of barred and unbarred disc galaxies in different gas fraction bins. Right column: Gas fraction and bar fraction as functions of stellar mass. The upper and lower bars of gas fraction represent the 25th and 75th percentiles in each bins. The left two column shows results of disc galaxies with stellar mass $M_*>10^{10.5}M_{\odot}$. Top row: gas fraction measured within twice of the stellar half mass radius, i.e. $2*r_{50}$; Bottom row: gas fraction measured within the stellar half mass radius $r_{50}$.}
\end{center}
\label{fig:bar_gas_frac}
\end{figure*}


\subsection{Gas fraction}
Firstly, we explore the dependence of bar fraction on the gas content in disc galaxies. We measure the gas fraction within twice of the stellar half mass radius $r_{50}$ as $f_{gas}(2r_{50})=\frac{M_{gas}(<2*r_{50})}{M_{gas}(<2*r_{50})+M_{star}(<2*r_{50})}$, where $M_{gas}(2r_{50})$ and $M_{star}(2r_{50})$ are the mass of gas and star particles within this region respectively. The top left panel in Fig.~\ref{fig:bar_gas_frac} presents the bar fraction as a function of the gas fraction of disc galaxies at redshift z=0 in the two simulations, while the top middle panel plots the number of galaxies in different bins of gas fraction. Fig.~\ref{fig:bar_gas_frac} indicates that the bar fraction generally increases as the gas fraction decreases. Namely, bars prefer to appear in gas-poor galaxies. Especially, the bar fraction grows rapidly when $f_{gas}(z=0)$ falls below 0.25-0.30. This trend holds for both simulations, and is consistent with the analysis of Illustris-1 in \cite{2019MNRAS.483.2721P} and that of TNG100 in \cite{2020MNRAS.491.2547R}. Note that, the definition of gas fraction in \cite{2020MNRAS.491.2547R} is somewhat different from our work. This result has been actually inferred from the idealized simulations that a higher gas fraction will hinder the presences of bars (e.g. \citealt{2013MNRAS.429.1949A}). On the other hand, we should be cautious in comparing with \citep{2013MNRAS.429.1949A}, as the value of gas fraction was put by hand in the initial conditions of their isolated simulations, rather than a naturally evolved result.

\begin{figure}[htbp]
\begin{center}
\includegraphics[width=0.65\columnwidth,trim=15 15 25 35,clip]{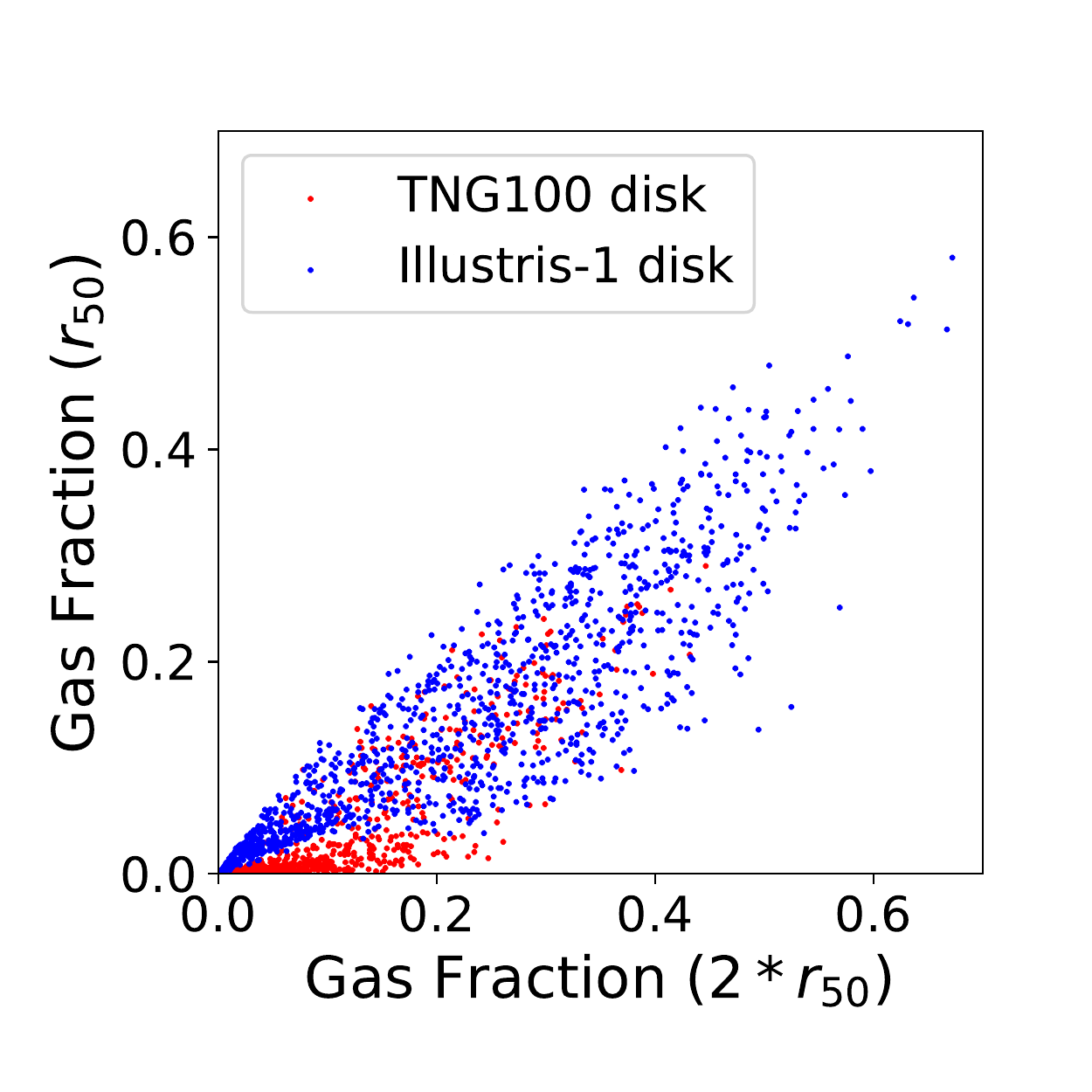}
\includegraphics[width=0.65\columnwidth,trim=15 15 25 35,clip]{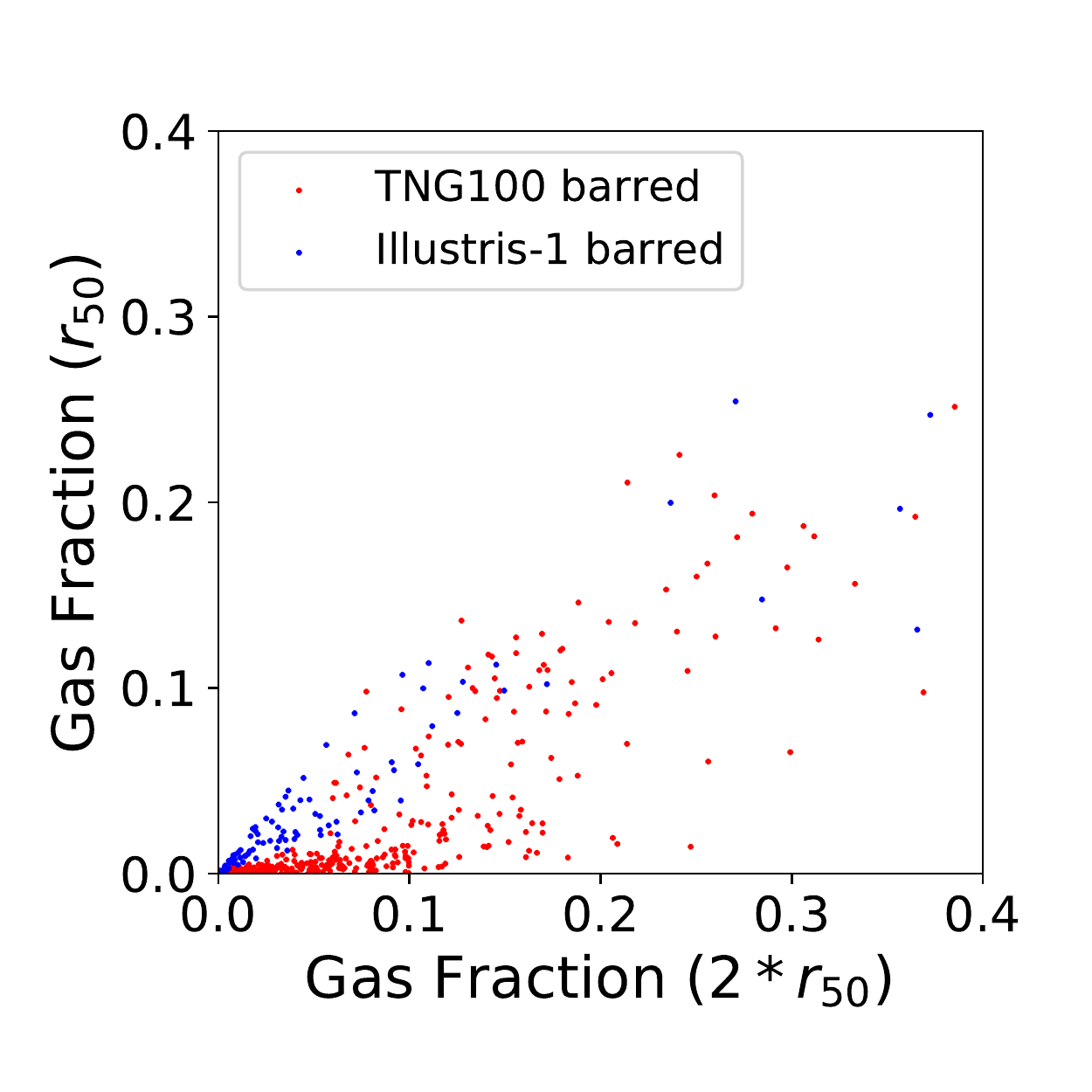}
\caption{Top: Gas fraction of disc galaxies measured within twice of half stellar mass radius against that measured within the half stellar radius; red and blue dots indicate galaxies in TNG100(Red) and Illustris-1(Blue) respectively. Bottom: same as the top panel, but only for barred galaxies.}
\end{center}
\label{fig:gasfrac_r50_2r50}
\end{figure}

The distribution of bar frequency over gas fraction in TNG100 differs from Illustris-1 in two aspects. First, the top middle panel of Fig.~\ref{fig:bar_gas_frac} shows that the gas fraction of disc galaxies in TNG100 is generally lower than disc galaxies in Illustris-1 at redshift z=0. About $\sim 67\%$ of the disc galaxies have a gas fraction lower than $0.054$ at z=0 in TNG100. In contrast, only $\sim 25\%$ of the disc galaxies in Illustris-1 have $f_{gas}<0.054$.
This feature is consistent with the statistics on gas fraction reported in \cite{2019MNRAS.486.4686K}. Given the trend mentioned above, the much lower gas fraction in TNG100 than Illustris-1 could be responsible for the higher bar fraction in TNG100. Second, we see that even in the same gas fraction bin, the bar fraction of disc galaxies in TNG100 is much higher than in Illustris-1. It suggests that in addition to gas fraction, there are some other factors that could promote the bar formation in TNG100. 

As shown in Fig.~\ref{fig:overall_bar_fraction}, the bar fractions as a function of stellar mass are different in the two simulations, TNG100 has much higher fractions than Illustirs-1 in the stellar mass range $\sim 10^{10.5-11.2} M_{\odot}$. This difference may be related to the distribution of gas fraction in different mass bins. The top right panel of Fig.~\ref{fig:bar_gas_frac} shows the median gas fraction as a function of stellar mass in the two simulations respectively. In Illustris-1, the median gas fraction decreases gradually from $\sim 40\%$ at $M_*=10^{10.5}M_{\odot}$ to $\sim 10\%$ at $M_*=10^{11.1}M_{\odot}$ and drops below $\sim 10\%$ for $M_*>10^{11.1}M_{\odot}$, which is reverse to the change of bar fraction. In TNG100, the median gas fraction is lower than $10 \%$ in all the mass bins, with relatively higher fractions at the high mass end, and larger scatters at the low mass end. The difference of bar fraction in different mass bins between the two simulations is consistent with the effect of gas fraction on bar presence. As Fig.~\ref{fig:overall_bar_fraction} includes galaxies with $10^{10.25}M_{\odot}<M_*<10^{10.5}M_{\odot}$, we also shown their median gas fractions in the top right panel of Fig.~\ref{fig:bar_gas_frac}. In Illustris-1, the gas fraction in systems less massive than $10^{10.5}M_{\odot}$ also decreases with increasing stellar mass, in agreement with the trend at $M_*>10^{10.5}M_{\odot}$. A similar trend is observed at the low mass end of TNG100, but with larger scatters.

As bars locate in the central region of galaxies, their formation and evolution may also correlate with the gas fraction in the more inner region of galaxies. We measure the gas fraction within $r_{50}$, $f_{gas}(r_{50})$, and show the dependence of bar fraction and stellar mass on it, as well as the distribution of galaxies in the bottom row of Fig.~\ref{fig:bar_gas_frac}. Bars are found in galaxies with $f_{gas}(r_{50})<0.3$, and the bar fraction decreases with increasing $f_{gas}(r_{50})$. Gas fractions within $r_{50}$ are lower than $f_{gas}(2r_{50})$ in both simulations, but the drop in TNG100 is more significant, which can be displayed by the scatter plot of $f_{gas}(2r50)$ against $f_{gas}(r50)$ in Fig.\ref{fig:gasfrac_r50_2r50}. This feature is likely to be caused by the more effective feedback from star formation and AGN accretion in the TNG simulations(\citealt{2018MNRAS.479.4056W}, \citealt{2018MNRAS.473.4077P}), which would expel gas more efficiently from the central region. Meanwhile, being more gas-poor in the central region could make the disc galaxies in TNG more favourable to form bar.

\begin{figure}[htbp]
\begin{center}
\includegraphics[width=1.0\columnwidth,trim=10 5 10 5,clip]{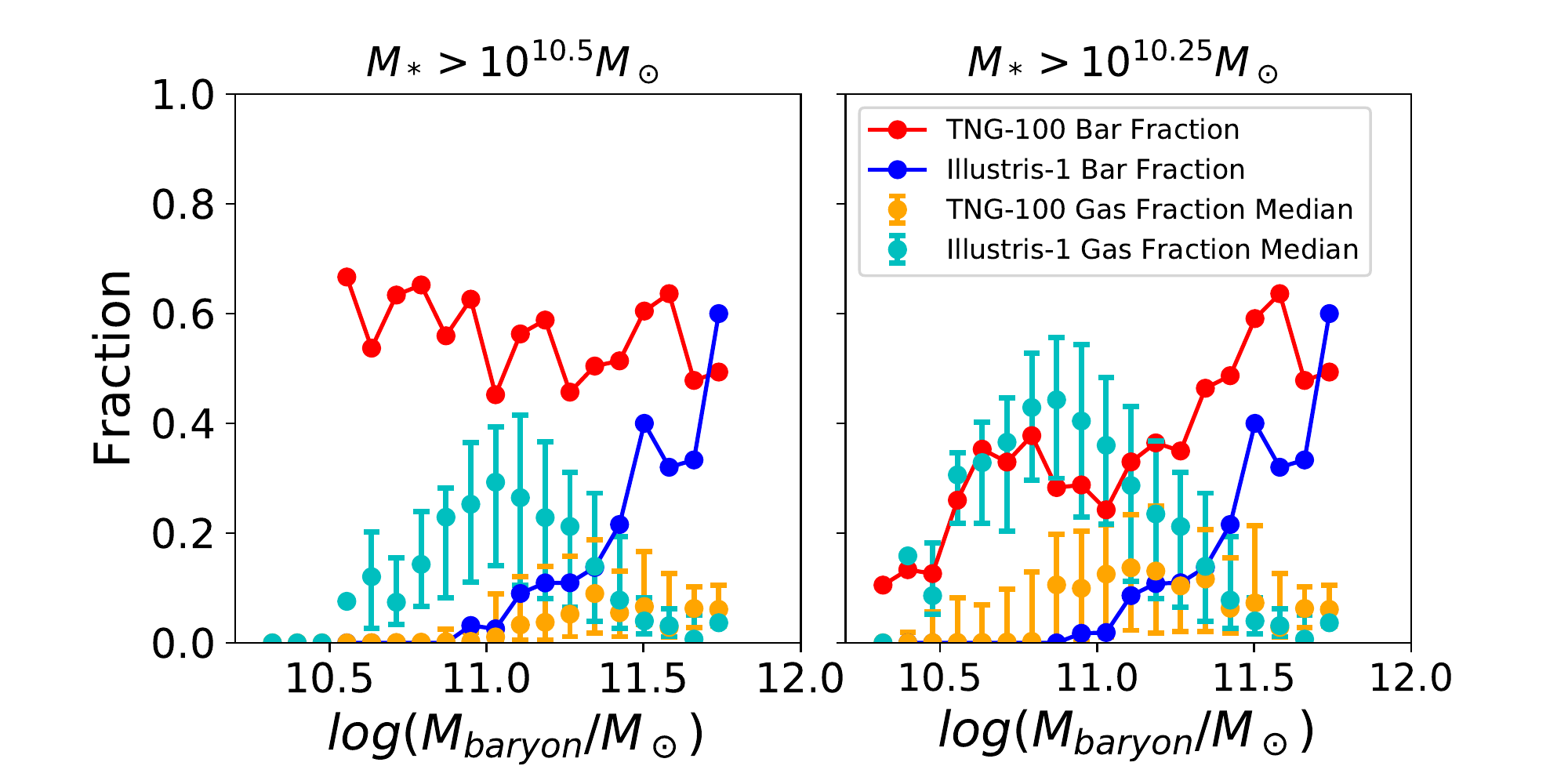}
\includegraphics[width=1.0\columnwidth,trim=10 5 10 5,clip]{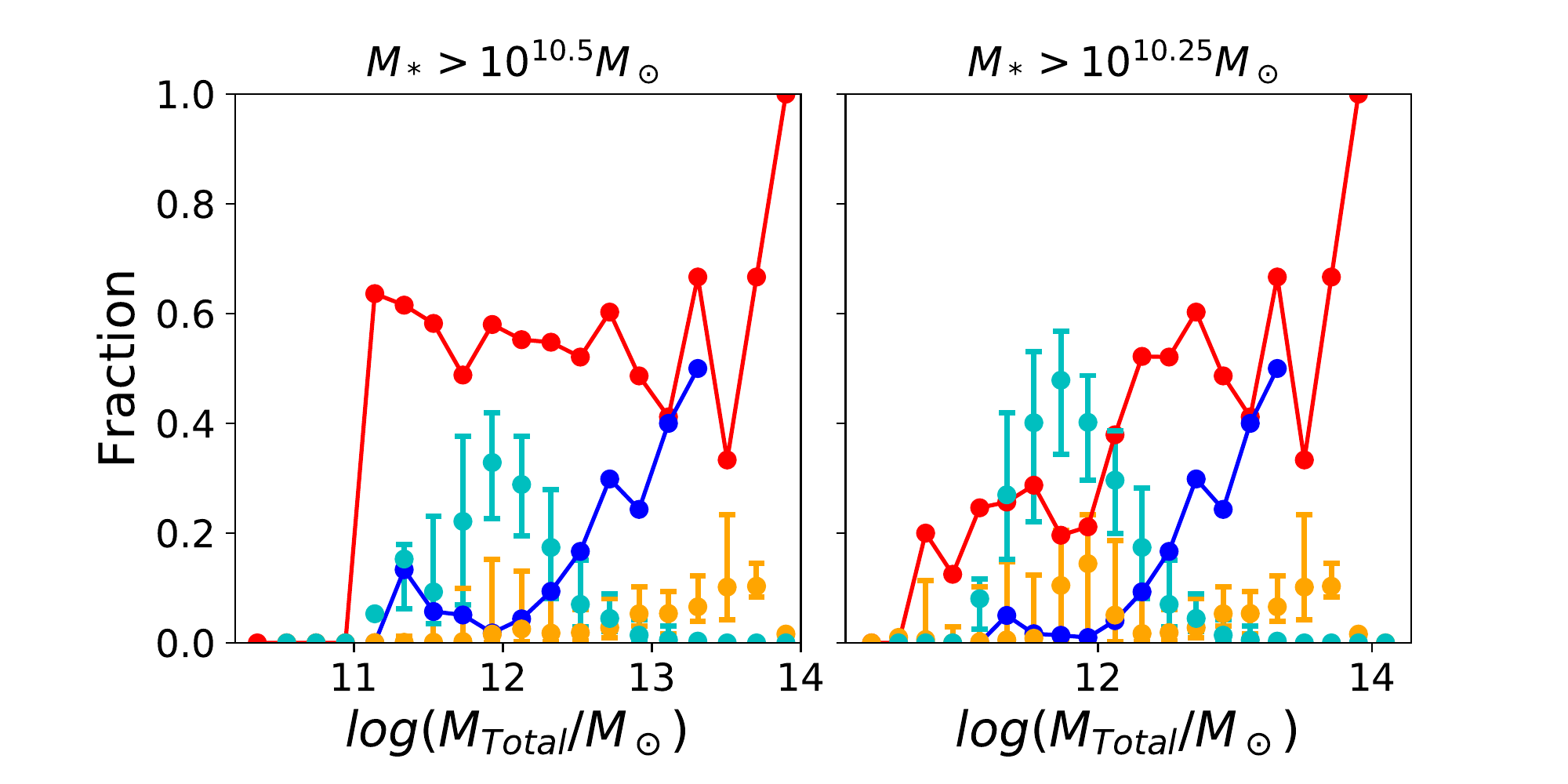}
\caption{Top and bottom plots indicate gas fraction and bar fraction as a function of baryon(gas+star) mass, of total(gas+star+dark matter) mass respectively. Left: Disc galaxies with stellar mass $M_*>10^{10.5} M_{\odot}$ are analysed. Right: Disc galaxies with stellar mass $M_*>10^{10.25} M_{\odot}$ are analysed.}
\end{center}
\label{fig:bar_fraction_and_gas_fraction}
\end{figure}

In Fig.~\ref{fig:bar_fraction_and_gas_fraction}, we plot the bar fraction and gas fraction as functions of the total baryonic mass i.e., sum of gases and stars, and of the total mass of baryonic and dark matter, within $2r_{50}$ at $z=0$, respectively. The trends are generally similar to functions of stellar mass shown in Fig.~\ref{fig:bar_gas_frac}. Galaxies with baryonic(total) mass less than $\sim 10^{11.5}(10^{12.7})M_{\odot}$ in Illustris-1 have much higher gas fractions and lower bar fractions than in TNG100. However, the gas fraction in Illustris-1 increase with increasing baryonic(total) mass at $M_{baryon}<10^{11.0}M_{\odot}$($M_{total}<10^{12.0}M_{\odot}$). This pattern is mainly due to the fact that we have placed a threshold of stellar mass for sample selection. The bins of galaxies with small baryonic/total mass would be biased by galaxies with low gas fraction. To demonstrate that, Fig.~\ref{fig:bar_fraction_and_gas_fraction} gives the results of two different thresholds of stellar mass. For instance, when the threshold decrease from $M_*=10^{10.5}M_{\odot}$ to $M_*=10^{10.25}M_{\odot}$, the median gas fraction of galaxies in the range $10^{10.5}M_{\odot}<M_{baryon}<10^{11.2}M_{\odot}$ are enhanced. This is because a lower threshold would include many gas rich galaxies with stellar mass of $10^{10.25}M_{\odot}<M_*<10^{10.5}M_{\odot}$ into the sample.

\begin{figure}[htbp]
\begin{center}
\includegraphics[width=0.85\columnwidth,trim=15 10 10 30,clip]{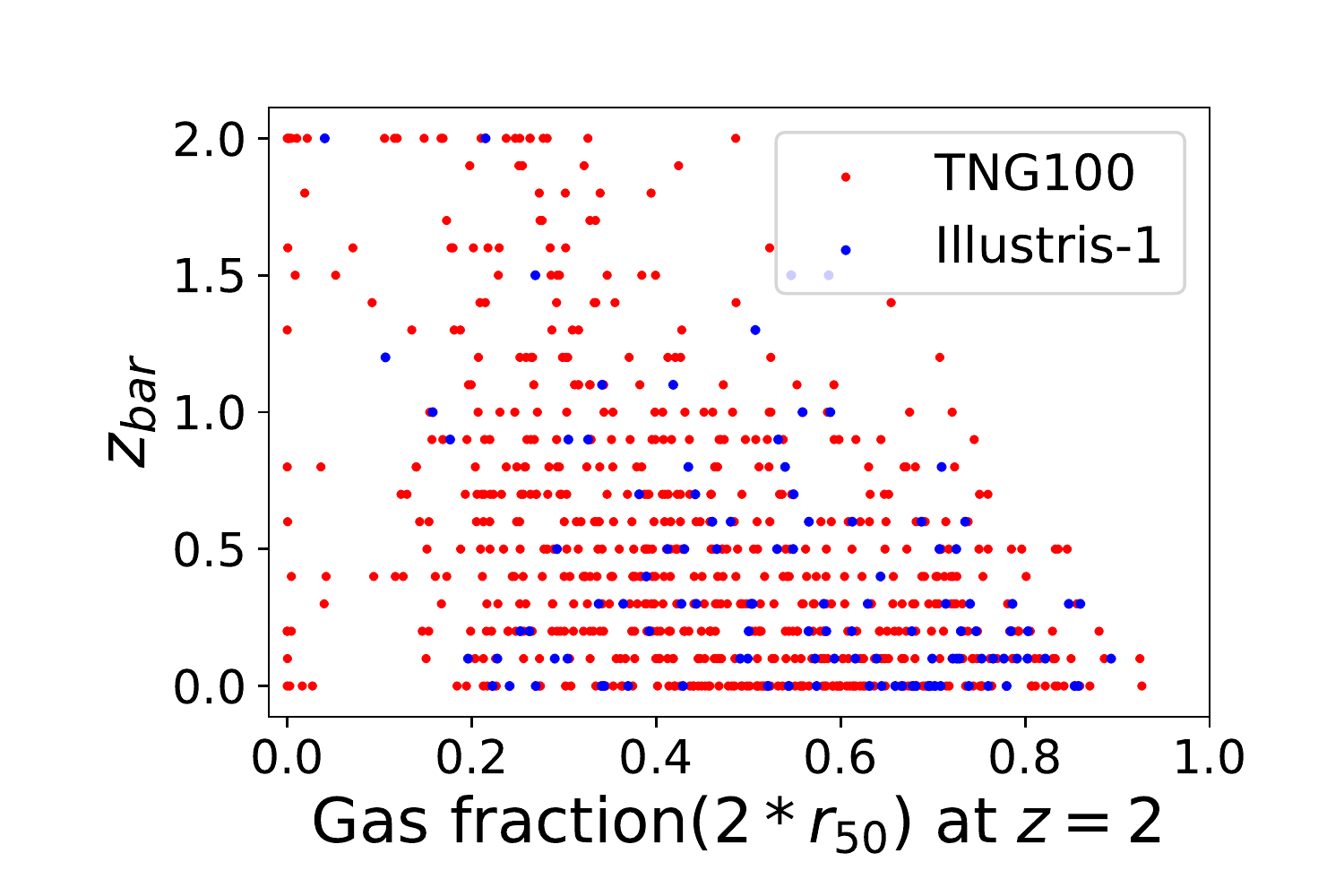}
\includegraphics[width=0.85\columnwidth,trim=15 10 10 30,clip]{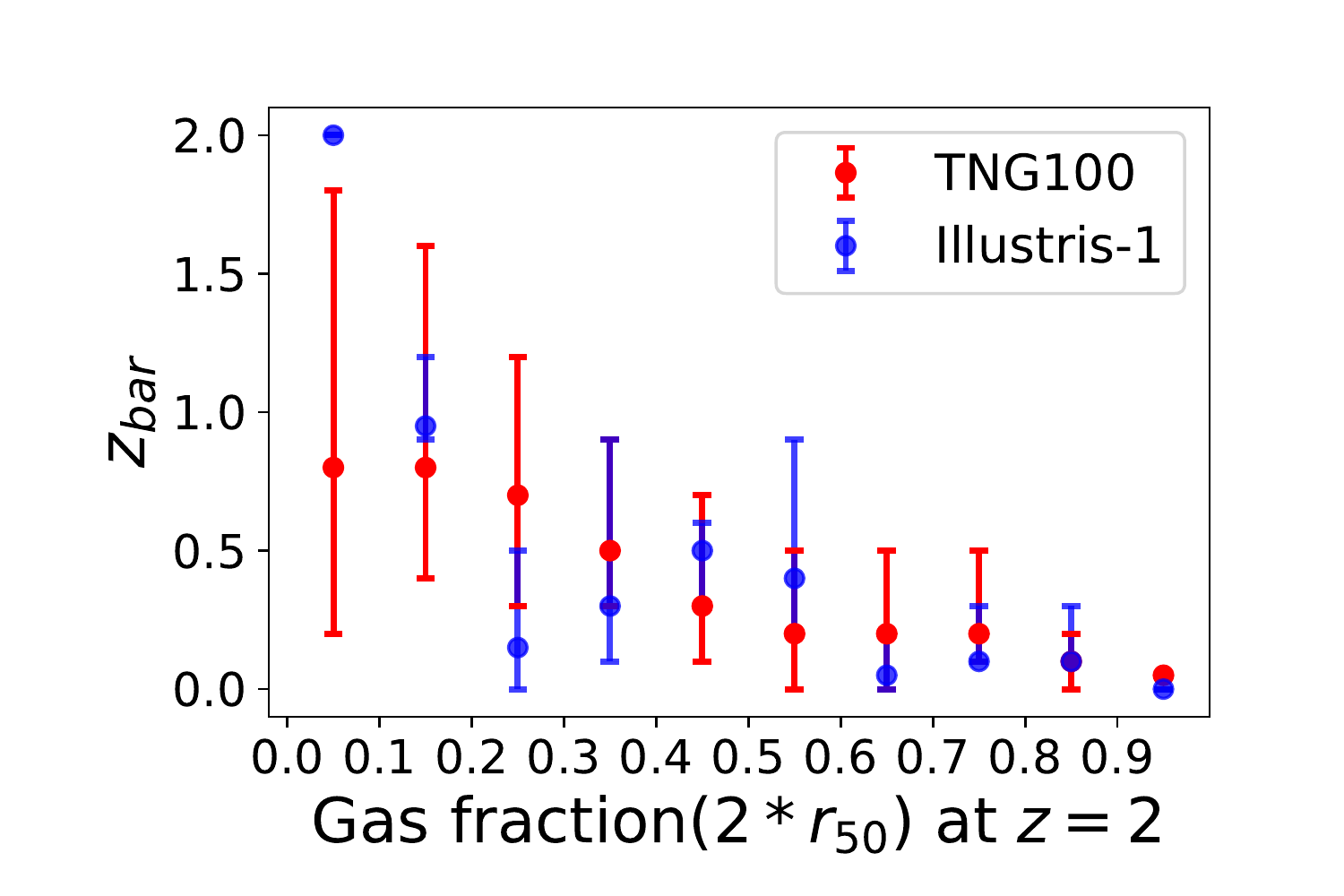}
\includegraphics[width=0.83\columnwidth,trim=0 0 10 30,clip]{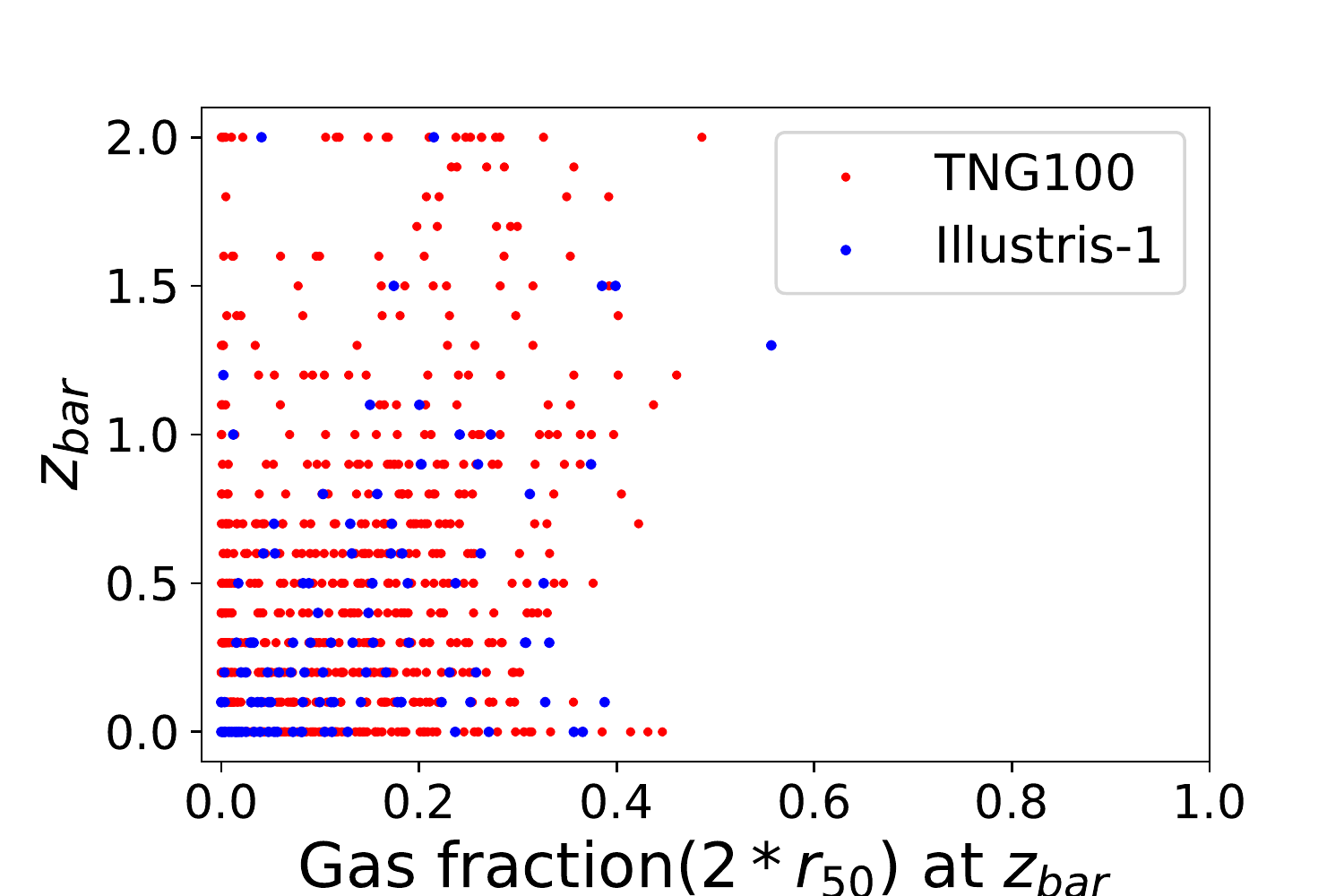}
\caption{Top: The gas fraction $f_{gas}(2r50)$ at z=2 of disc galaxies that have a bar at z=0 against the bar formation redshift. Middle: Filled circles indicates the median redshift of bar formation in different bins of gas fraction at z=2; the upper and lower bars represent the 25th and 75th percentiles in each bins. Bottom: The gas fraction at the bar formation redshift of galaxies that have a bar at $z=0$.}
\end{center}
\label{fig:zbar_scatter}
\end{figure}

To further probe the impact of gas on bar formation, we trace the evolution of gas fraction in disc galaxies toward high redshifts. For those galaxies developed bars at z=0, Fig.~\ref{fig:zbar_scatter} shows the bar formation redshift against their gas fraction $f_{gas}(2r50)$ at $z=2$, since when most of the bars appear. We can see that galaxies with a relatively higher $f_{gas}(2r50)$ at $z=2$ will form a bar at relatively lower redshifts, and vice versa. For those galaxies with $f_{gas}(2r50)>0.5$ at $z=2$, the median redshift of bar formation are lower than $0.4$ and depend weakly on gas fraction in both simulations, shown by the middle panel. But the median redshift of bar fraction increases from $z \sim 0.4-0.5$ for $0.3<f_{gas}(2r50)<0.4$ to $z \sim 0.75-0.80$ for $0.0<f_{gas}(2r50)<0.1$ at $z=2$. This trend agrees with \cite{2013MNRAS.429.1949A}. We also measure the gas fraction of barred galaxies at their bar formation redshifts and present the result in the bottom plot of Fig.~\ref{fig:zbar_scatter}. For most of the barred galaxies, the gas fractions $f_{gas}(2r50)$ were lower than 0.4 at the epoch when their bars can be identified by our algorithm. This feature would be helpful to explain the enhanced bar fractions in galaxies with $f_{gas}(2r50)<0.3-0.4$ at $z=0$ as shown in Fig.~\ref{fig:bar_gas_frac}.

\begin{figure}[htbp]
\begin{center}
\includegraphics[width=0.82\columnwidth,trim=5 0 10 40,clip]{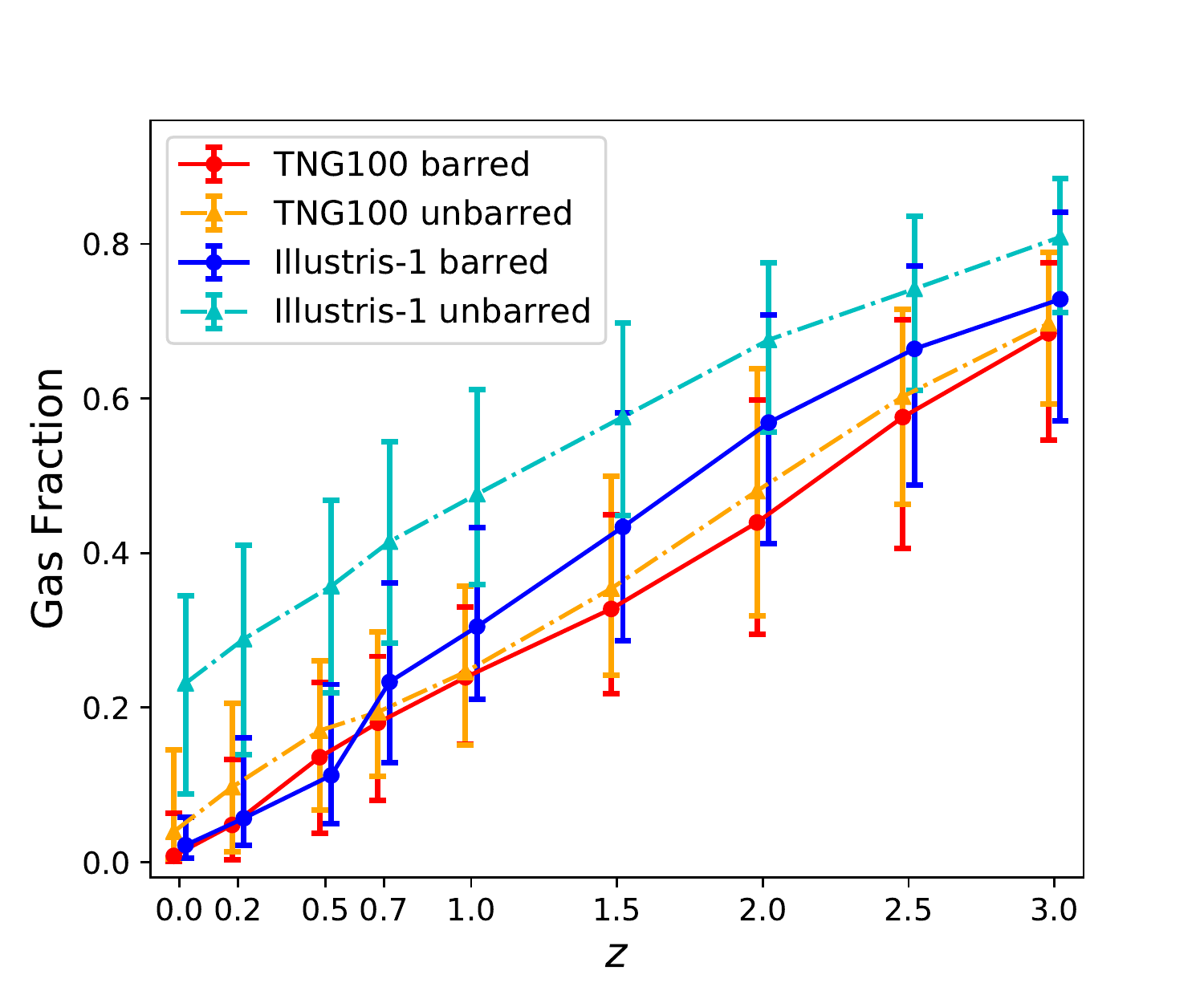}
\includegraphics[width=0.82\columnwidth,trim=5 0 10 40,clip]{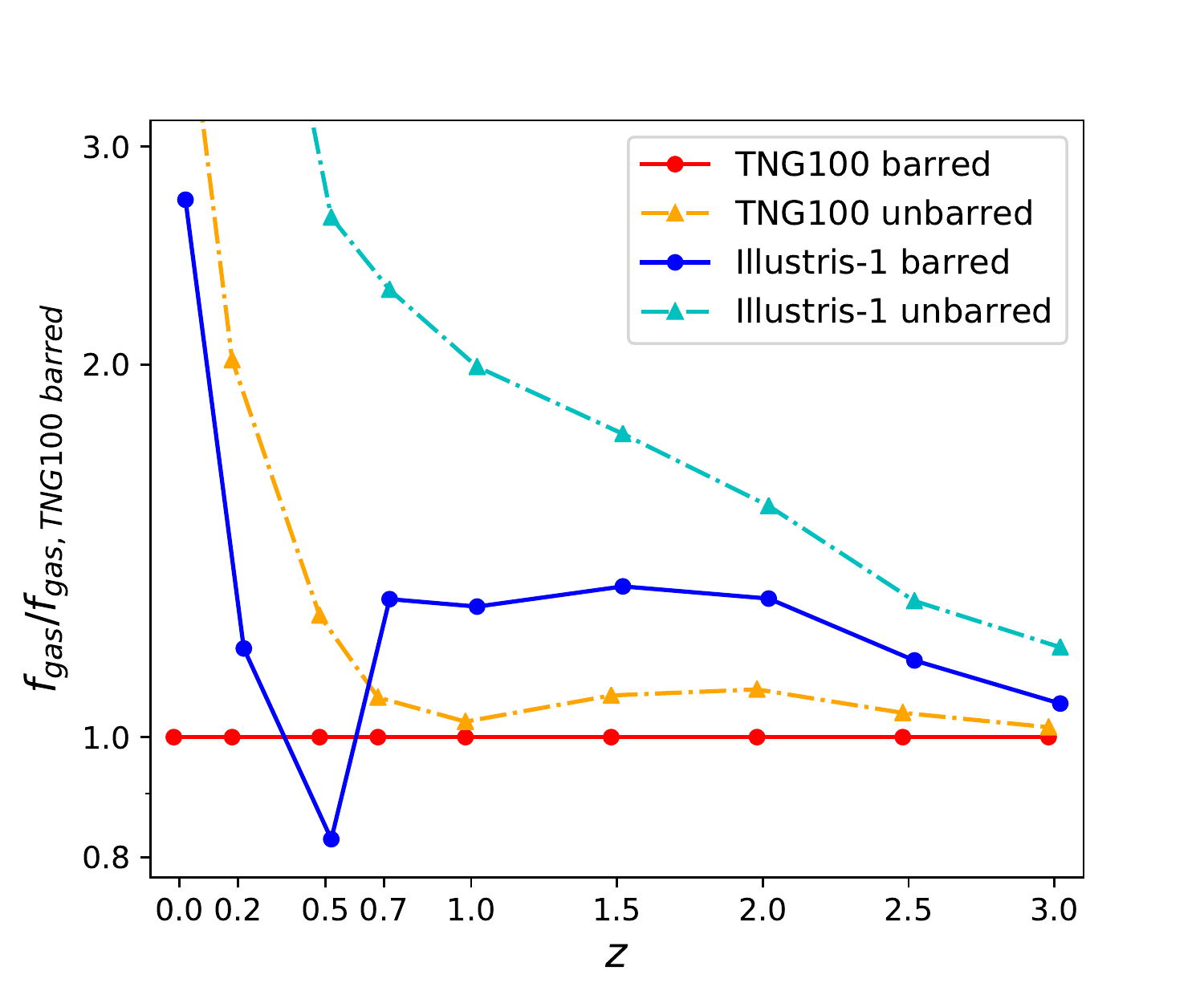}
\caption{Top: Evolution of median gas fraction $f_{gas}(2r50)$ in disc galaxies in TNG100 and Illustris-1. Solid(dashed-dotted) lines indicate galaxies that have(does not have) a bar at z=0. The upper and lower bars represent the 25th and 75th percentiles in each bins for each category. Red and orange colors represent galaxies samples in TNG100. Blue and cyan colors represent samples in Illustris-1. Bottom: The ratio of median gas fraction of galaxies in four categories over that of TNG100 barred galaxies data.}
\end{center}
\label{fig:bar_unbar_gas_fraction}
\end{figure}

The top panel of Fig.~\ref{fig:bar_unbar_gas_fraction} shows the evolution history of the median gas fraction for the disc galaxies more massive than $10^{10.5}M_{\odot}$ that found at z=0 in the two simulations. The bottom panel presents the ratio of median gas fraction in each category, normalized by the median gas fraction of TNG100 barred galaxies. Generally, the gas fraction in all the galaxies decreases with decreasing redshift. The systematic differences of gas fraction between z=0 barred and unbarred galaxies, and between the two simulations can be traced back to high redshifts up to $z\sim 3$. Again, this feature agrees with the scenario that gas component would delay the formation and growth of bars. In addition, the difference between barred and unbarred galaxies in Illustris-1 is more significant than in TNG100. The gas fraction of barred galaxies at z=0 in Illustris-1 decreases sharply at $z<0.7$ and then are getting close to their counterparts in TNG100. At $z<=0.5$, the median gas fraction of TNG100 barred galaxies is small, resulting in fluctuations in the ratio plot.


\begin{figure}[htbp]
\centering
\begin{center}
\includegraphics[width=0.85\columnwidth,trim=0 0 0 0,clip]{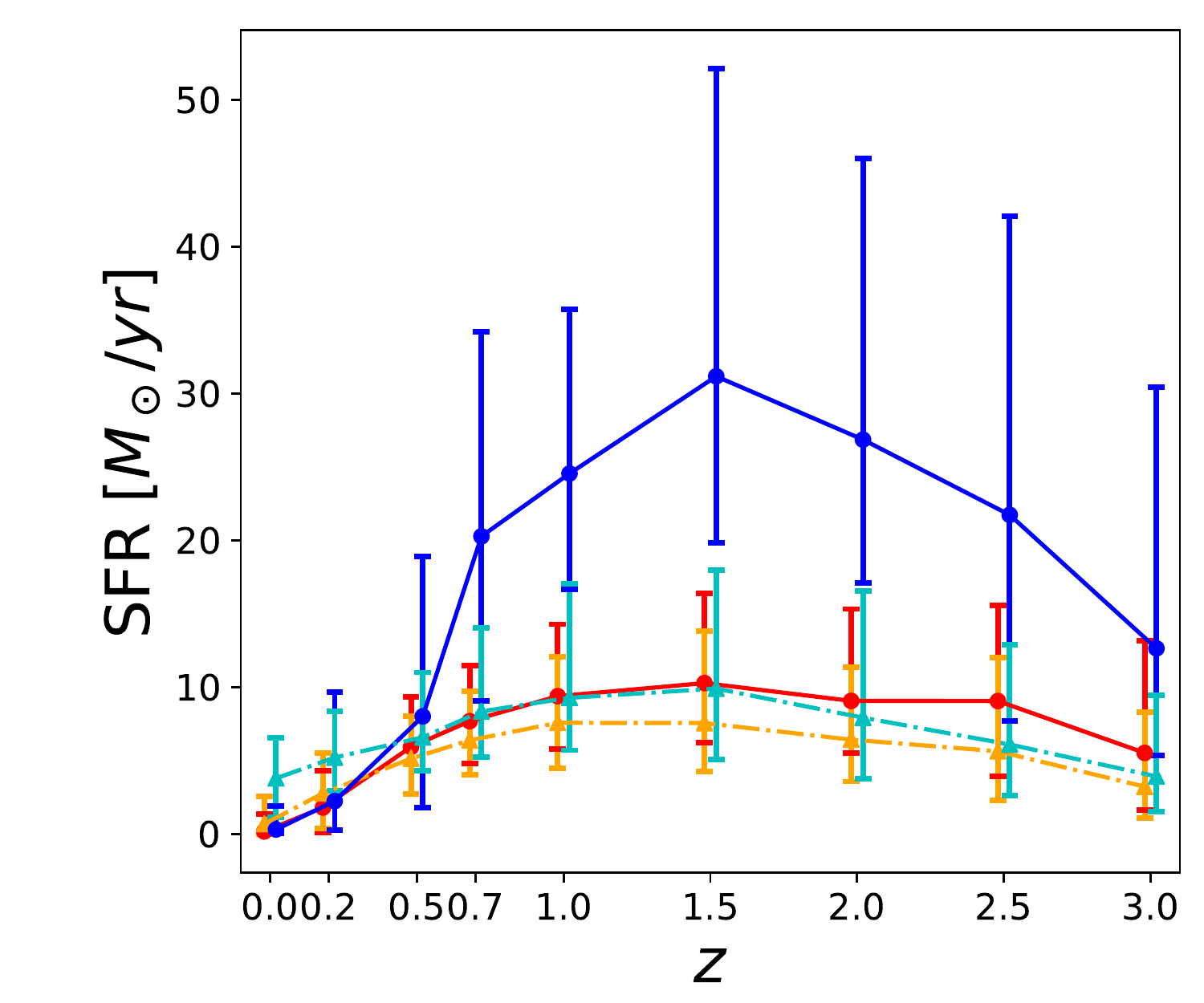}
\includegraphics[width=0.85\columnwidth,trim=0 0 0 0,clip]{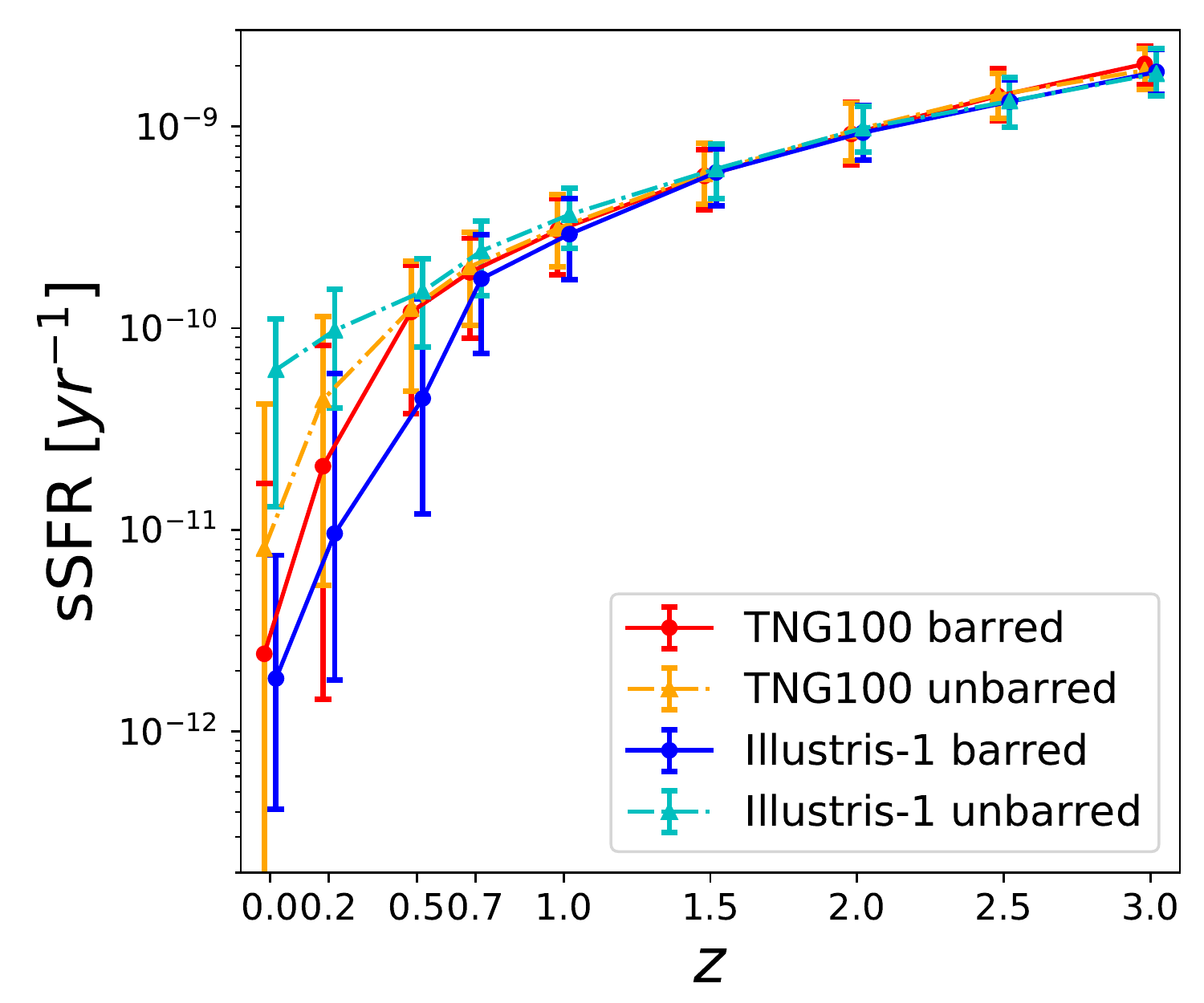}
\includegraphics[width=0.85\columnwidth,trim=0 0 0 0,clip]{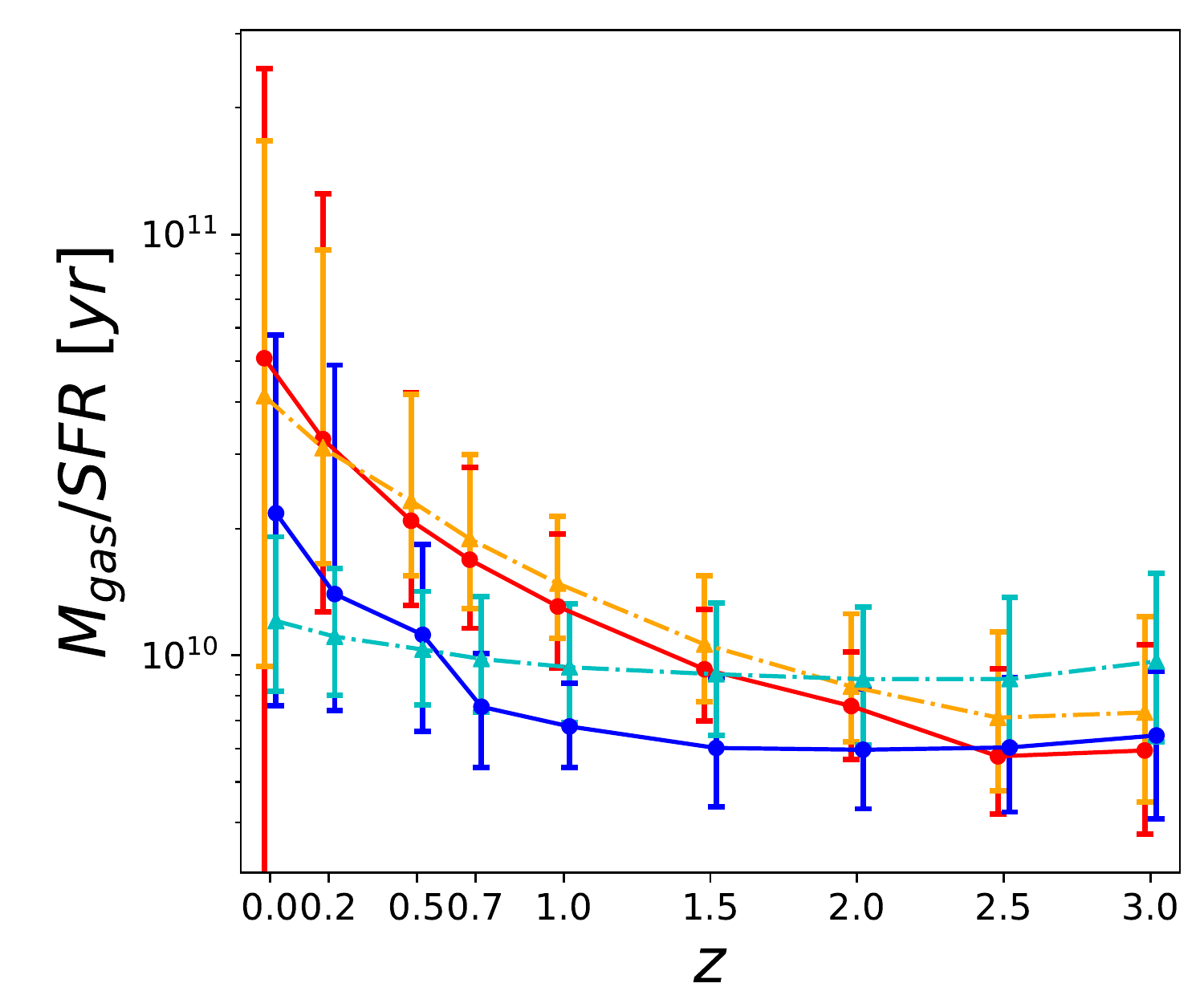}
\caption{Top, middle and bottom panels show the evolution of median star formation rate, specific star formation rate and gas depletion time in disc galaxies respectively. The meaning of lines and bars are the same as in Fig.~\ref{fig:bar_unbar_gas_fraction}}.
\end{center}
\label{fig:global_sfr}
\end{figure}


\subsection{Star formation}
An important physical process related to gas component is star formation. Also, feedback from massive stars and supernovae can affect the distribution of gas and stars, and influence the bar formation. In previous studies, the link between star formation activity and bar formation has been investigated. However, whether bar can suppress or enhance the star formation activity is still a controversial issue. Nevertheless, it has been shown that more effective stellar feedback can help to build stronger and longer bars(e.g. \citealt{2019MNRAS.488.1864Z}). We now explore the star formation activity in our samples. Fig.~\ref{fig:global_sfr} presents the evolution of star formation rate(SFR) in the simulated disc galaxies with $M_*>10^{10.5}M_{\odot}$ at z=0. In both simulations, the SFR generally increases from $z=3.0$ to $z\sim 1.5$ and then declines as redshift decreases, where barred galaxies fall faster than unbarred galaxies. The SFRs of barred and unbarred galaxies in Illustris-1 are generally higher than their counterparts in TNG100. The barred galaxies in Illustris-1 have much higher SFR than other three categories at $z \geq 0.5$, which may be because the barred galaxies in Illustris-1 are more massive, as shown by Fig.~\ref{fig:overall_bar_fraction}. SFR of barred galaxies in TNG100 is higher than unbarred galaxies in Illustris-1 before $z=1.5$, but the situation turn over since then. Above $z\sim0.3-0.5$, the star formation rate of barred galaxies are higher than unbarred galaxies in both simulations. The situation is reversed below $z\sim 0.3-0.5$, which is close to the median redshift of bar formation. \cite{2020MNRAS.491.2547R} reported similar result for the TNG100 disc galaxies in the stellar mass range $10^{10.4-11} M_{\odot}$ and they argue that the presence of bars can promote quenching in the galaxy central region. 

The middle panel of Fig.~\ref{fig:global_sfr} shows the evolution of specific star formation rate(sSFR) in the disc galaxies samples. At redshifts higher than $0.7$, the sSFRs are almost the same for all sub-samples.  After $z=0.7$, when most of the bars began to appear, the sSFR in barred galaxies drop more rapidly than unbarred galaxies.  At redshifts lower than $0.5$, unbarred and barred galaxies in Illustris-1 have the highest and lower median sSFR respectively. Multiple factors may account for the drop of sSFR in barred galaxies after bar formation. The formation of bar could stabilize the gas disc and inhibit star formation, as suggested by \cite{2018A&A...609A..60K}. Alternatively, it is also likely that the progenitors of barred galaxies, which have the higher star formation rates than unbarred galaxies, have consumed more gas before the bar formation(\citealt{2017ApJ...845...93K}). A detail investigation is needed to justify this issue, which however is out of the scope of this work.

The bottom panel of Fig.~\ref{fig:global_sfr} shows the depletion time of gas, which is denoted as $\tau_g=M_{gas}/SFR$ and is the inverse to star formation efficiency(SFE), of disc galaxies. Barred galaxies have shorter gas depletion time, i.e., higher SFE, than unbarred galaxies at $z>0.2$ in TNG100 and at $z>0.5$ in Illustris-1. At $z<\sim1.5-2.0$, the star formation efficiency of galaxies in TNG100 is lower than Illustris-1. At $z>2.0$, the SFE in TNG100 is moderately higher than Illustris-1. For these massive disc galaxies with $M_*>10^{10.5}M_{\odot}$ at $z=0$, the suppression of star formation in TNG100 is probably more efficient than in Illustris-1 at $z<\sim 2$, but less efficient at $z>2$. Actually, the amplitude of galaxy stellar mass function in TNG100 is higher than that in Il within the range $M_*\sim 5\times 10^{9}-7\times10^{10} M_{\odot}$ at $z=2$, and in the range $M_*>\sim 5\times 10^{9} M_{\odot}$ at $z>=3$(see Figure 14 in \citealt{2018MNRAS.475..648P}). A relatively higher SFE at high redshifts could help the disc galaxies in TNG100 to consume more gas, inject more stellar feedback energy and hence lower down the disc gas fraction. On the other hand, combining with the results of SFR,  energy injected into massive disc galaxies from the stellar feedback in TNG100 is probably no more than in Illustris-1 at $z<\sim1.5-2.0$. We will further discuss the possible impact of stellar feedback to gas content and bar formation in the next subsection, together with AGN feedback.

\subsection{Growth of super massive black hole and feedback}
\begin{figure}[htbp]
\centering
\begin{center}
\includegraphics[width=0.85\columnwidth,trim=25 10 10 15,clip]{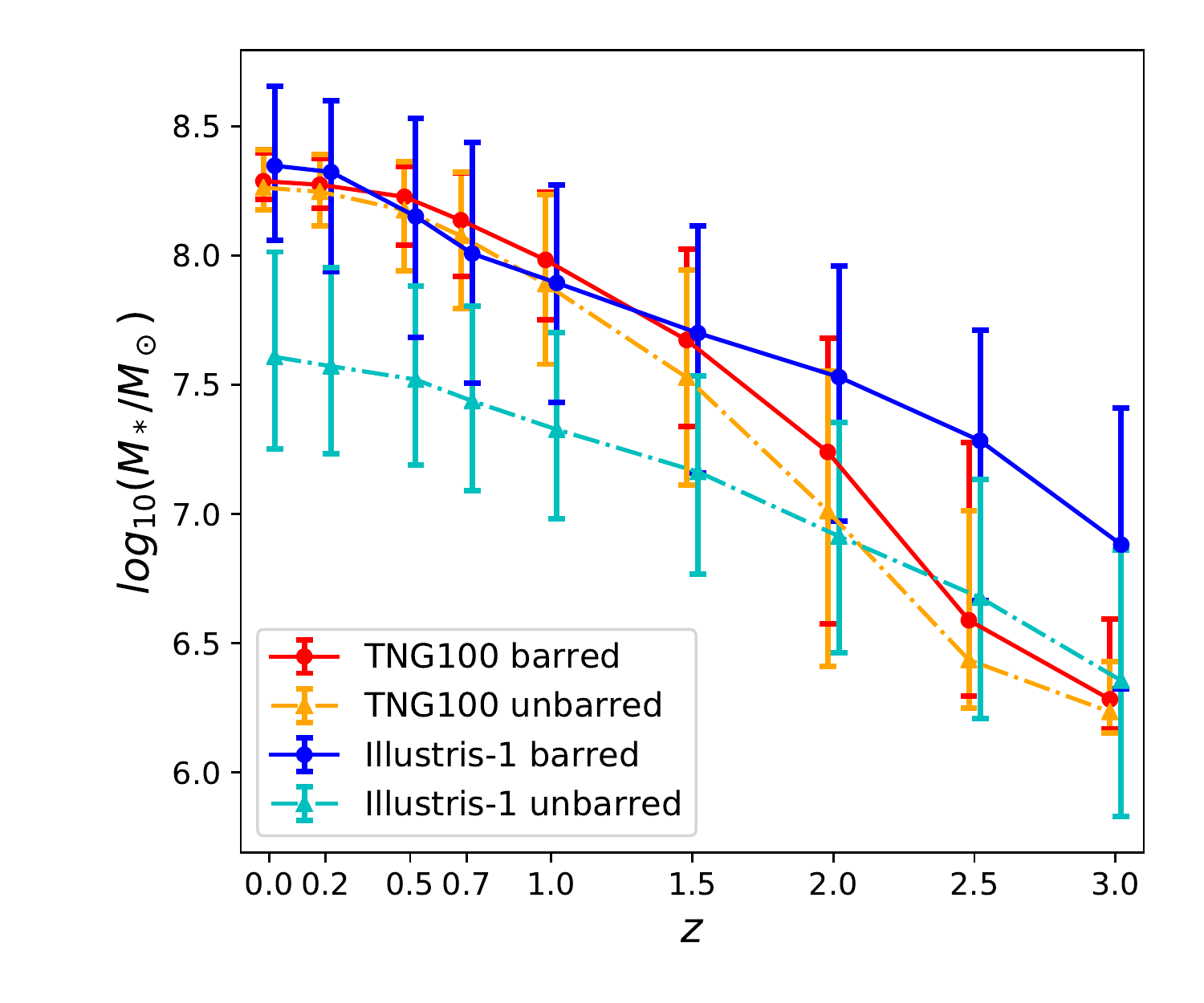}
\includegraphics[width=0.85\columnwidth,trim=25 10 10 15,clip]{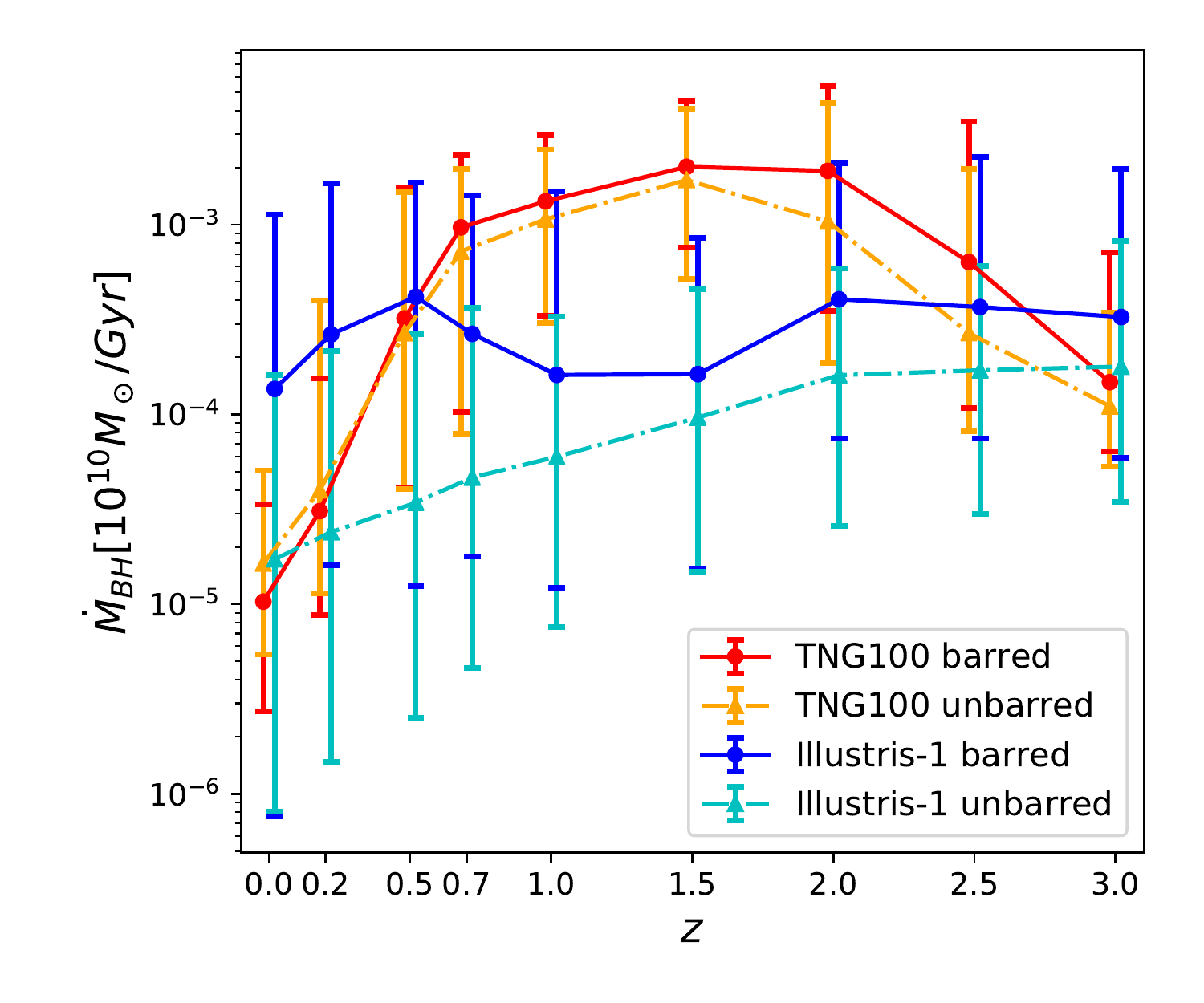}
\caption{Top(Bottom): The evolution of median mass(accretion rate) of super massive black hole in disc galaxies in the two simulations. The meaning of lines and bars are the same as in Fig.~\ref{fig:bar_unbar_gas_fraction}}
\end{center}
\label{fig:bh_mass}
\end{figure}
In the previous subsections, it has been concluded that the gas fraction is strongly correlated with the development of bars. In addition to stellar processes, i.e., star formation and feedback, the growth of super massive black hole and AGN feedback may also play important roles in regulating the gas component and influencing growth of bars. We trace the evolution of black hole mass in each disc galaxy with $M_*>10^{10.5}M_{\odot}$. Fig.~\ref{fig:bh_mass} shows that barred galaxies in the two simulations have similar black hole masses since $z<1.5$. Before that, the black holes in the Illustris-1 barred galaxies are more massive. For TNG100, the black hole mass in barred galaxies is slightly higher than unbarred ones at $z<0.7$, and this mass gap is moderate at higher redshifts. However, in Illustris-1, the median SMBH mass in the barred galaxies are much higher than unbarred galaxies since $z=3.0$.

The bottom plot of Fig.~\ref{fig:bh_mass} presents the mass accretion rate of SMBH in disc galaxies. In the redshift range $0.5<z<2.5$, the TNG100 barred galaxies have the highest median accretion rate, followed by the TNG100 unbarred, the Illustris-1 barred, and the Illustris-1 unbarred in decreasing order. At lower redshifts, i.e., $z<0.5$, the median accretion rates in the TNG100 barred and unbarred galaxies become comparable, decline sharply with time and is lower than that in the Illustris-1 barred galaxies. 

\begin{figure}[htbp]
\centering
\begin{center}
\includegraphics[width=0.78\columnwidth,trim=5 5 5 5,clip]{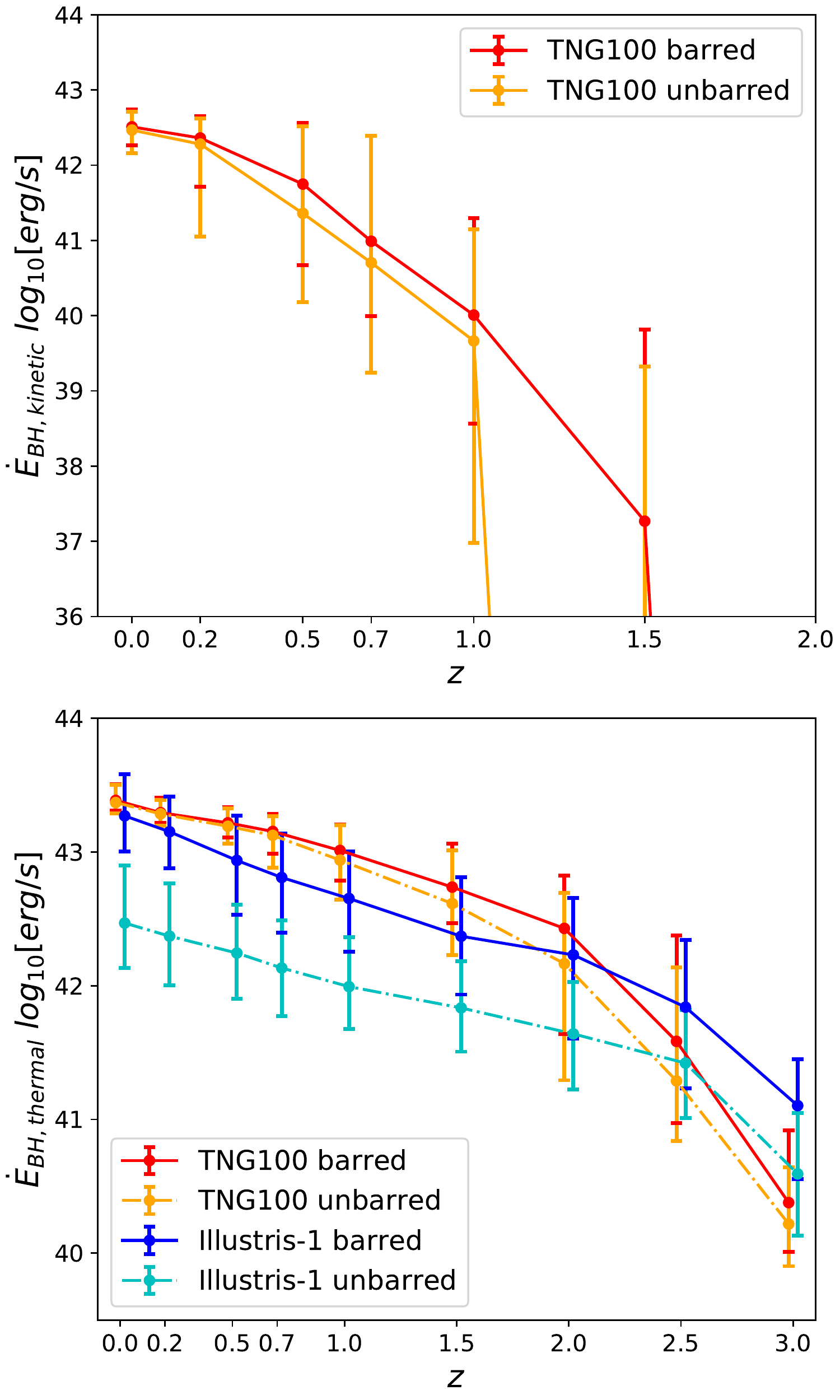}
\caption{Top: The evolution of median kinetic feedback energy of BH particles in disc galaxies in TNG100. Bottom: The evolution of thermal feedback energy of BH particles in disc galaxies in two simulations. The upper and lower bars represent the 25th and 75th percentiles in each bins for each category.}
\end{center}
\label{fig:bh_agn_engy}
\end{figure}


\begin{figure*}[htbp]
\begin{center}
\includegraphics[width=0.32\textwidth,trim=0 0 15 10,clip]{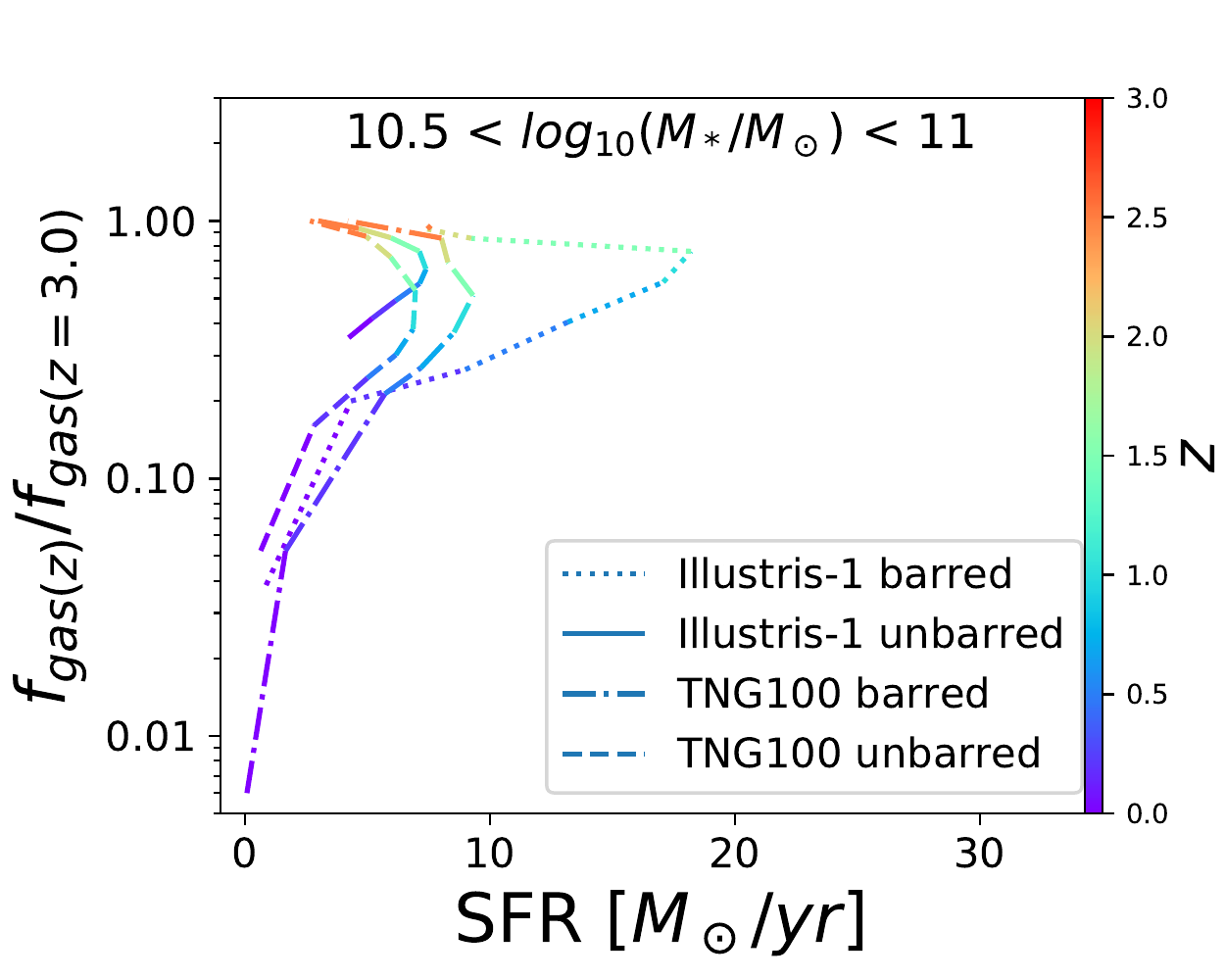}
\includegraphics[width=0.32\textwidth,trim=0 0 15 10,clip]{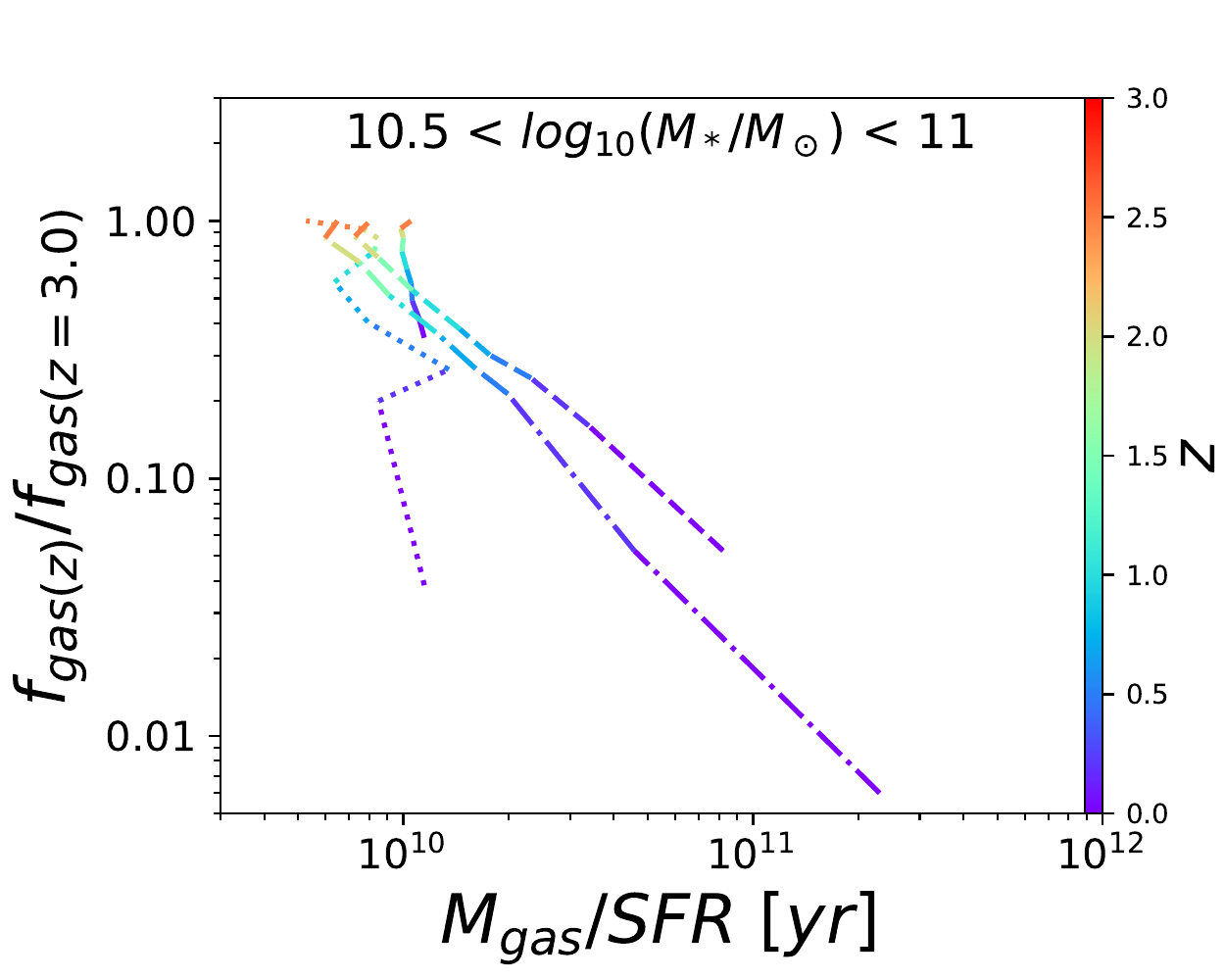}
\includegraphics[width=0.32\textwidth,trim=0 0 15 10,clip]{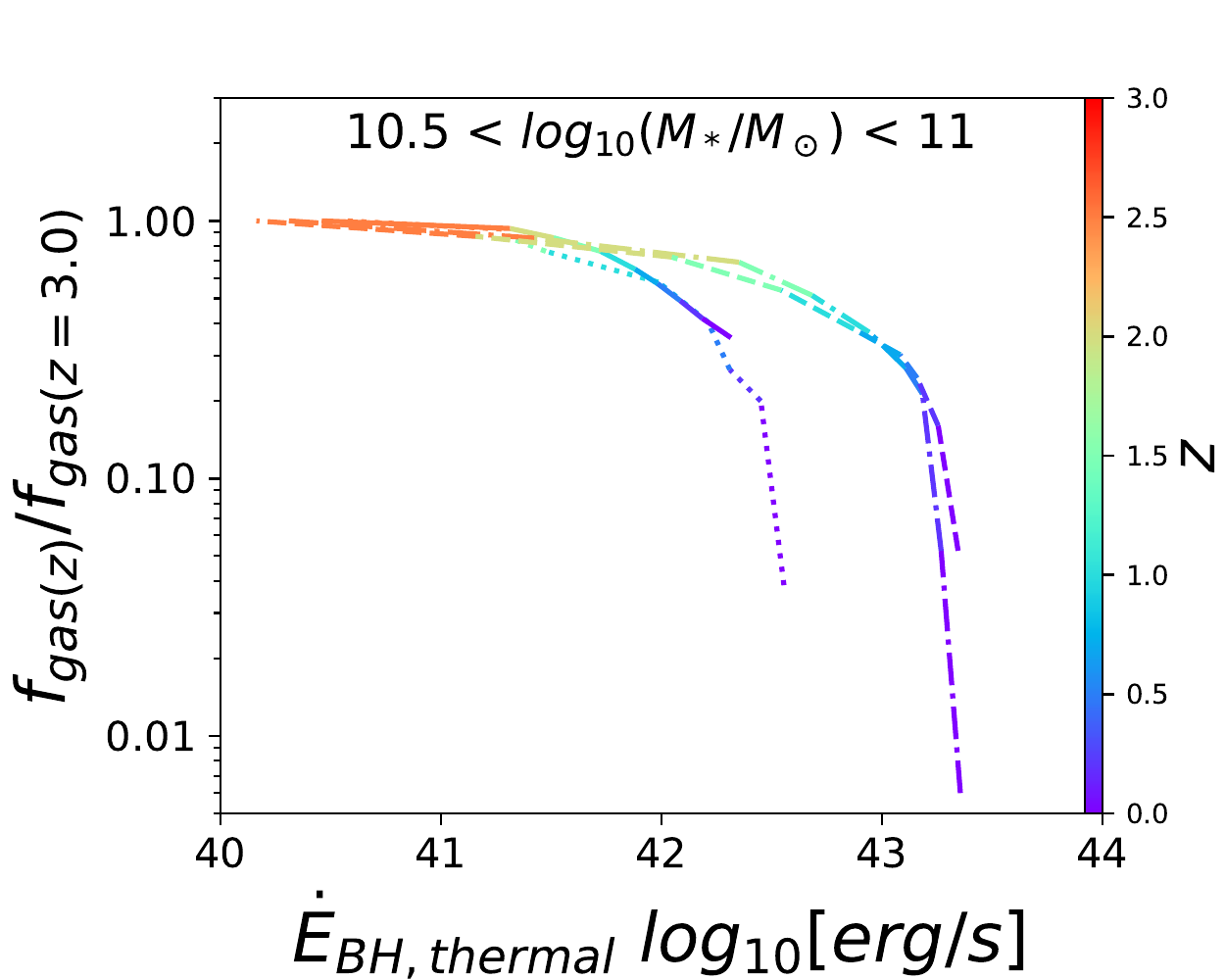}
\includegraphics[width=0.32\textwidth,trim=0 0 15 10,clip]{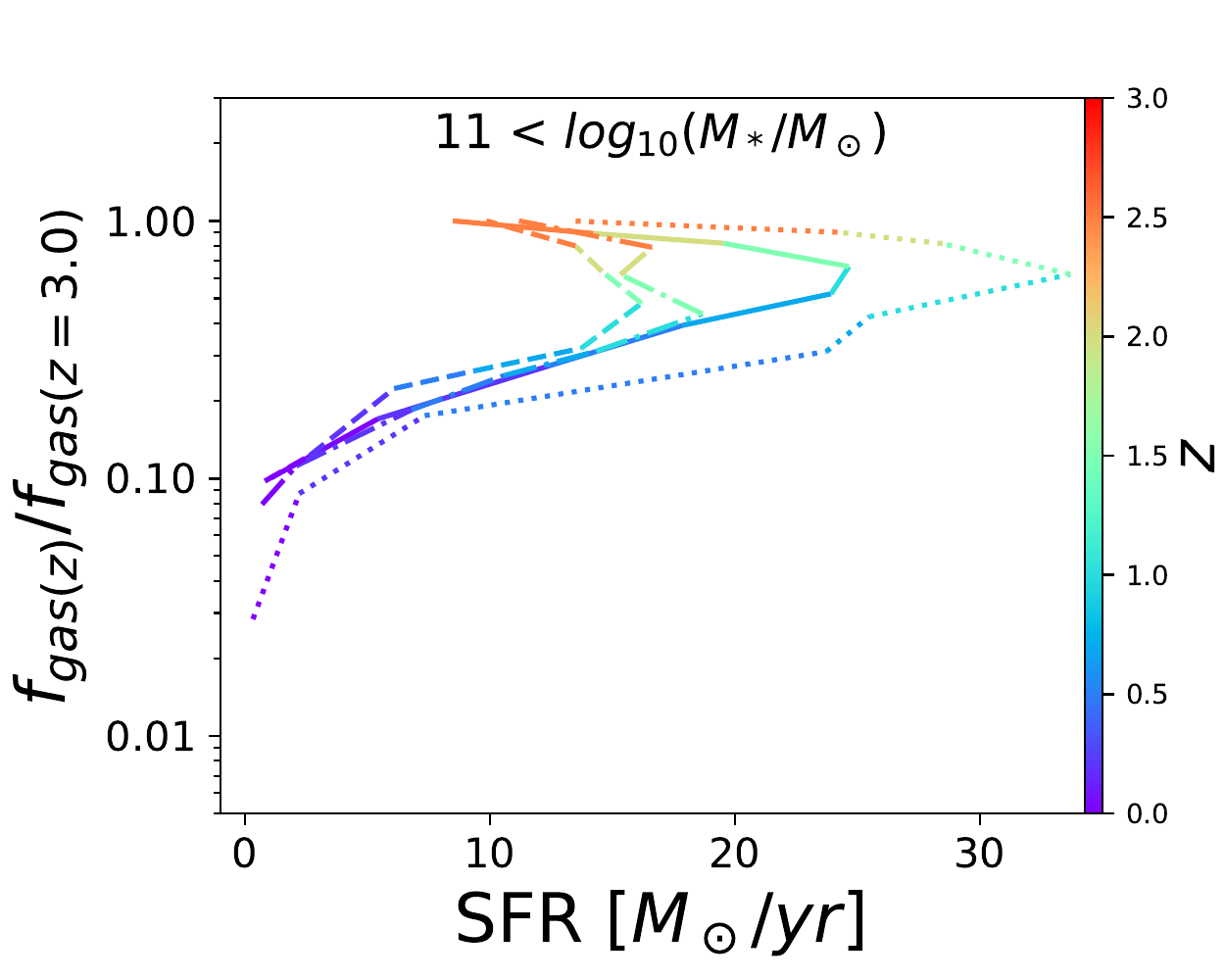}
\includegraphics[width=0.32\textwidth,trim=0 0 15 10,clip]{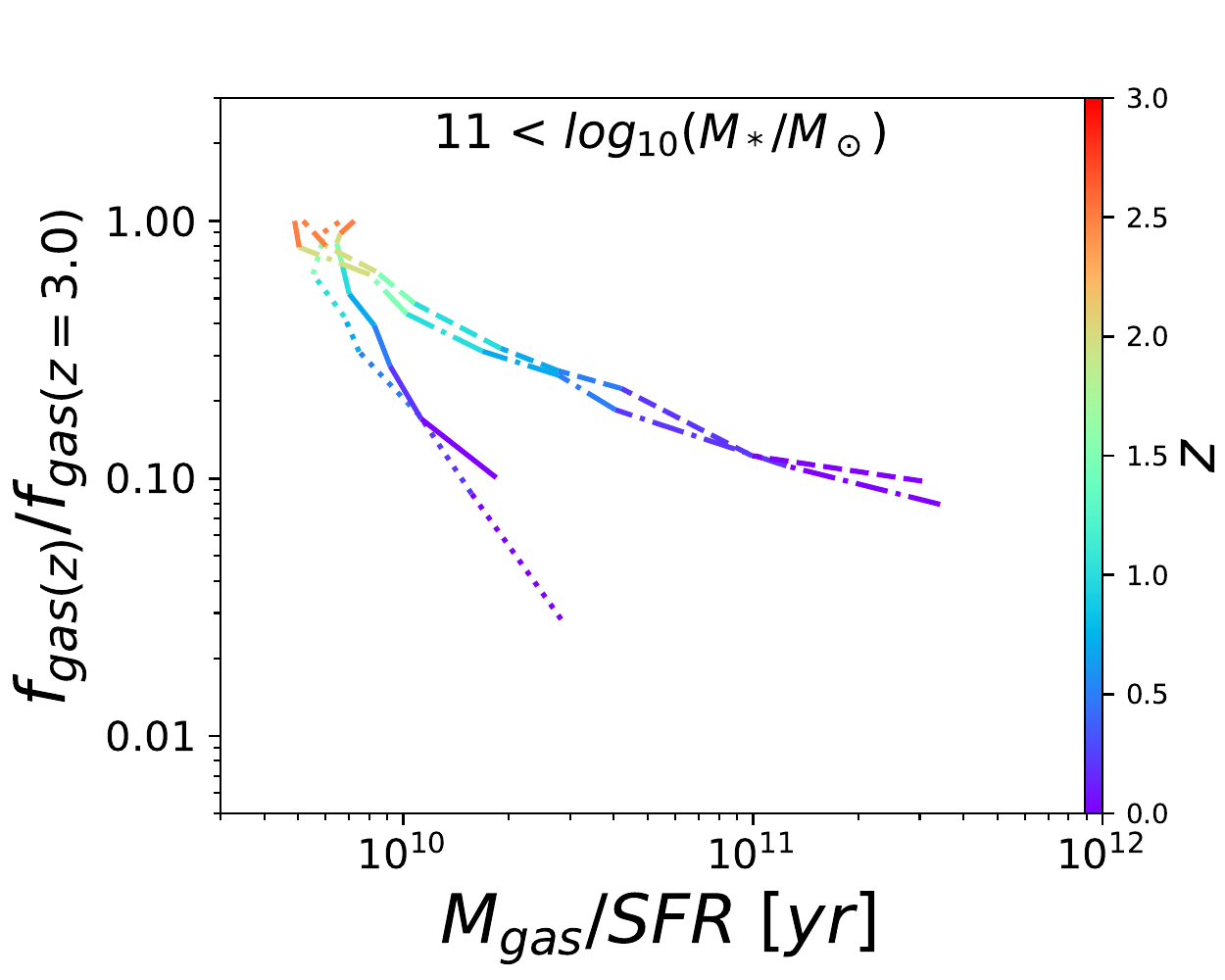}
\includegraphics[width=0.32\textwidth,trim=0 0 15 10,clip]{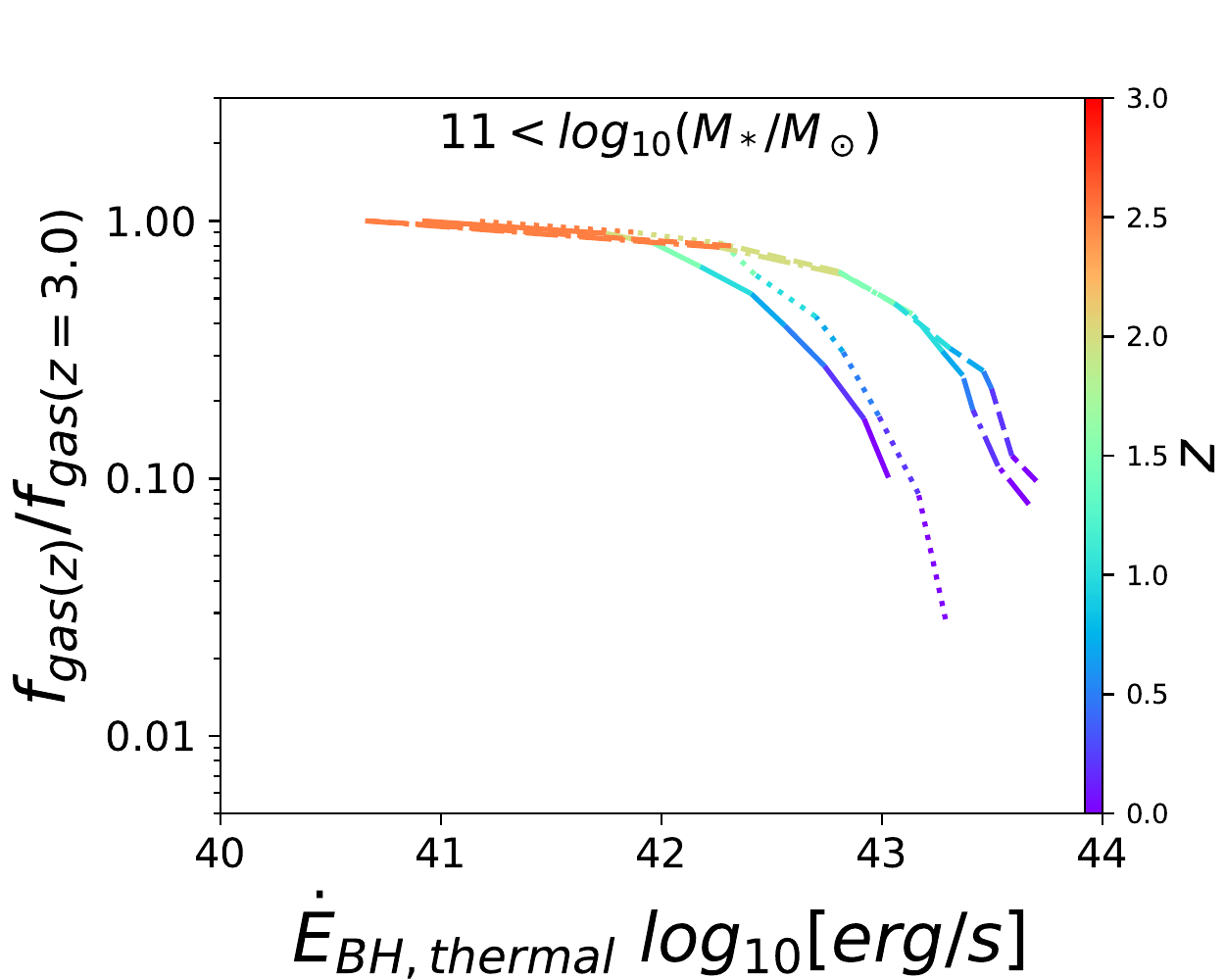}
\caption{Left, middle, right column illustrate the ratio of the median gas fraction at redshift $z$ over the median gas fraction at $z=3$ for four categories of disc galaxies, against the median star formation rate, gas depletion time and thermal AGN feedback energy rate respectively. The color bar indicates the redshift z. Top and bottom row show results of galaxies with $z=0$ stellar mass $10^{10.5}M_{\odot}<M_*<10^{11.0}M_{\odot}$, and $M_*>10^{11.0}M_{\odot}$ respectively. }
\end{center}
\label{fig:fgas_drop_driving}
\end{figure*}

The recipes of AGN feedback in the two simulations are different in several aspects(\citealt{2017MNRAS.465.3291W}). To further explore the differences of AGN feedback in galaxies, we track the energy released by SMBH accretion in the two simulations. The bottom plot in Fig.~\ref{fig:bh_agn_engy} gives the rate of thermal energy injected by SMBH, $\dot{E}_{BH, thermal}$. At $z>2.0$, there is more thermal energy injected to gas component in Illustris-1 than TNG100, and the situation is reversed at $z<=2.0$. In TNG100, $\dot{E}_{BH, thermal}$ in barred galaxies is slightly higher than unbarred galaxies at $z>0.5$, i.e. the median redshift of bar formation. The median $\dot{E}_{BH, thermal}$ in the Illustris-1 barred galaxies are lower than that in the TNG barred galaxies by a factor of $40\% \sim 100\%$, but is higher than the Illustris-1 unbarred galaxies by a factor of 3-4 at $z<2$. Note that, \cite{2018MNRAS.479.4056W} illustrate that a large amount of the thermal AGN feedback energy would be radiated away immediately by dense star-forming gas around SMBH. Only the left amount of injected thermal energy would actually regulate the gas component.

In the TNG simulations, the AGN feedback is partially injected into gas via kinetic mode, which has not been implemented in Illustris-1. The top plot in Fig.~\ref{fig:bh_agn_engy} shows the rate of kinetic energy released by BH, denoted as $\dot{E}_{BH, kin}$, in TNG100. $\dot{E}_{BH, kin}$ increases rapidly since $z\sim1.5$. The median rate grows from $\sim 10^{37} \rm{erg/s}$ at $z=1.5$ to $\sim 10^{42.5} \rm{erg/s}$ at $z=0$. The median rate in barred galaxies is higher than unbarred galaxies by $\sim 70\% $ at $z>0.2$. Combining the thermal and kinetic channels, massive disc galaxies in TNG100 experience stronger AGN feedback than in Illustris-1 at $z<\sim 2$. The evolution of AGN feedback energy in two simulations are coincident with the evolution of star formation rate and efficiency presented in the last subsection. At $z<\sim2$, SFR and SFE in the TNG100 samples drop more rapidly than in Illustris-1. This coincidence agrees with results of TNG300 in \cite{2018MNRAS.479.4056W}. Note that, our galaxies samples have $M_*>10^{10.5}M_{\odot}$ at z=0. For galaxies with similar mass in the TNG300 simulation, \cite{2018MNRAS.479.4056W} found that their star formation rate is significantly reduced since $z\sim 2$, when the kinetic AGN feedback became the dominant feedback energy channel.

One particularly important question is that how star formation, stellar feedback, BH growth and AGN feedback have influenced the bar formation and led to the different properties of bar in these two simulations. The influence may have been exerted through many aspects. One aspect that is closely related to our investigation is the disc gas fraction. More specifically, lowering down the gas fraction in discs, either by more effective expulsion with stronger feedback or more consumption by relatively higher star formation rate, or by both means, could create a more favourable condition for bar formation.  As for the TNG100 and Illustris-1 simulations concerned here, the relatively higher star formation efficiency and hence more effective stellar feedback at $z>2$ in TNG100 may had result in proto discs with lower gas fraction than Illustris-1 at high z, as shown in Fig.~\ref{fig:bar_unbar_gas_fraction}. Then at $z<\sim2$, the stronger AGN feedback(Fig.~\ref{fig:bh_agn_engy}), in combination with stellar feedback, helps the massive disc galaxies in TNG100 to get their gas fraction drop more rapidly than disc galaxies in Illustris-1. 

The AGN feedback may has served as the primary factor that drive the gas fraction of galaxies in TNG100 declining more rapidly at $z<2$. We illustrate this point by Fig.~\ref{fig:fgas_drop_driving}. From $z=3$ to $z=0$, the median gas fraction drop by factors of 3, 20, 25, and 160 for Illustris-1 unbarred, Illustris-1 barred, TNG100 unbarred, and TNG100 barred disc galaxies with $z=0$ stellar mass of $10^{10.5}-10^{11.0} M_{\odot}$ respectively. The fastest decline happens at $z<2$, when the SFR and SFE in TNG100 is lower than those in Illustris-1, as shown in the left and middle columns. But the AGN feedback energy rate in TNG100 is higher than in Illustris-1, presented in the right column. Only the thermal feedback energy is shown in this plot, but we remind that TNG100 also implement kinetic AGN feedback but Illustris-1 does not. For more massive galaxies with stellar mass of $10^{11.0}-10^{11.5} M_{\odot}$, the decline rate of gas fraction in two simulations are more closer, although the differences on SFR and SFE between two simulations are comparable to the case in less massive galaxies. The difference on AGN feedback energy rate, however, are narrowed down with respect to the galaxies with lower stellar masses. In addition, as SMBHs sit at the center of galaxies, their feedback could have relatively more significant influence on the gas fraction of central region, and could partly contribute to the relation between $f_{gas}(2*r50)$ and $f_{gas}(r50)$ shown in Fig.~\ref{fig:gasfrac_r50_2r50}.


\subsection{Properties of host dark matter halos}
\begin{figure}[htbp]
\begin{center}
\includegraphics[width=0.9\columnwidth,trim=25 0 5 10,clip]{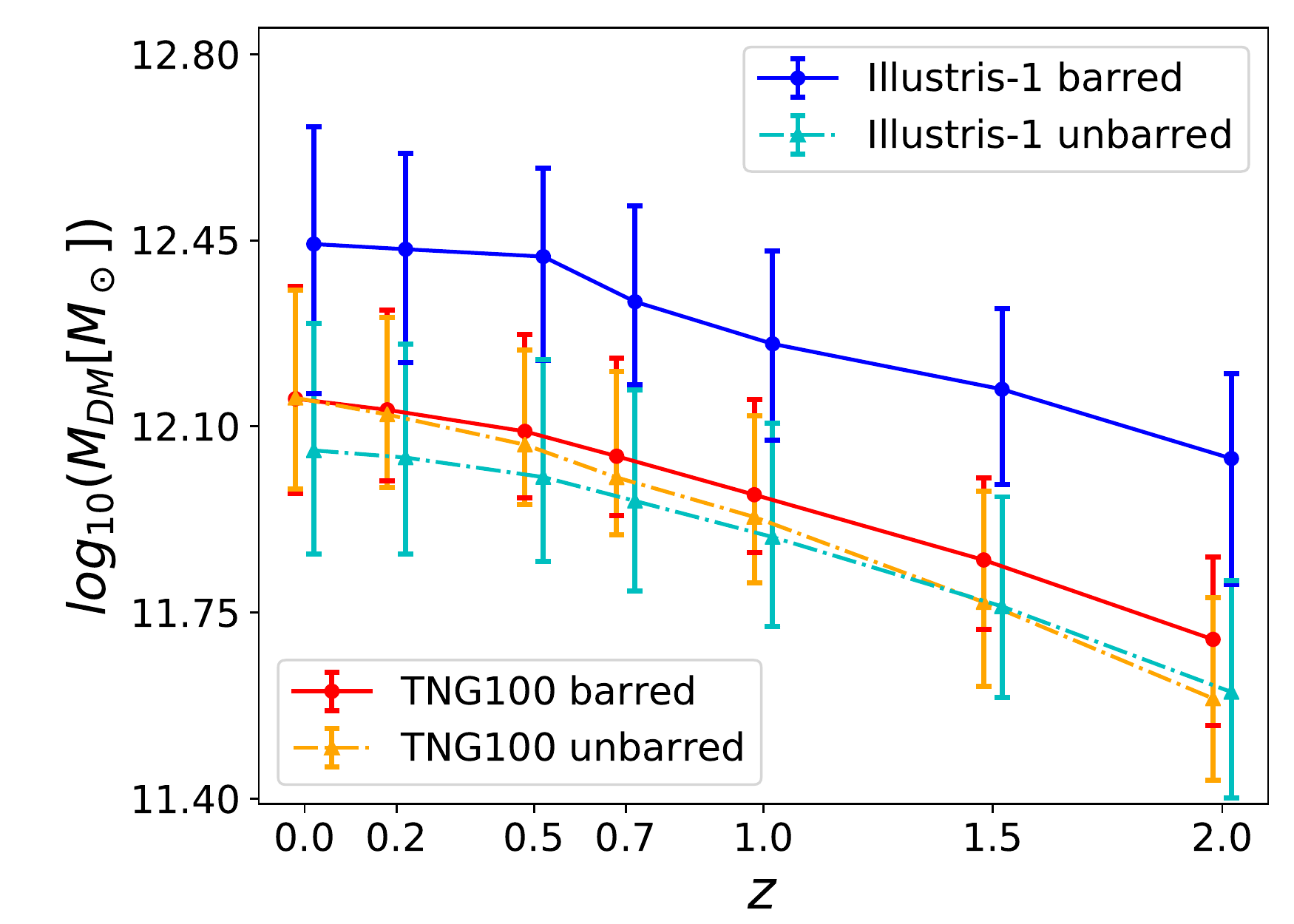}
\includegraphics[width=0.9\columnwidth,trim=25 0 5 10,clip]{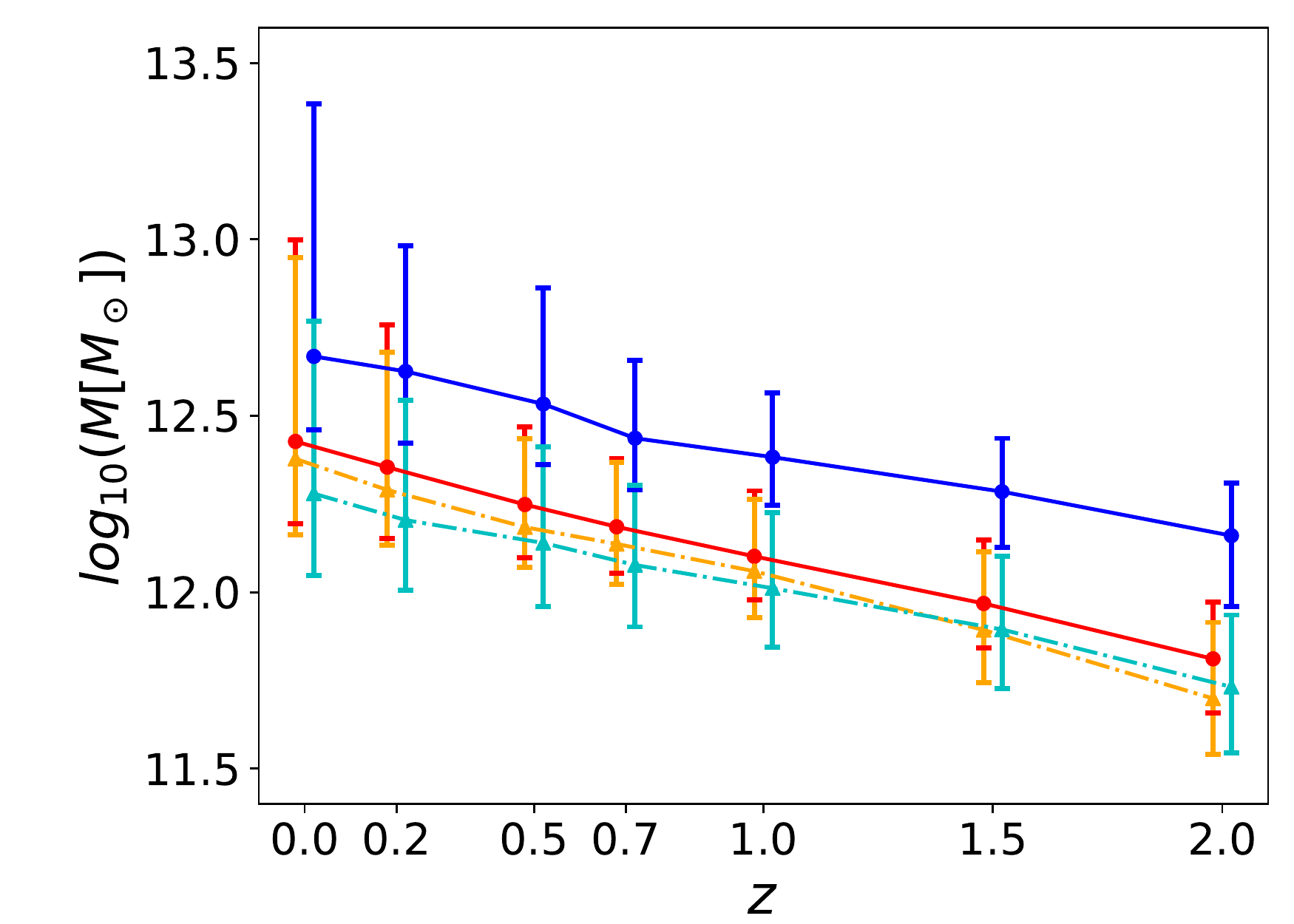}
\caption{Top: The median dark matter mass of subhalo; Bottom: The median total mass within $R_{200}$ of host halo, including dark matter, stellar and gas. }
\end{center}
\label{fig:subhalo_halo_mass}
\end{figure}

\begin{figure}[htbp]
\begin{center}
\includegraphics[width=0.9\columnwidth,trim=5 0 0 15,clip]{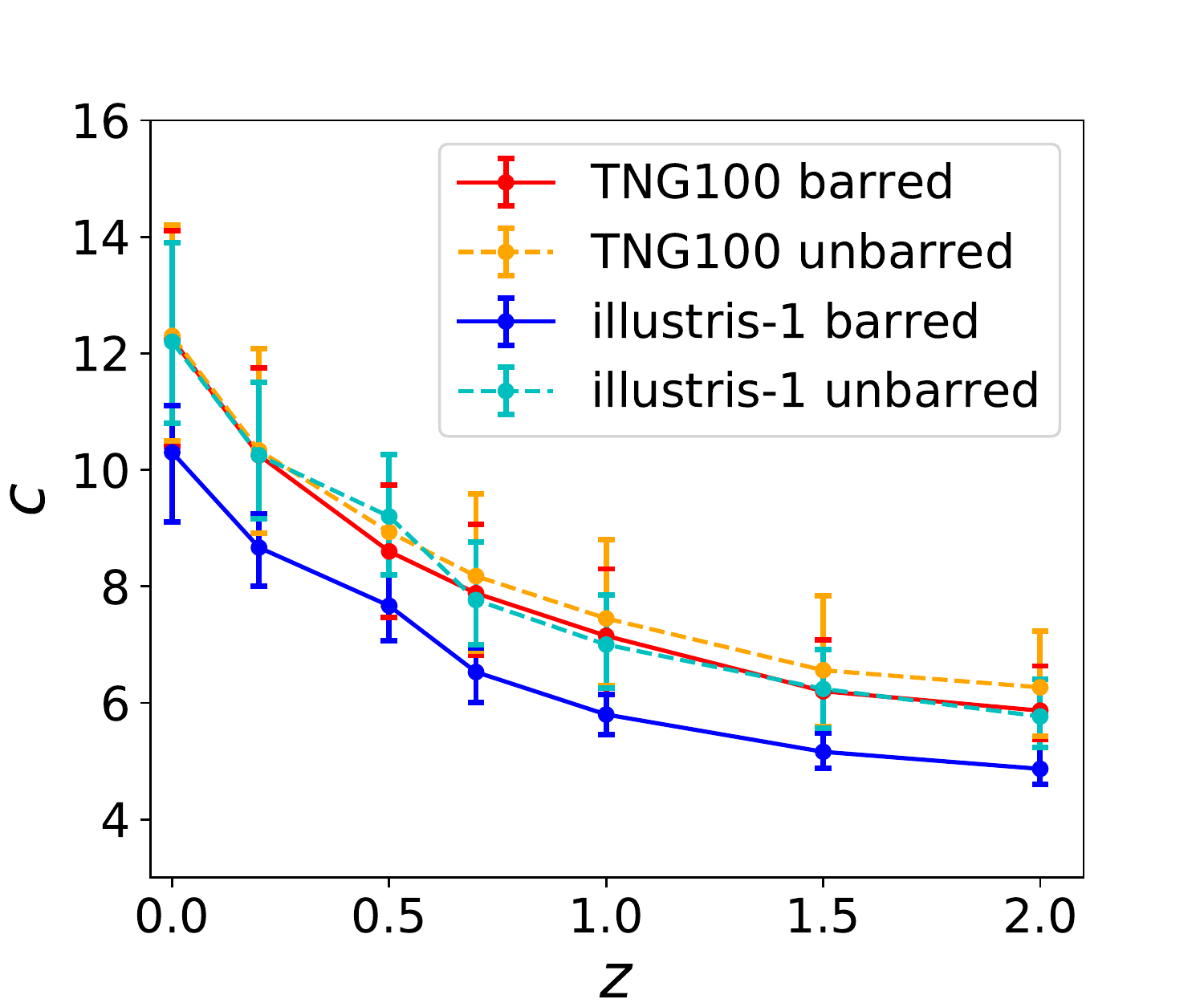}
\includegraphics[width=0.9\columnwidth,trim=5 0 0 15,clip]{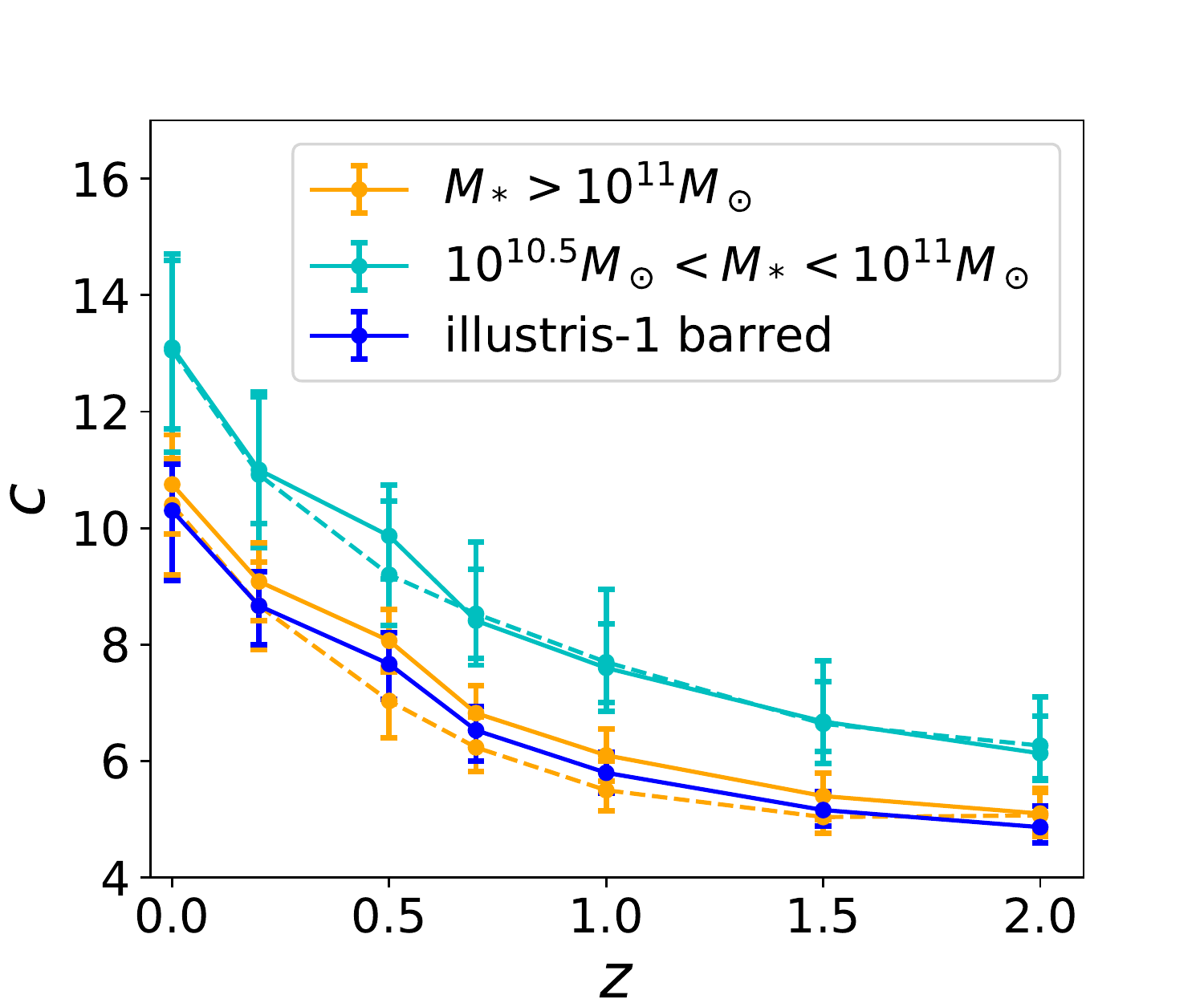}
\caption{Top: The median concentration parameter, c,  of  dark matter halo hosting disc galaxies in both simulations. The meaning of bars are the same as previous pictures.
Bottom: Yellow and cyan lines indicate concentration parameter of halos hosting disc galaxies with $M_*<10^{11}M_{\odot}$ and $M_*<10^{10.5}M_{\odot}M_*<10^{11}M_{\odot}$ respectively. Blue line is the concentration parameter of z=0 barred galaxies in Illustris-1. Solid lines represent Illustris-1 halos and dashed lines represent TNG100 halos.}
\end{center}
\label{fig:halo_conc}
\end{figure}

\begin{figure}[htbp]
\begin{center}
\includegraphics[width=0.9\columnwidth,trim=20 0 5 5,clip]{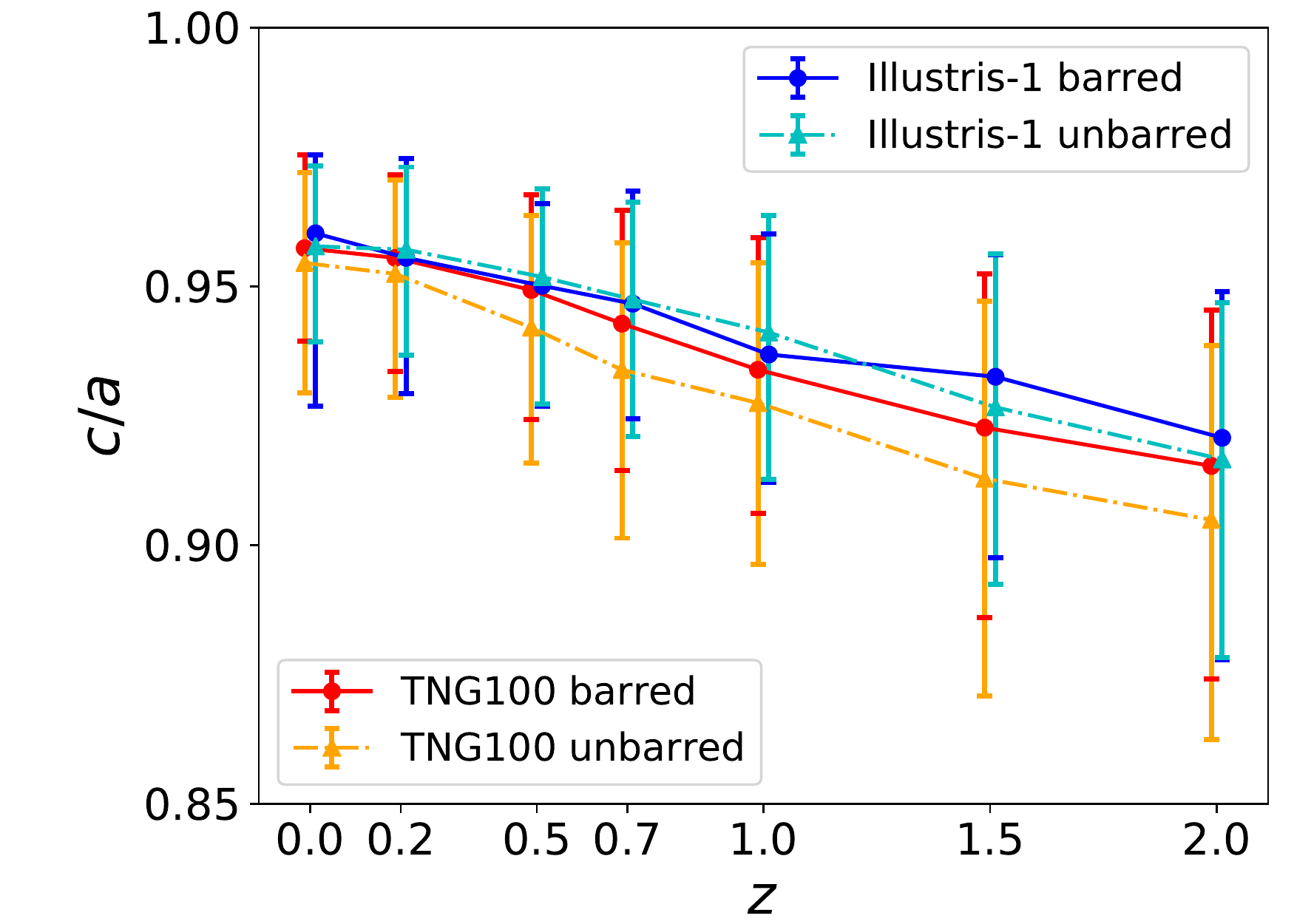}
\includegraphics[width=0.9\columnwidth,trim=20 0 5 5,clip]{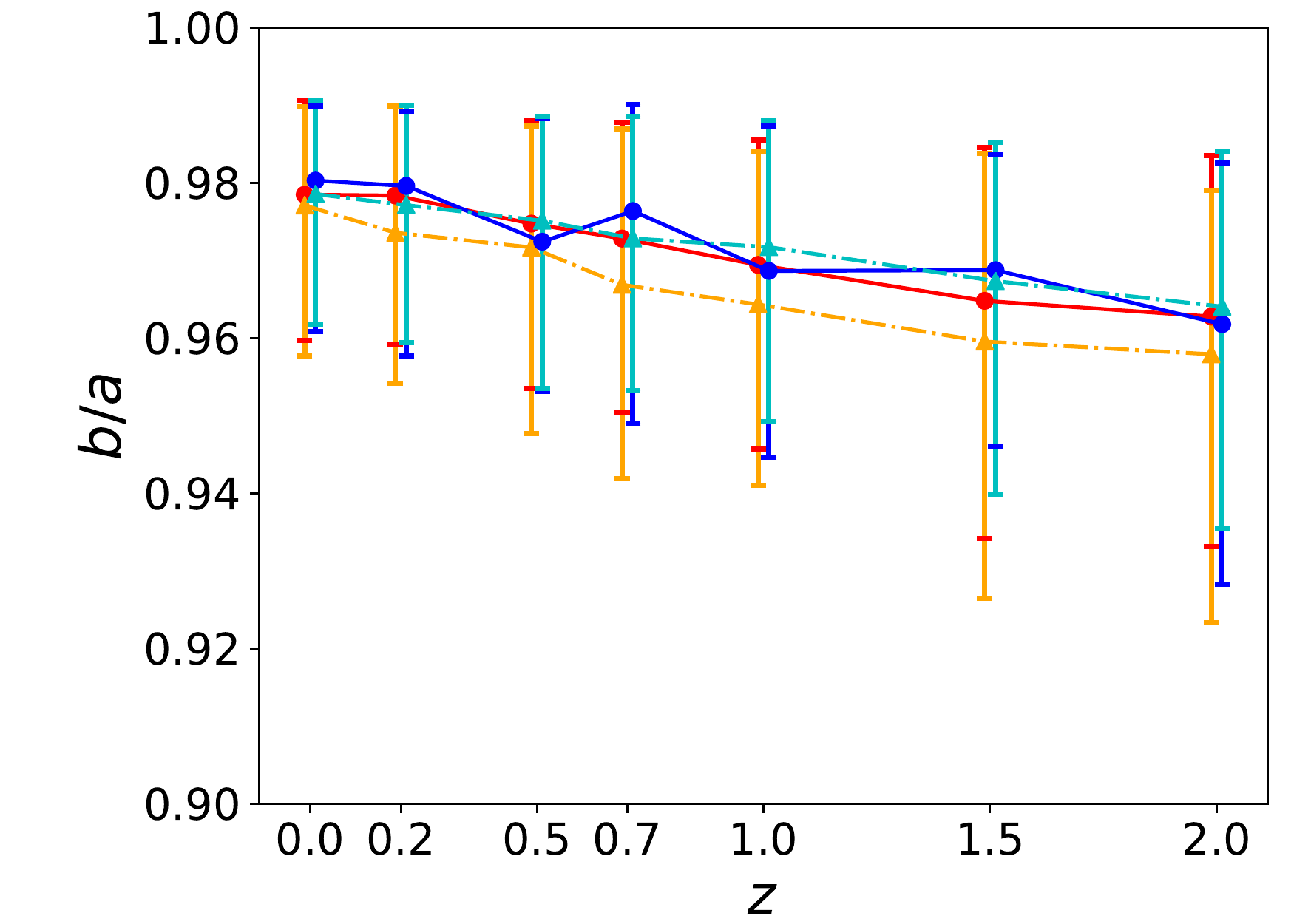}
\caption{The median axial ratio of dark matter halo hosting disc galaxies in both simulations. Top: the median halo flatness(c/a). Bottom: the halo axial ratio b/a. The meaning of bars are the same as in Fig.~\ref{fig:bar_unbar_gas_fraction}  }
\end{center}
\label{fig:axis_ratio}
\end{figure}

Isolated simulations show that the properties of dark matter halo have important effects on bar formation(e.g. \citealt{2002MNRAS.330...35A}, \citealt{2003MNRAS.341.1179A}). Here we investigate the connection between halo properties and the presence and strength of bars in our samples, focusing on the halo mass, concentration, and triaxiality of sub halos.
The top plot in Fig.~\ref{fig:subhalo_halo_mass} traces the mass evolution of dark matter subhalos. The masses of subhalos hosting Illustris-1 unbarred and all the TNG100 disc galaxies are similar. Host subhalos of barred galaxies in Illustris-1 are about two to three times as massive as other subhalos hosting disc galaxies in the two simulations. This feature should result from the bars in Illustris-1 appear in more massive galaxies, as suggested by Fig.~\ref{fig:overall_bar_fraction}. In TNG100, subhalos of barred galaxies are slightly more massive than unbarred galaxies. The bottom plot in Fig.~\ref{fig:subhalo_halo_mass} shows the mass of host halos within $R_{200}$, exhibiting similar trend as the mass of subhalos, but the gap between barred galaxies in Illustris-1 and other samples is narrowed. At $z=0$, the median halo mass within R200 are about $2-5\times 10^{12} M_{\odot}$.

The top panel of Fig.~\ref{fig:halo_conc} presents the concentration parameter, c, of halos hosting disc galaxies in the two simulations.  Except for barred galaxies in Illustris-1, the median values of c for dark matter halos are similar in both simulations. The barred galaxies in Illustris-1 have relatively lower value of c than others. 
This feature is also related to the halo mass. For each simulation, we divide the disc galaxies into two categories by the stellar mass threshold of $10^{11.0} M_{\odot}$. Then we show the concentration parameter of these two categories in the bottom panel of Fig.~\ref{fig:halo_conc}. The concentration of halos hosting barred galaxies in Illustris-1 is close to that of halos hosting galaxies with stellar mass $M_*>10^{11.0} M_{\odot}$ in the both simulations. For galaxies with the same stellar mass, the concentration of host halos are very similar either in Illustris-1 or in TNG100, and also either in barred or unbarred galaxies. Therefore, the halo concentration should have negligible effect on the discrepancy of bar frequency between the two simulations. 

Fig.~\ref{fig:axis_ratio} shows the evolution of halo axial ratio in disc galaxies in the both simulations. Generally, all host halos in our disc galaxy sample have a considerable round shape at high redshifts. The axial ratios $c/a$ and $b/a$ are larger than 0.86 and 0.92 respectively at $z=2$ for most of the halos. All the halos have been becoming more rounder gradually with redshift decreasing. There are barely any differences between barred and unbarred galaxies in Illustris-1, except for slight scatters in barred samples which should be caused by the limited number of samples. Overall, the halos in TNG100 show stronger triaxiality than Illustris-1. Furthermore, halos of unbarred galaxies in TNG100 show a little bit stronger triaxiality than barred galaxies. Overall, the halo shape should have minor effect on the discrepancy of bar frequency between TNG100 and Illustris-1. 


\section{evolution of bars in matched halos}
The results described above give the statistical views of bars in the TNG100 and Illustris-1 simulations, including differences between two simulations and related physical factors. Since the initial conditions are basically the same, it allows us to compare galaxies evolved from the similar initial conditions and environments in the two simulations. This section is to examine the differences in the bar structure and related physical factors between galaxies hosted
by analogue subhalos in the two simulations.

\subsection{Halo match algorithm and matched galaxies pairs}
We apply the Lagrangian-region matching algorithm proposed by \cite{2014MNRAS.439..300L} to identify pairs of galaxies hosted by analogue subhalos across the two simulations. This algorithm first finds candidates of subhalo pairs by comparing their positions, and then traces their particles back to the initial conditions to determine whether overlapping Lagrangian patches are matched or not. The initial density distribution and gravitational potentials of halo particles are used as key indicators in this algorithm.

We select all disc galaxies with more than 40000 stellar particles from one simulation, and try to match their counterparts in another simulation. For disc galaxies with $M_*>10^{10.5}M_{\odot}$ in TNG100, 813 out of 1269 are found to have counterparts in Illustris-1. Meanwhile, 748 out of 1232 disc galaxies with $M_*>10^{10.5}M_{\odot}$ in Illustris-1 have analogues in TNG100. A total number of 1079 matched pairs of galaxies are compiled by cross-checking. The morphology correlation of these galaxy pairs in the two simulations are listed in Table ~\ref{tab:match_morphology}.

\begin{table}[htbp]
\caption{Morphology of matched galaxies pairs hosted by analogue halos in two simulations: Bar/unbarred indicate disc galaxies with/without bars. Others indicates non-disc galaxies.}
\begin{center}
\begin{tabular}{|c|c|c|c|c|}
\hline
\multicolumn{2}{|l|}{\multirow{2}{*}{}} & \multicolumn{3}{c|}{Illustris-1} \\ \cline{3-5} 
\multicolumn{2}{|l|}{}                  & bar    & unbarred    & others    \\ \hline
\multirow{3}{*}{TNG100}   & bar        & 27     & 243         & 167   \\ \cline{2-5} 
      & unbarred   & 12     & 200         & 164       \\ \cline{2-5} 
                & others     & 24     & 242         & 0         \\ \hline
\end{tabular}
\end{center}
\label{tab:match_morphology}
\end{table}
We can see that most of the matched galaxies pairs have different morphological types at $z=0$. Only a fraction of about $21\%$ of the matched pairs have the same morphology. Moreover, for 331/813 of the disc galaxies in TNG100, their counterparts in Illustris-1 are non-disc galaxies. The corresponding fraction is 266/748 for the Illustris-1 disc galaxies. This result suggests that for an individual galaxy hosted in similar halo environments, the differences in baryonic physics can lead to significant discrepancy in galaxy properties. In the literature, differences in stellar and AGN feedback models, and initial conditions have been found to lead to notable differences in the properties of simulated galaxies(e.g., \citealt{2019MNRAS.482.2244K}, \citealt{2019ApJ...871...21G} and references therein).

\begin{figure*}[htbp]
\begin{center}
\hspace{-0.8cm}
\includegraphics[width=0.34\textwidth,trim=5 2 10 25,clip]{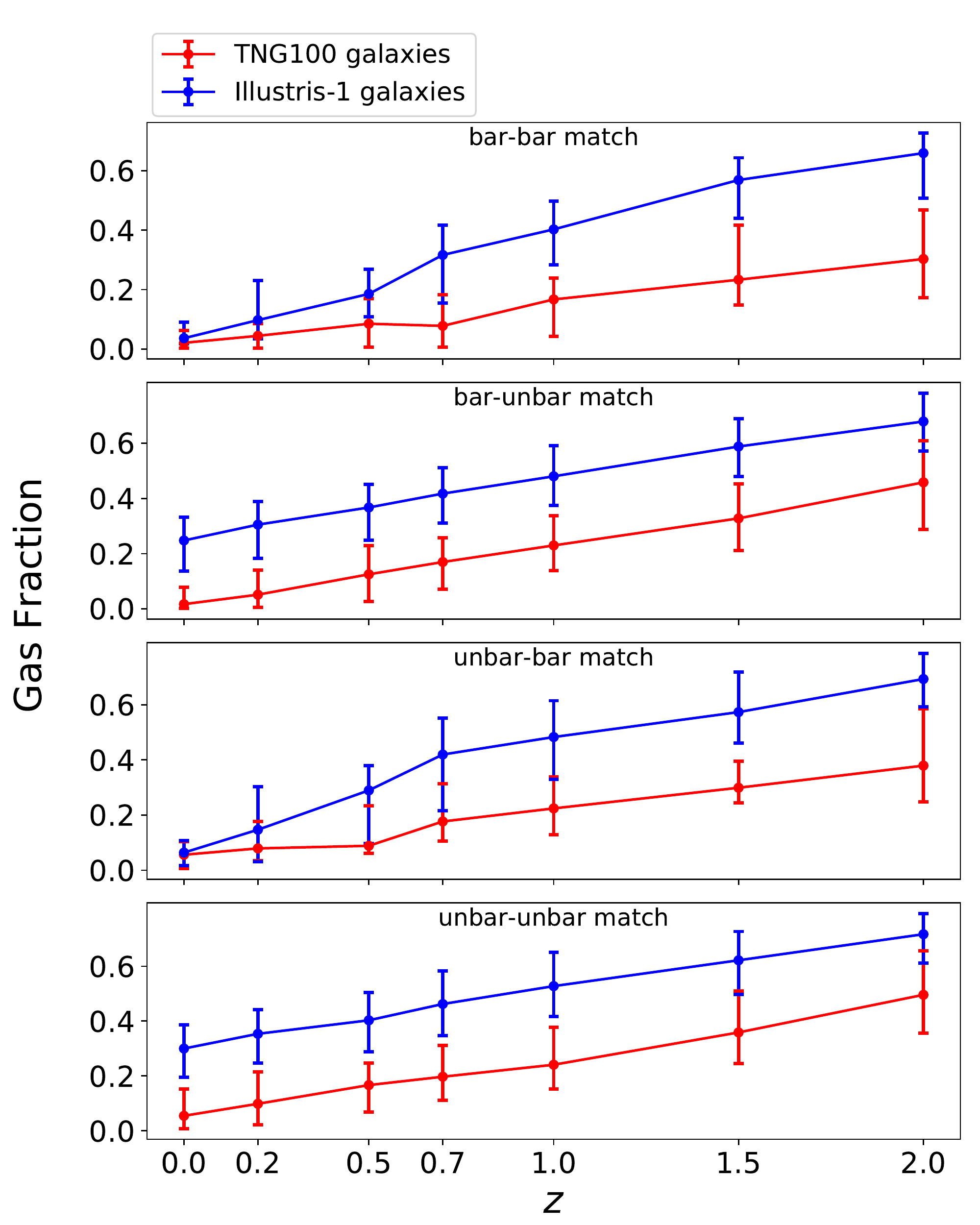}
\includegraphics[width=0.34\textwidth,trim=10 2 10 25,clip]{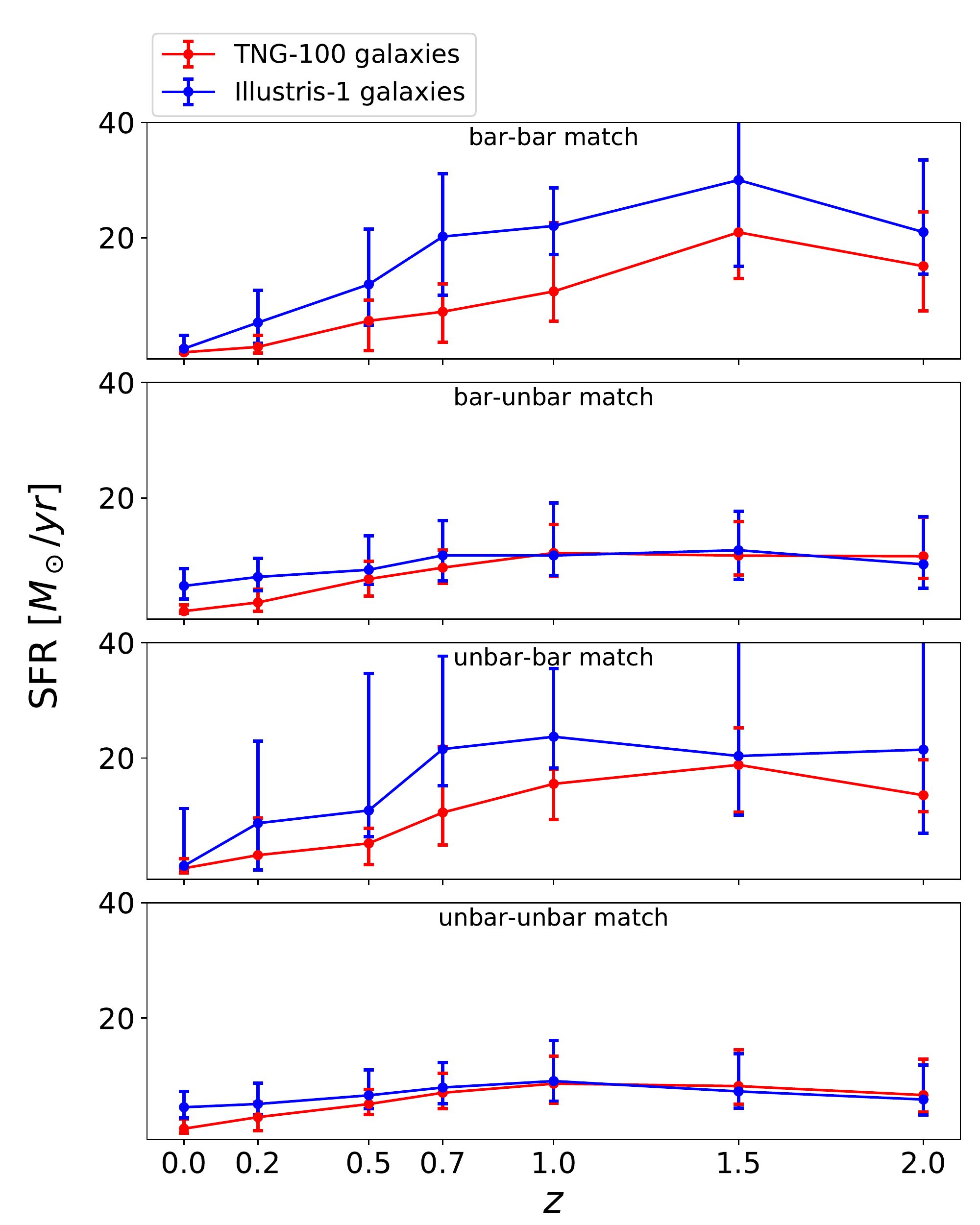}
\includegraphics[width=0.34\textwidth,trim=10 2 10 25,clip]{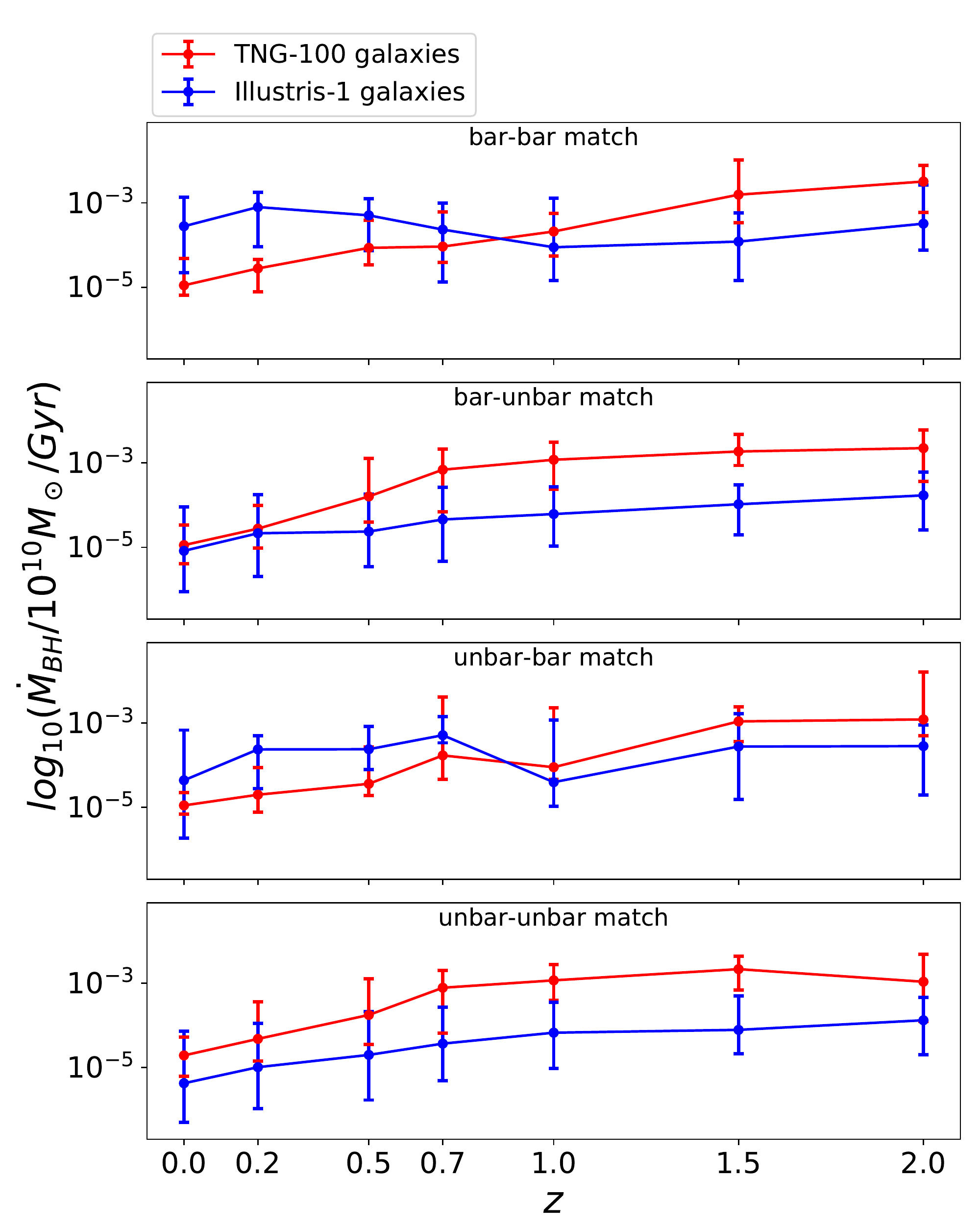}
\caption{Left, middle and right panel present the gas fraction, star formation rate and BH mass accretion rate of analogue galaxies pairs in TNG100 and Illustris-1 as a function of redshift. Matched galaxies pairs  are divided into four categories, bar-bar, bar-unbar, unbar-bar, and unbar-unbar(from top to bottom).The tag before and after dash in the category name indicate the morphology of galaxies in TNG100 and Illustris-1 respectively. }
\end{center}
\label{fig:match_Gas_SFR_BHdot}
\end{figure*}

\subsection{Comparison of analogue disc galaxies}
We then analyze the properties of galaxies pairs that are both disc galaxies in the two simulations, focusing on gas fraction, star formation rate and super massive black hole. We carry out controlled comparisons by dividing these pairs into four categories marked as bar-bar, bar-unbar, unbar-bar, and unbar-unbar. The tags before and after dash in the category name indicate the morphologies in TNG100 and Illustris-1 respectively. The evolution of gas fraction, star formation rate and SMBH mass accretion rate are shown in Fig.~\ref{fig:match_Gas_SFR_BHdot}. The SMBH mass accretion rate is used here to approximately indicate the strength of AGN feedback. Illustris-1 includes thermal AGN feedback only, but TNG100 has implemented both kinetic and thermal AGN feedback, and the feedback energy injected by these two channels can not be simply added together. As shown in section 4.3, the SMBH mass accretion rate can roughly characterize the relative strengths of AGN feedback. On the other hand, we should keep in mind that, the kinetic AGN feedback energy may dominate in the massive disc galaxies of TNG100 at low z(\citealt{2018MNRAS.479.4056W}).


Overall, the discrepancies in gas fraction, star formation rate and black hole mass accretion rate between galaxies in TNG100 and Illustris-1 are similar in the four categories. Namely, the galaxies in TNG100 basically have lower gas fraction, star formation rate, and higher black hole mass accretion rate at $z<=2.0$ in each category, except that the BH mass accretion rate of barred galaxies in Illustris-1 being able to catch up with or exceed their counterparts in TNG100 at redshift $z<0.5$. Both the star formation and SMBH activity in Illustris-1 unbarred galaxies are relatively weak. These differences are in keeping with the overall systematic differences shown in Section 4. We draw particular attention to the bar(TNG100) - unbar(Illustris-1) category, which has 243 pairs of galaxies and has contributed significantly to the discrepancy of bar frequency between the two simulations. The star formation rates of galaxies in this category compiled from both simulations are comparable at $z>=0.5$, and are relatively inefficient, while growth of SMBH and AGN feedback in TNG100 samples is much stronger since $z=2$. Hence, AGN feedback is likely a real game changer in this category.

\begin{figure*}[htbp]
\begin{center}
\vspace{-1.0cm}
\includegraphics[width=0.32\textwidth,trim=5 5 15 10,clip]{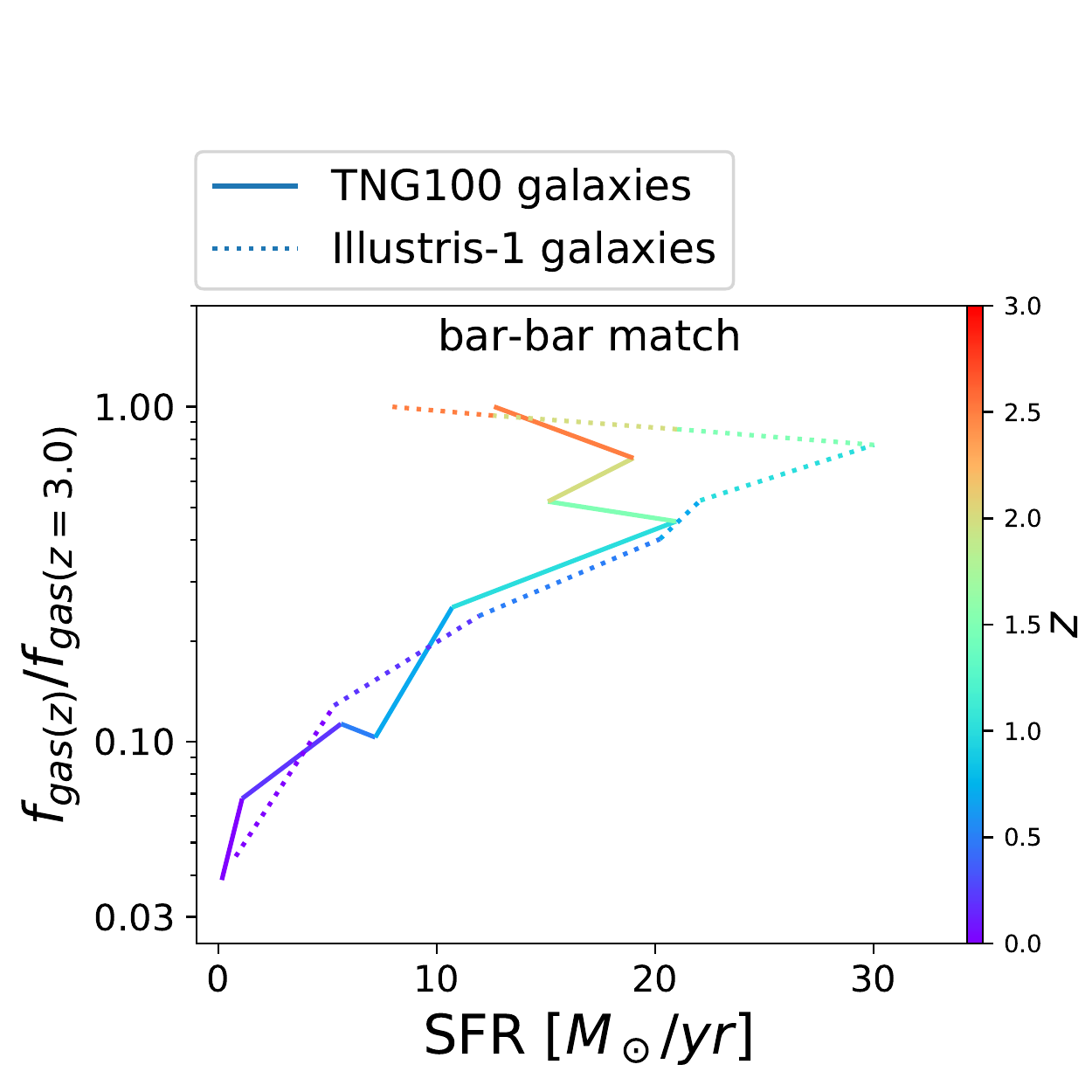}
\includegraphics[width=0.32\textwidth,trim=5 5 15 10,clip]{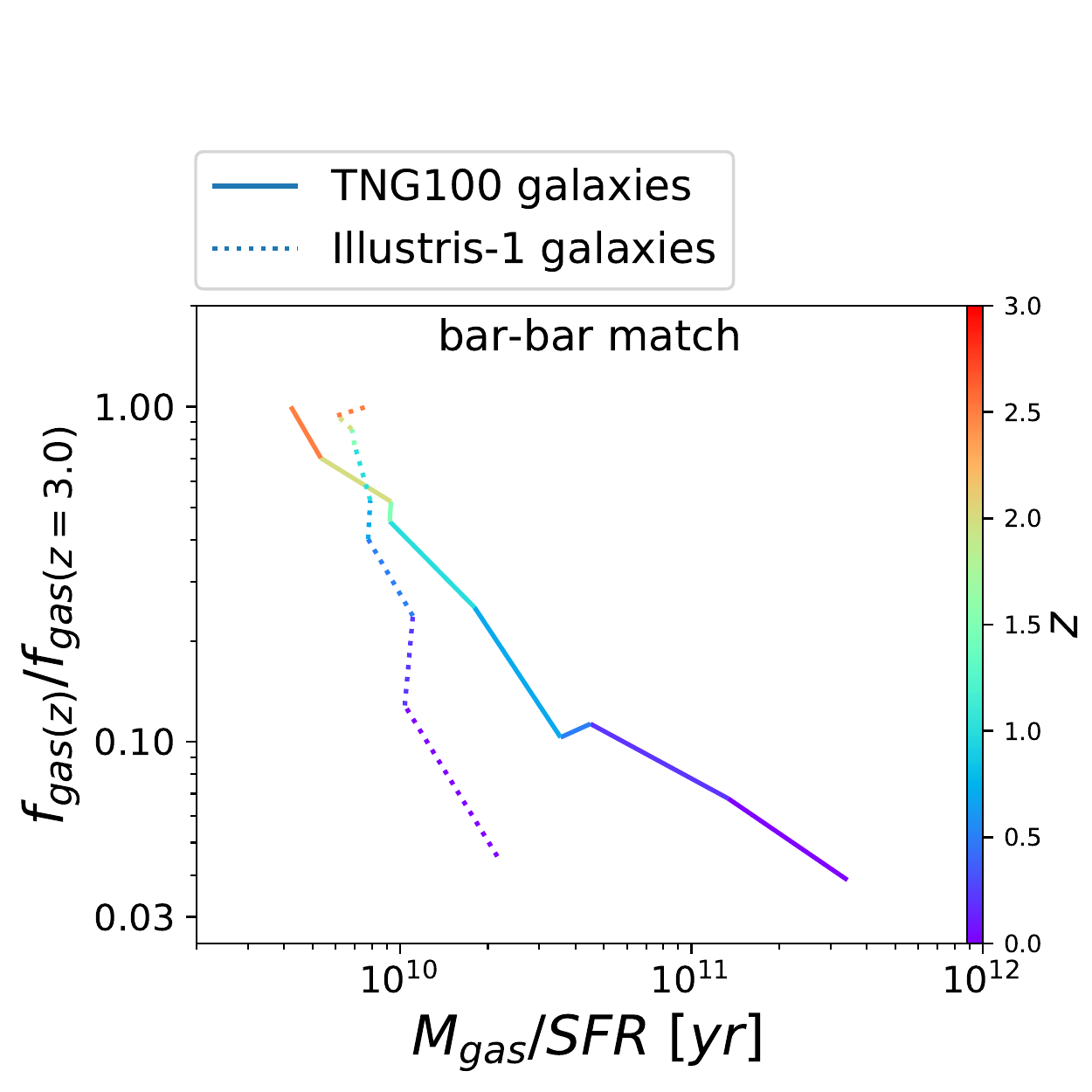}
\includegraphics[width=0.32\textwidth,trim=5 5 15 10,clip]{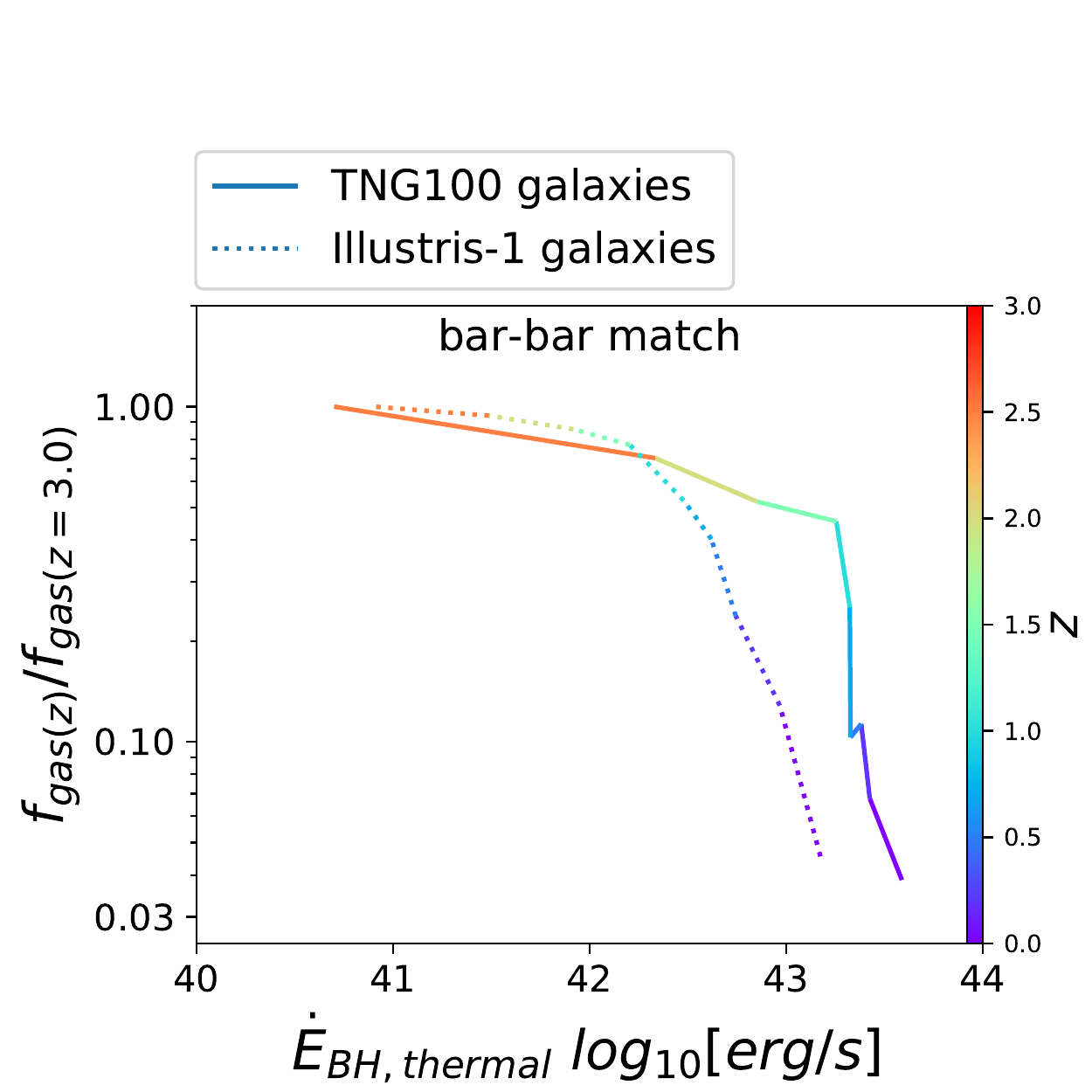}
\includegraphics[width=0.32\textwidth,trim=5 5 15 10,clip]{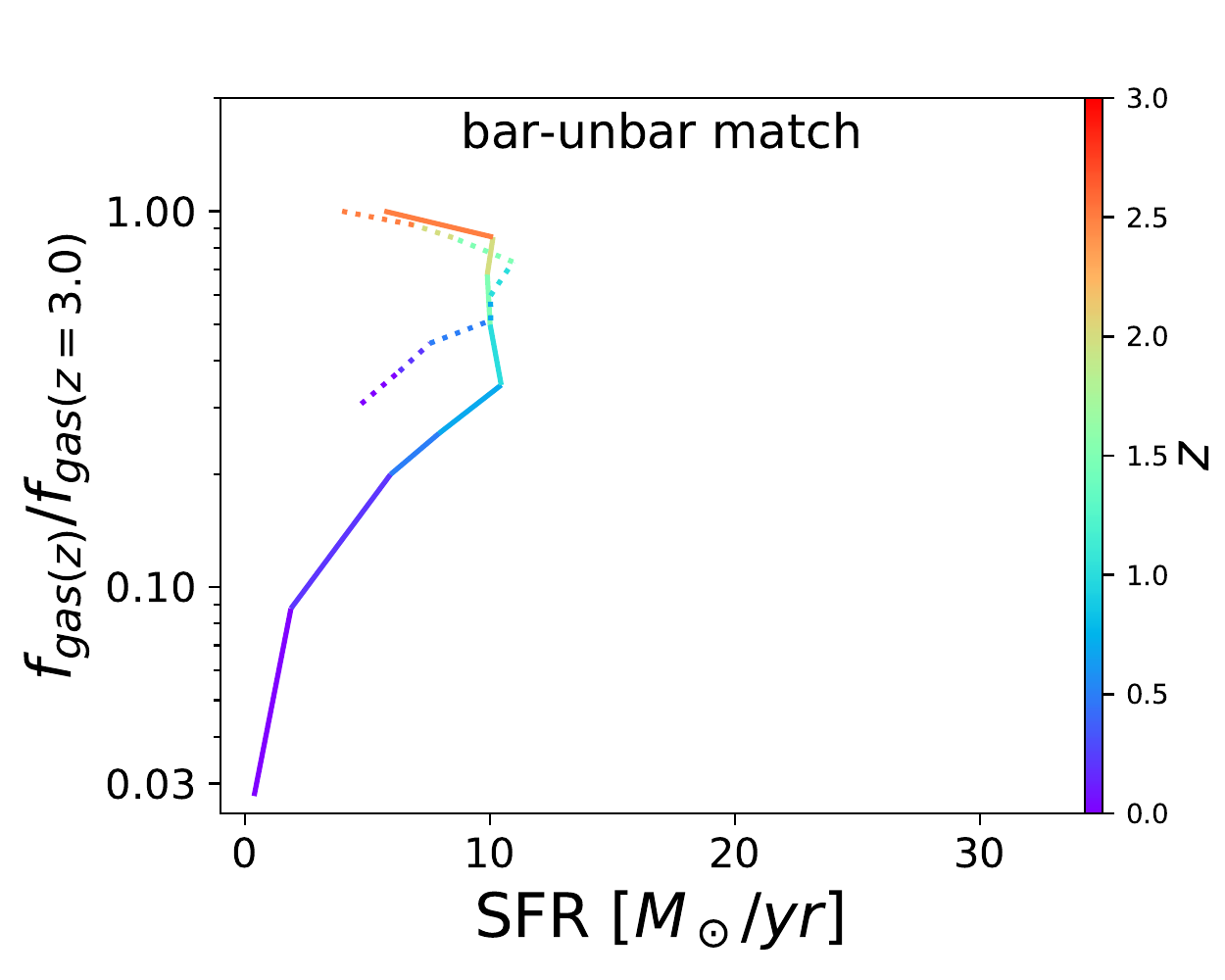}
\includegraphics[width=0.32\textwidth,trim=5 5 15 10,clip]{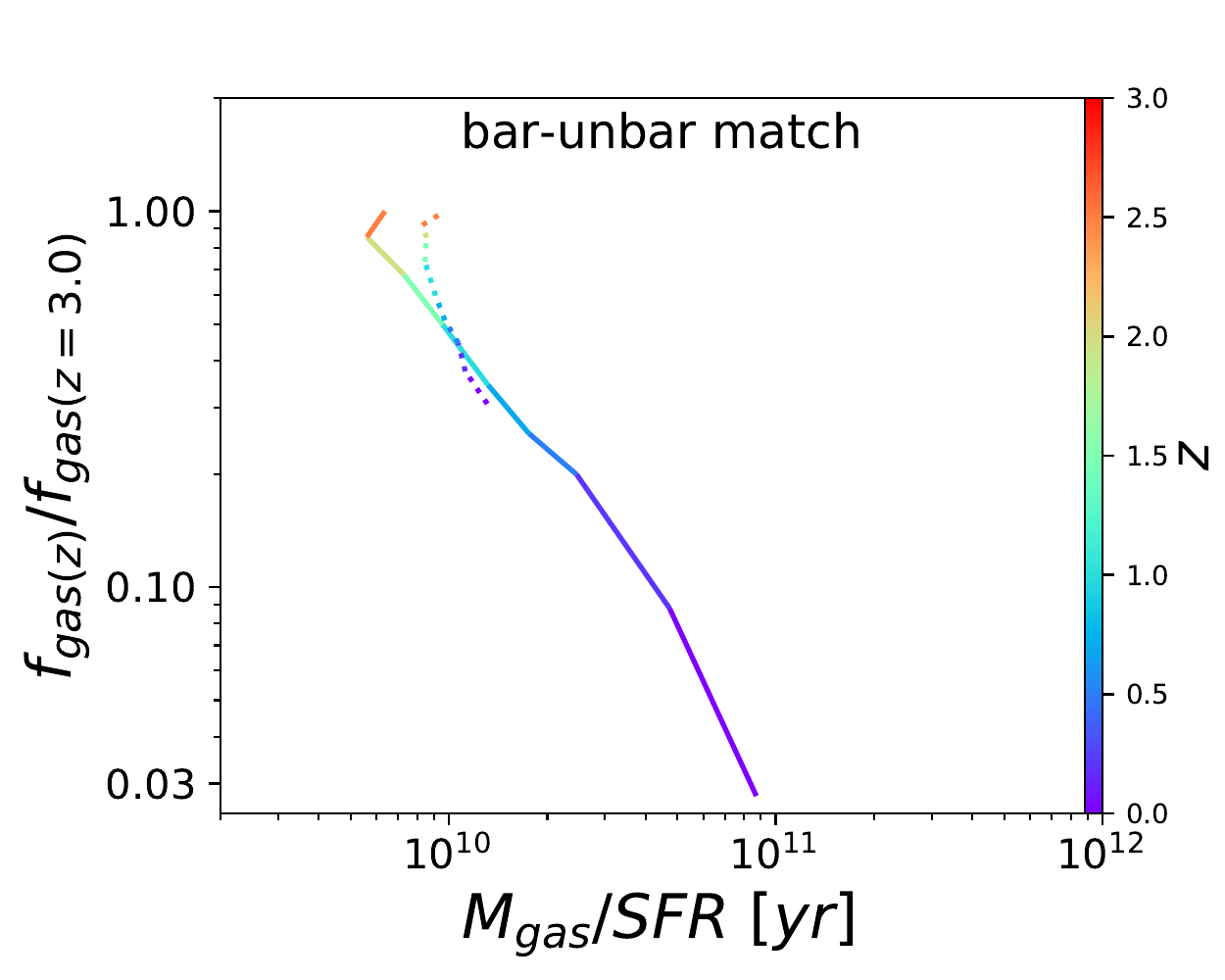}
\includegraphics[width=0.32\textwidth,trim=5 5 15 10,clip]{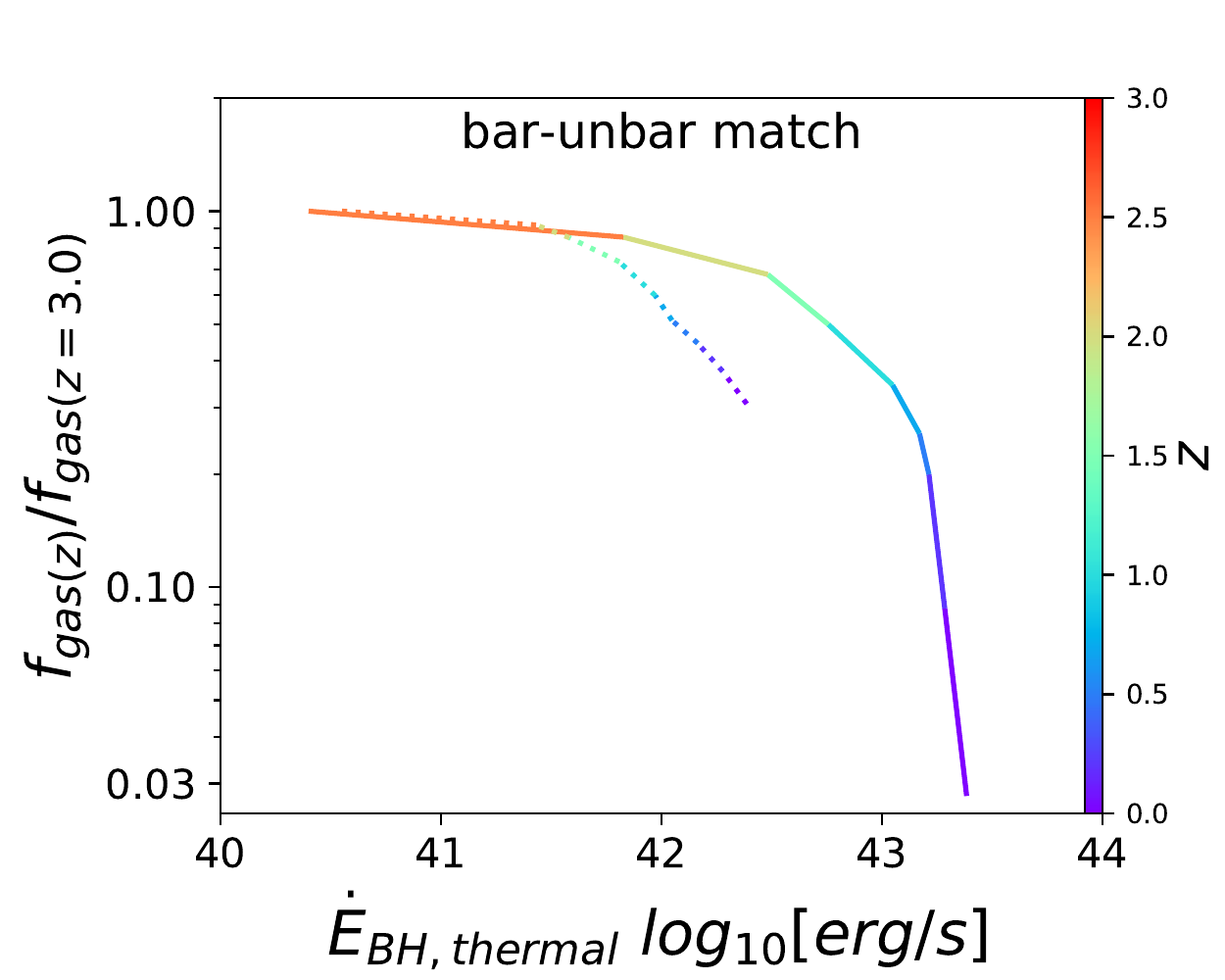}
\includegraphics[width=0.32\textwidth,trim=5 5 15 10,clip]{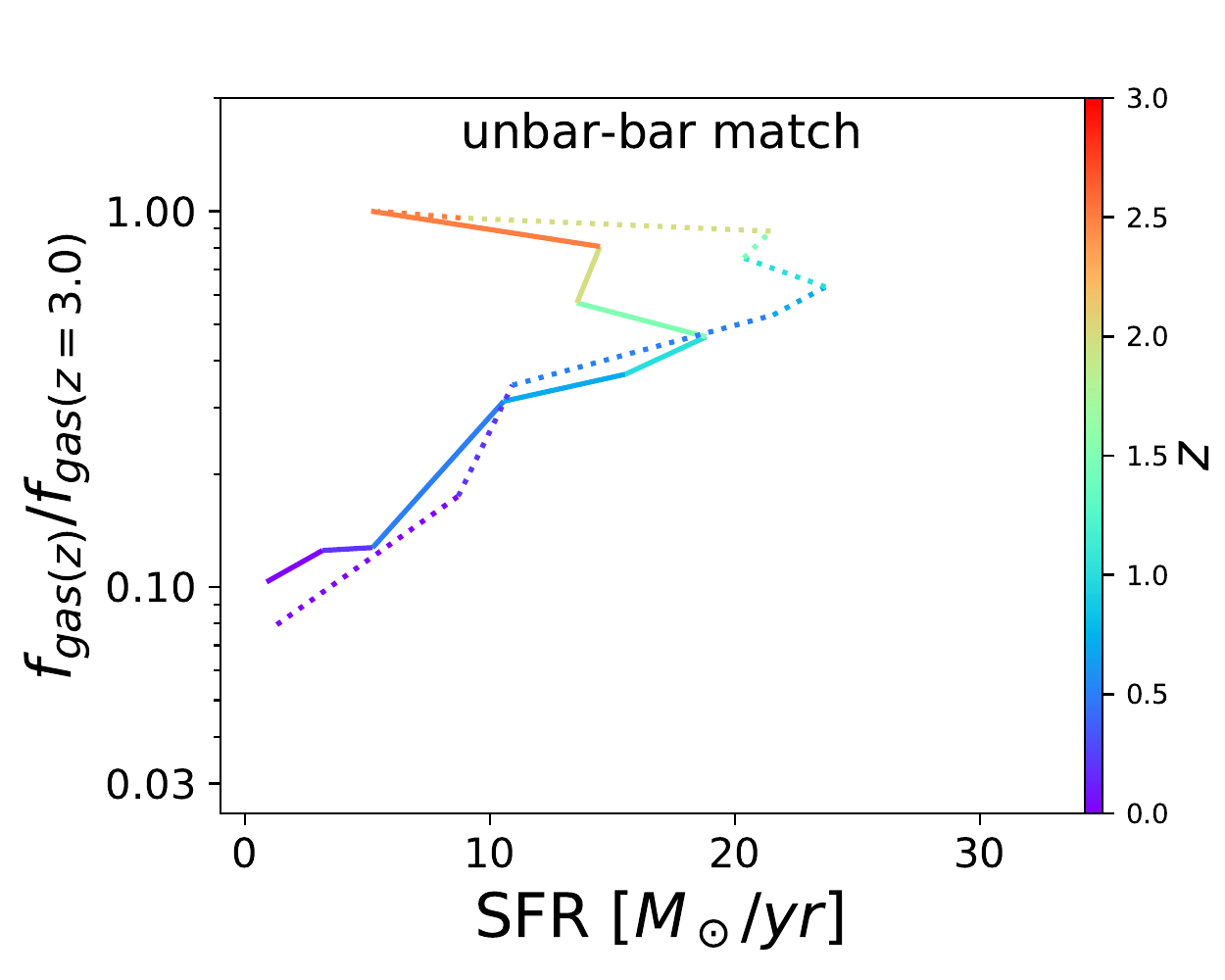}
\includegraphics[width=0.32\textwidth,trim=5 5 15 10,clip]{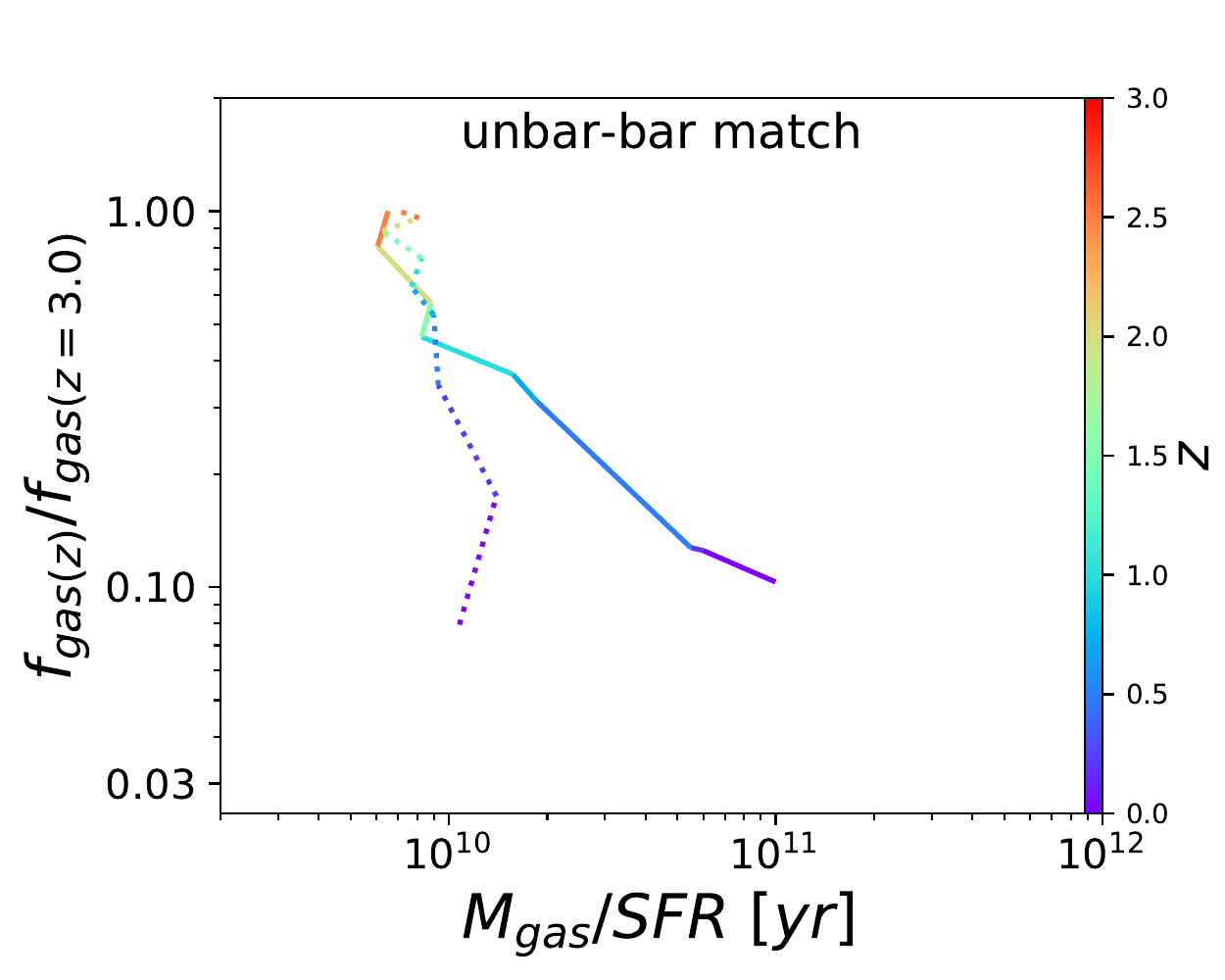}
\includegraphics[width=0.32\textwidth,trim=5 5 15 10,clip]{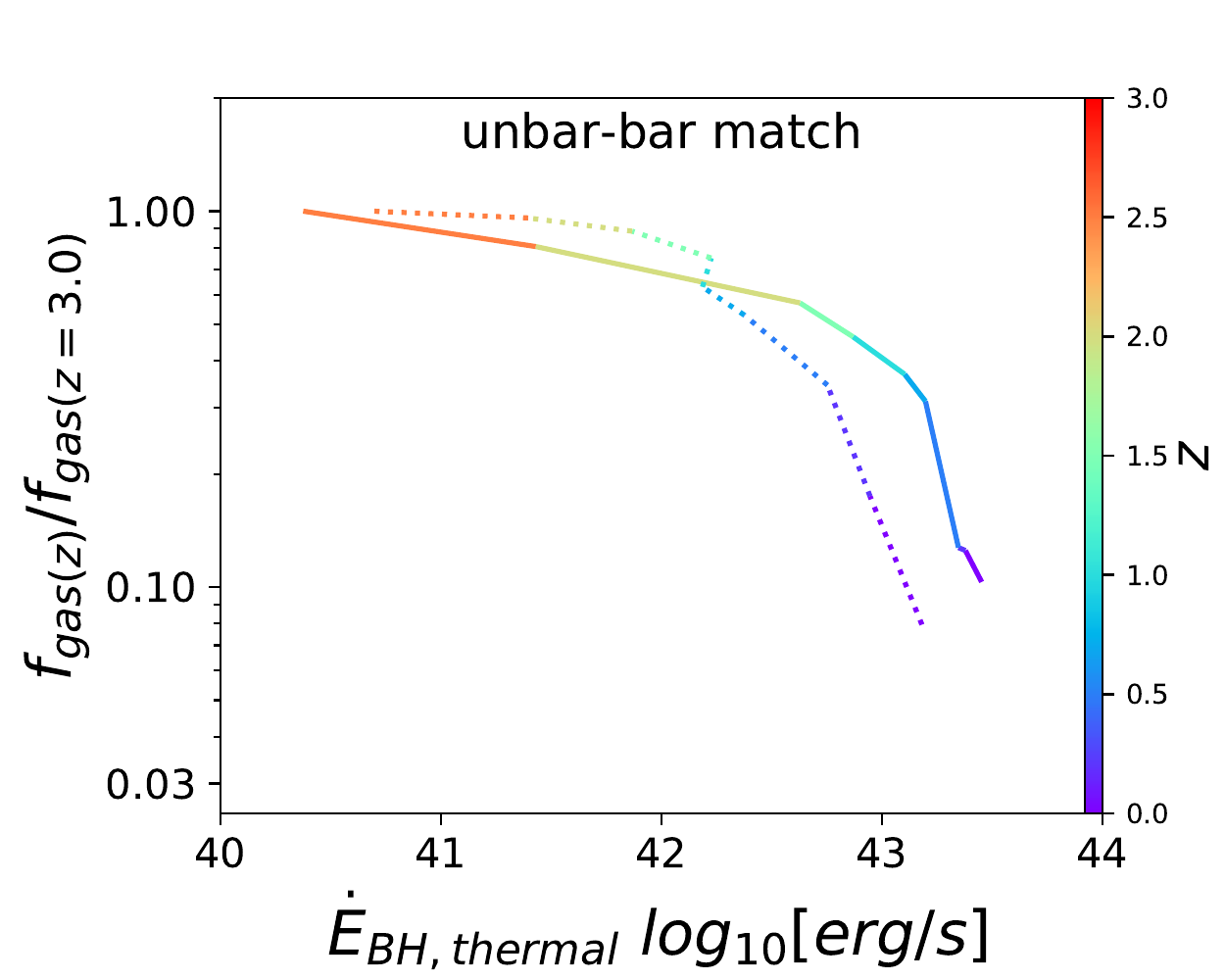}
\includegraphics[width=0.32\textwidth,trim=5 5 15 10,clip]{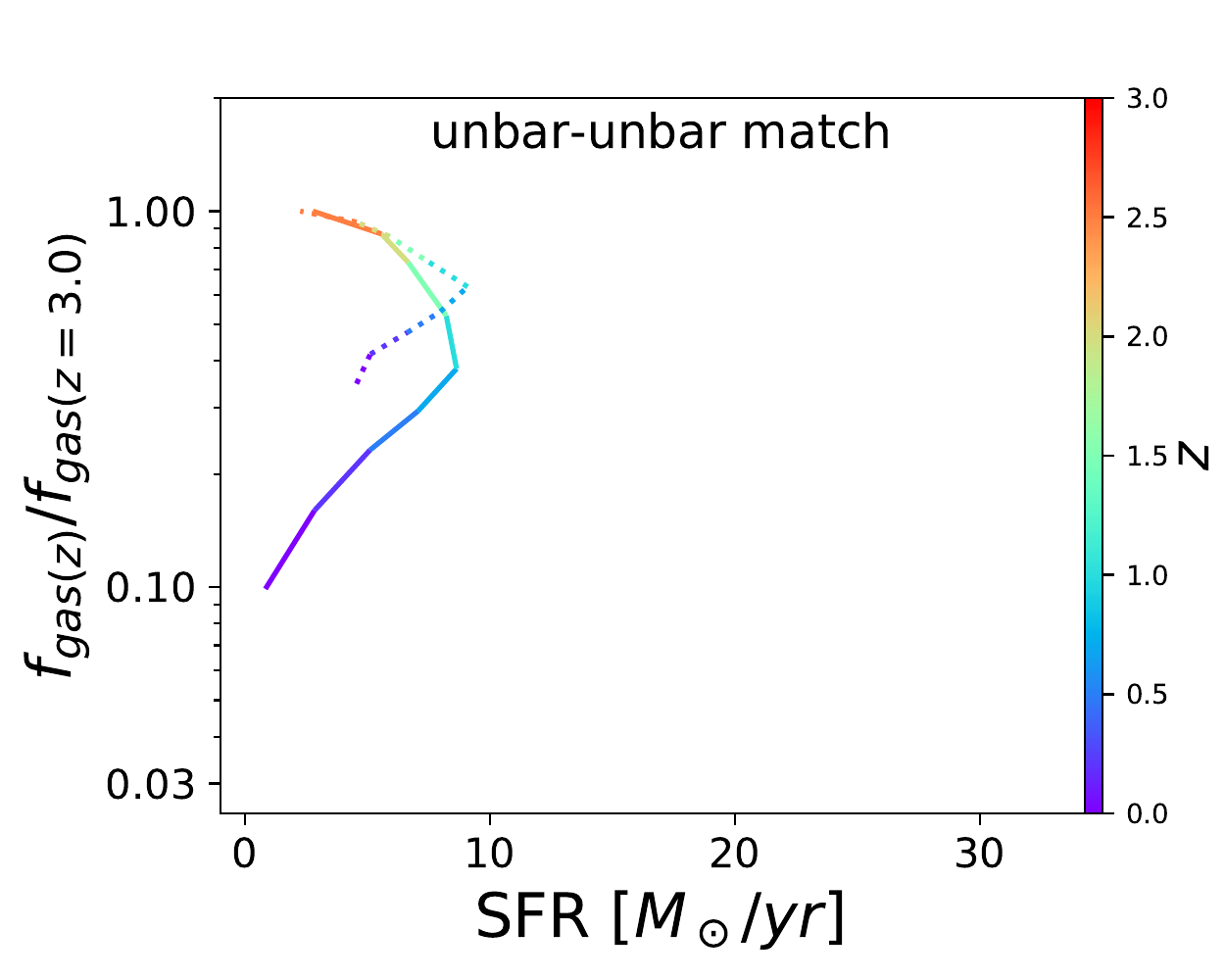}
\includegraphics[width=0.32\textwidth,trim=5 5 15 10,clip]{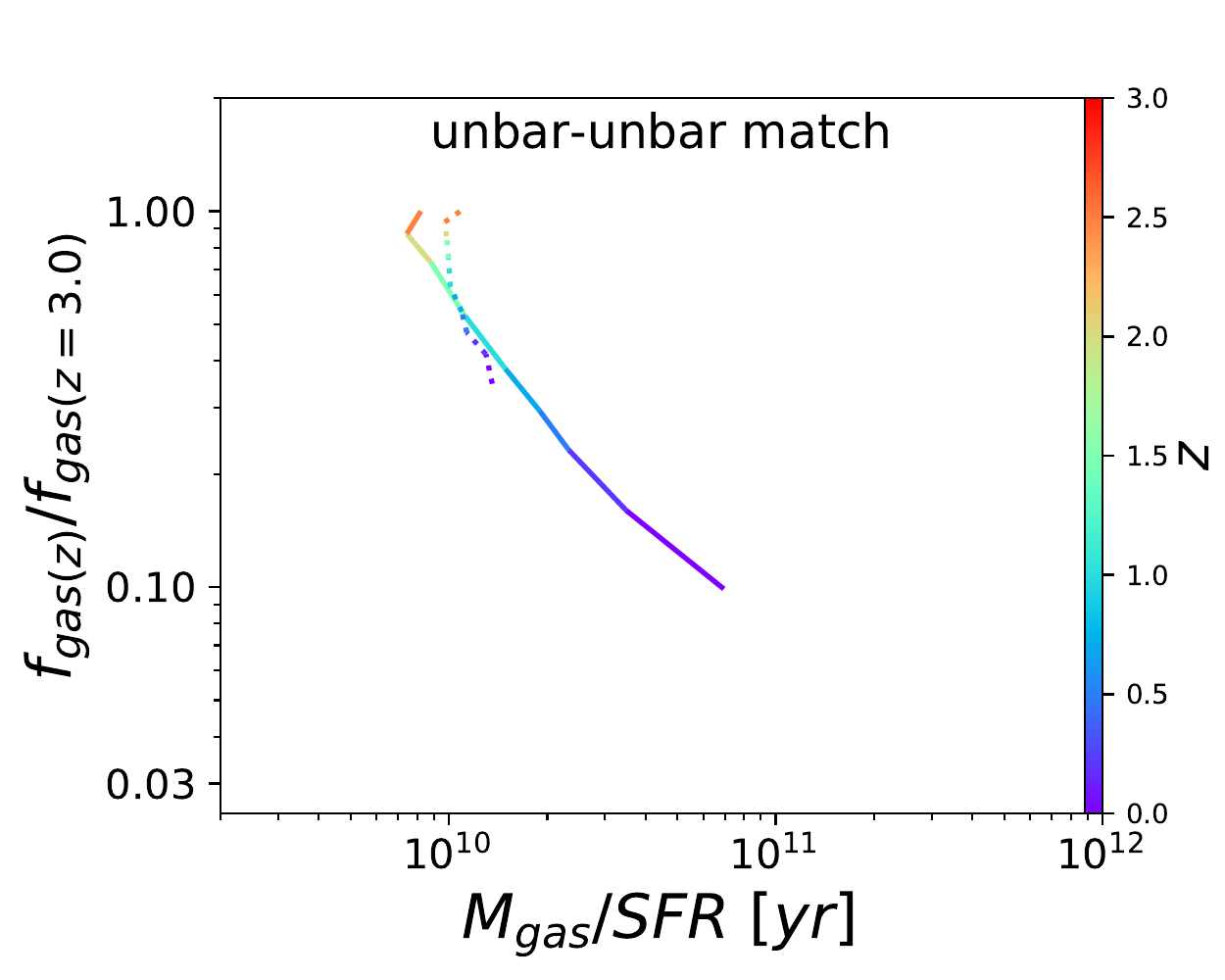}
\includegraphics[width=0.32\textwidth,trim=5 5 15 10,clip]{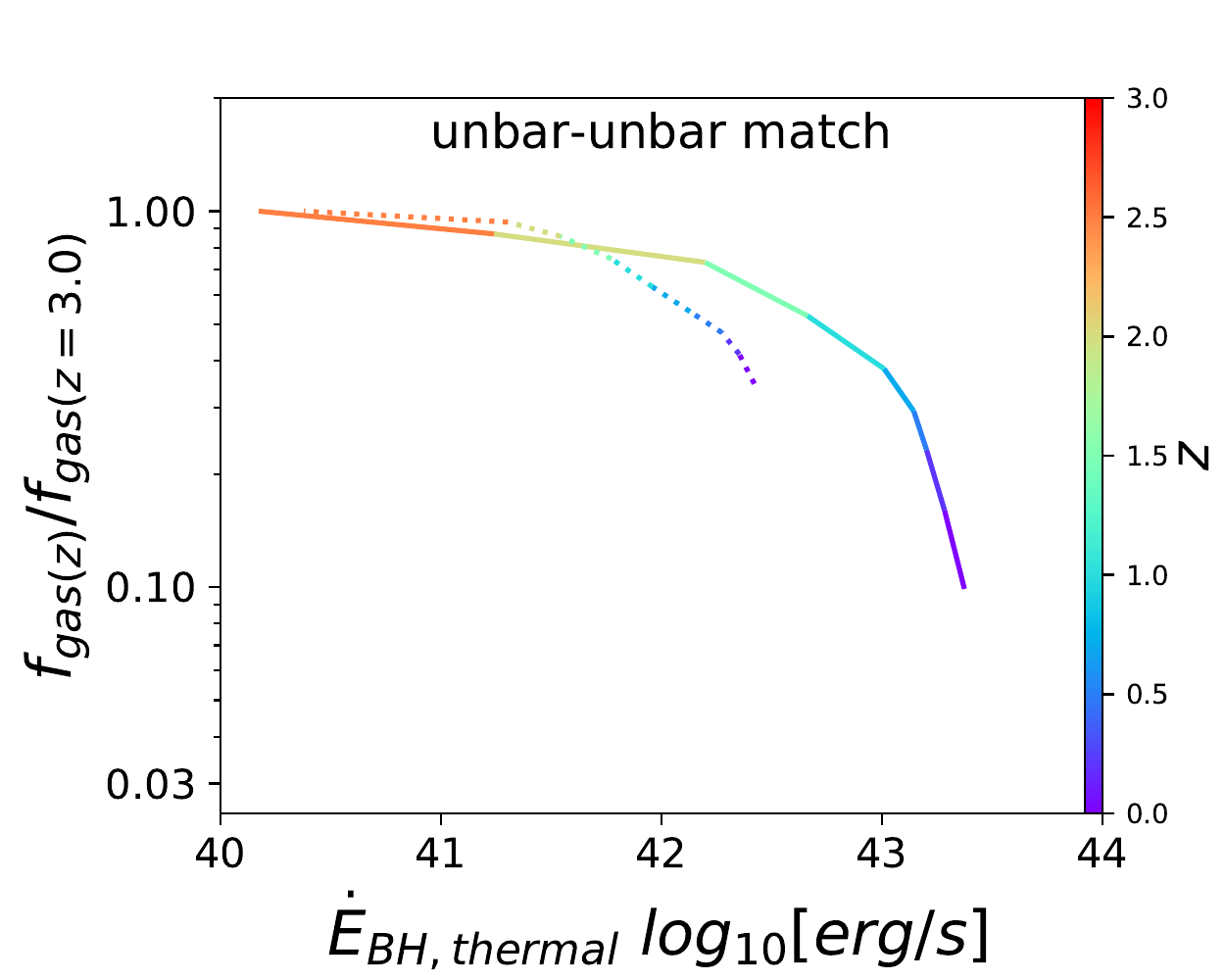}
\caption{The same as Fig.~\ref{fig:fgas_drop_driving}, but shows results of matched galaxies pairs hosted by analogue halos in four categories: bar-bar, bar-unbar, unbar-bar, and unbar-unbar(from top to bottom). The tag before and after dash in the category name indicate the morphology of galaxies in TNG100 and Illustris-1 respectively.}
\end{center}
\label{fig:fgas_drop_match}
\end{figure*}

We also show the decline of normalized gas fraction since $z=3$ against star formation rate, star formation efficiency, and thermal AGN feedback energy rate in the four matched categories as displayed in Fig.~\ref{fig:fgas_drop_match}. The differences between pairs in four categories again are similar with the overall difference between two simulations shown in section 4.3. The star formation efficiency of massive disc galaxies in TNG100 are comparable to or higher than that in Illustris-1 at high redshifts, and are suppressed more effectively since $z=2$. Coincidentally, the gas fraction in TNG100 decreases more significantly and the BH mass accretion rate in all the TNG disc galaxies are relatively higher at $z<\sim2$. The bar-unbar category is a good example of these differences.

The similarity among each category on the divergence of galaxy properties across two simulation also suggests that the morphology of individual galaxy is a combination result of environment and internal multiple baryonic physics, and is often not predictable. The higher bar fraction in TNG100 disc galaxies benefits from more favorable conditions for bar growth, such as the lower gas fraction in the discs, which may result from the combination of enhanced star formation and stellar feedback efficiency at high redshifts, as well as stronger AGN feedback at $z<2$.

\section{Summary and Discussions}
In this work, we have carried out a comparison study on the bar structure in the simulations Illustris-1 and TNG100. Based on methods described in the literature, we identify 1269 and 1232 disc galaxies with stellar mass more massive than $10^{10.5}M_{\odot}$ in TNG100 and Illustris-1 respectively. By examining bar structures in these galaxies, we find a much higher bar fraction in TNG100, and further study the correlation between bar structures and galaxies properties including the gas component, star formation, AGN feedback and dark matter halo. This attempt is to understand the underlying baryonic physics that may have led to the different bar frequency in the two simulations. We summarize our major findings as follow:

\begin{enumerate}
\item At redshift $z=0$, the overall bar fraction is $55\%$ and $8.9\%$ for galaxies with stellar mass $M_*>10^{10.5}M_{\odot}$ in TNG100 and Illustris-1, i.e. a number of 698 and 110 barred galaxies, respectively. In TNG100, bar fraction grows from $\sim 30\%$ at $M_*=10^{10.5}M_{\odot}$ to $\sim 50\%$ at $M_*=10^{10.75}M_{\odot}$, and keeps flat to around $M_*=10^{11.25}M_{\odot}$. In the stellar mass range $M_*=10^{10.66-11.25}M_{\odot}$, the bar fraction in TNG100 agrees well with the results of local survey $S^4G$ \citep{2016A&A...587A.160D}. In Illustris-1, bar fraction grows from $\sim 0\%$ at $M_*=10^{10.5}M_{\odot}$ to $\sim 10 \%$ at $M_*=10^{11.0}M_{\odot}$, and then to $\sim 30-40\%$ for $M_*>10^{11.25}M_{\odot}$. In Illustris-1, merge, $48.2\%$, $40.9\%$ and $10.9\%$ of the bars are associated with merge, flyby and secular evolution respectively. The corresponding fractions in TNG100 are $57.3\%$, $17.0\%$,and $25.7\%$ respectively. The median redshifts of bar formation is $z\sim 0.4-0.5$ in TNG100,  and is $z\sim 0.25$ in Illustris-1. 

\item In both simulations, bars are much easier to form in galaxies with less gas in the disc. Namely, the bar fraction increases as the gas fraction decreases. At $z=0$, the disc galaxies in TNG100 generally have much lower gas fraction than those in Illustris-1. This systematic discrepancy can be traced back to a redshift as high as $z\sim 3$. Moreover, if a z=0 barred galaxy had a higher gas fraction at high redshift, it will tend to form a bar later, which holds for both simulations. For most of the barred galaxies, their disc gas fractions were lower than $0.4$ at the time their bar were just formed.  

\item  The star formation efficiency in TNG100 disc galaxies is higher than that in Illustris-1 at $z>\sim 1.5-2.0$, but the situation is reversed thereafter. The thermal AGN feedback energy injected into disc galaxies in TNG100 is lower than in Illustris-1 at $z>2$, however becomes higher at $z<\sim2$, when the gas fraction in TNG100 disc galaxies drop more significantly than in Illustris-1. In addition, the kinetic feedback in TNG100 increase rapidly at $z<1.5$, which is not implemented in Illustris-1. In each of the two simulations, barred galaxies have relatively higher star formation rate and efficiency than unbarred galaxies before the median redshift of bar formation. Once bars are formed, the star formation rate and efficiency in barred galaxies decline more significantly. Also, the AGN feedback in barred galaxies is enhanced with respect to unbarred galaxies in both simulations.

\item By and large, the properties of halos hosting disc galaxies are found to be comparable in the two simulations, and have little contributions to the discrepancy of bar frequency. The mass of host dark matter halos of barred galaxies in TNG100 are similar to those of unbarred galaxies in TNG100 and Illustris-1, but are less massive than barred galaxies in Illustris-1, this is mainly because that the latter have higher stellar mass. There is little difference between the two simulations on the concentration of halos hosting disc galaxies. All these halos in the two simulations are quite rounder. There are only slight difference on halo shape between the two simulations, and further between the barred and unbarred galaxies. 

\item  A large fraction of galaxies pairs hosted by analogue halos i.e., have the similar initial conditions and evolution environment, across two simulations can have striking different morphology at $z=0$. The morphology of individual galaxies are subject to combined effects of environment and internal baryonic physics, and are often not predictable. Based on the morphological types in TNG100 and Illustris-1, the matched galaxies pairs are divided into four sub categories, bar-bar, bar-unbar, unbar-bar, unbar-unbar. We find the differences in star formation rate and AGN feedback between samples in each categories are similar with the overall differences between TNG100 and Illustris-1. The bar (TNG100) - unbar (Illustris-1) category, which contributes significantly to discrepancy in bar fraction, showing evident difference on AGN feedback between two simulations.

\end{enumerate}

Our results about the bar structure in TNG100 and Illustris-1, including the bar fraction and formation times, agree with previous investigation reported in \cite{2019MNRAS.483.2721P} and \cite{2020MNRAS.491.2547R}. The number of our samples are larger than \cite{2020MNRAS.491.2547R} as we study samples in a wider range of stellar mass of galaxies and use different sample selection method. The star formation rate in barred galaxies is found to decline more significantly than unbarred galaxies after bar formation, which also agrees with \cite{2020MNRAS.491.2547R} and the previous observational study(e.g. \citealt{2015A&A...580A.116G}). However, this should not be taken as direct evidence as bar quenching, considering that the star formation rate in barred galaxies is higher than in unbarred galaxies before bar formation. The trend that bars are more common in more massive and gas poor galaxies is consistent with the findings of \cite{2019MNRAS.483.2721P} and \cite{2020MNRAS.491.2547R}. Furthermore, the result that a high level of gas fraction will suppress the growth of bar agrees well with result of idealized isolate simulations(e.g., e.g., \citealt{2004MNRAS.347..220B}; \citealt{2005MNRAS.364L..18B}; \citealt{2007ApJ...666..189B}; \citealt{2013MNRAS.429.1949A} ). 

The trend that massive and gas poor disc galaxies are more favourable to host bars is in line with findings in many observations(e.g. \citealt{2011MNRAS.411.2026M}, \citealt{2013ApJ...779..162C}, \citealt{2015A&A...580A.116G}, \citealt{2017ApJ...835...80C}, \citealt{2020MNRAS.492.4697N}), However, we note that there is a divergence of views on this point between different observations. There are still some studies implying the opposite picture(e.g.\citealt{2008ApJ...675.1194B},\citealt{2010ApJS..186..427N}). More recently, \cite{2018MNRAS.474.5372E} reports that the bar frequency peaks at $M_*=10^{9.7}M_{\odot}$, and shows barely any dependence on gas content, using samples of the Spitzer Survey of Stellar Structure in Galaxies. On the other hand, factors including halo properties seems to have minor contributions to the discrepancy on the overall bar fraction in the two simulations. It's because that these factors are statistically similar for disc galaxies with similar stellar masses across the two simulations. This is not surprising, since the initial conditions are quite similar in these two simulations. However, this result doesn't indicate that the halo is not important on bar growth for galaxies individuals. 

Our investigation indicates that the much higher bar fraction in the TNG100 simulation probably arises from a joint effect of multiple physics that are more favorable for bar formation. Particularly, one factor could be the lower gas fraction in massive disc galaxies, which may result from the more effective stellar and AGN feedback in TNG. Massive disc galaxies in TNG100 have relatively higher star formation efficiency at $z>2$, which could lead to lower gas fraction than that in Illustris-1 at $z\sim 2-3$. At $z<\sim2$, stronger AGN feedback through both thermal and kinetic channels, in combination with stellar feedback, will help the TNG100 disc galaxies to make their gas fraction falling below 0.4 more rapidly than their counterparts in Illustris-1. At $z<\sim2$, the AGN feedback is probably the primary factor that cause the gas fraction of massive disc galaxies in TNG100 decreasing dramatically, given the evolution history of star formation efficiency, and AGN feedback energy in disc galaxies with stellar mass more massive than $10^{10.5}M_{\odot}$ in these two simulations. Actually,  \cite{2018MNRAS.479.4056W} show that for galaxies with $z=0$ stellar mass $M_*>10^{10.5} M_{\odot}$ in TNG300, whenever the AGN feedback, especially the kinetic channel, became a dominant feedback channel at redshift $z\sim 2$, the star formation rate in those galaxies were suppressed significantly.

In addition to lowering down the gas fraction, the more effective stellar and AGN feedback models in TNG100 may be helpful in increasing the bar fraction by other means, such as changing the galaxy size, the bulge to disc ratio, the angular momentum of gas and stellar particles, which make the disc galaxies in TNG100 more favourable for bar growth. 
A careful investigation on the stellar properties and dynamical evolution of the bars in the two simulations, as have been done in the previous isolated and cosmological zoom-in simulations(e.g. \citealt{2002MNRAS.330...35A}, \citealt{2016MNRAS.459.2603B}, \citealt{2019MNRAS.488.1864Z}), is urged in the future to obtain a more direct view of the roles of stellar and AGN feedback on bar structure in those simulations of galaxy formation within cosmic volume.

Last but not least, our work indicates that the bar frequency in TNG100 is in well agreement with the observational result obtained from the Spitzer Survey of Stellar Structure in Galaxies ($S^4G$). With respect to Illustris-1, the TNG models has made a substantial improvement, and our work offers a complement to the literature that showing the TNG models can overcome the main shortcomings of Illustris-1 in confrontation with observations(e.g., \citealt{2018MNRAS.473.4077P}, \citealt{2018MNRAS.475..624N}). On the other hand, both Illustris-1 and TNG100 reproduce the $M_{BH} - M_*$ correlation, but are tighter than the observations, especially in TNG100. This is because the simulated galaxies hosts over-massive black holes (\citealt{2019arXiv191000017L}). This needs further improvement, and may change the bar fraction somewhat. Meanwhile, it is noted that the bar fraction in the EAGLE simulation also agrees with the observations(\citealt{2017MNRAS.469.1054A}), and yet EAGLE adopts a quite different modeling of stellar and AGN feedback comparing with the TNG model. For instance, stellar and AGN feedback are injected only in thermal channel in EAGLE(\citealt{2015MNRAS.446..521S}). Consequently, we still need to be very cautious about the agreement of bar fraction between simulations and observations. More reliable results from observations are expected to ensure proper interpretation and application of simulations results, to make sure the sub-grid physics are implemented appropriately in different simulations and converge with each other.

\section*{Acknowledgements}
We would like to thank the anonymous referee for the helpful
comments and suggestions in improving the manuscript.
This work is supported by 
the Key Program of the National Natural Science Foundation of
China (NFSC) through grant 11733010. W.S.Z. is supported by
the NSFC grant 11673077. W. Y. is supported by the NSFC grant 11803095 and the Fundamental Research Funds for the Central Universities. F.L.L. is supported by the NSFC grant 11851301.





\bibliography{ref}




\end{document}